\documentclass[a4paper,onecolumn,11pt,unpublished,nopdfoutputerror]{quantumarticle}
\usepackage[english]{babel}
\usepackage[T1]{fontenc}
\usepackage{amsmath,amssymb,amsfonts}
\usepackage{amsthm}
\usepackage{dsfont}
\usepackage{array}
\usepackage{tabularx}
\usepackage{booktabs}
\usepackage{enumitem}
\usepackage{float}
\usepackage{algpseudocode}
\floatstyle{ruled}
\newfloat{algorithm}{htbp}{loa}
\floatname{algorithm}{Algorithm}
\floatstyle{plain}
\usepackage{graphicx}
\usepackage[numbers,sort&compress]{natbib}
\usepackage{bibunits}
\usepackage{xurl}
\usepackage[breaklinks]{hyperref}
\usepackage{xcolor}
\usepackage[most]{tcolorbox}
\usepackage{tikz}
\usetikzlibrary{arrows.meta, positioning, fit, backgrounds, calc, matrix, shapes.geometric}

\graphicspath{{./}}

\hypersetup{colorlinks=true, linkcolor=blue!60!black, citecolor=blue!60!black}

\newcommand{\GF}{\mathbb{F}_2}
\newcommand{\LER}{p_{L}}
\newcommand{\ind}{\mathds{1}}
\DeclareRobustCommand{\Gmodel}[1]{\ensuremath{\mathsf{G_{#1}}}}
\DeclareRobustCommand{\Gsearch}[1]{\ensuremath{\mathsf{G_{#1}^{+}}}}
\DeclareRobustCommand{\Ddata}[1]{\ensuremath{\mathsf{D_{#1}}}}
\newcommand{\eqnref}[1]{Eq.~\eqref{#1}}
\newcommand{\LPtwenty}{\ensuremath{\mathrm{LP}_{20}^{3,5}}}
\newcommand{\LPsixteen}{\ensuremath{\mathrm{LP}_{16}^{3,7}}}
\newcommand{\BBgross}{\ensuremath{\mathrm{BB}_{144}}}

\newcommand{\BBtwogross}{\ensuremath{\mathrm{BB}_{288}}}
\newcommand{\GBeighteen}{\ensuremath{\mathrm{GB}_{180}}}
\newcommand{\GBnine}{\ensuremath{\mathrm{GB}_{900}}}
\newcommand{\M}{\mathrm{Mitten}}

\newcommand{\nn}[1]{\mathsf{#1}}

\newcommand{\pin}{\mu}
\newcommand{\pinf}{\mathrm{free}}

\newcolumntype{L}[1]{>{\raggedright\arraybackslash}p{#1}}

\makeatletter
\newcommand{\bibnamespace}[1]{%
  \def\@extra@b@citeb{.#1}%
  \def\@extra@binfo{.#1}%
}
\makeatother

\newtcolorbox{keyidea}{
  colback=blue!3!white,
  colframe=blue!35!black,
  fonttitle=\bfseries,
  title=Key idea,
  boxrule=0.4pt,
  arc=2pt,
  left=5pt,
  right=5pt,
  top=4pt,
  bottom=4pt,
}

\newtcolorbox{slogan}{
  colback=gray!5!white,
  colframe=gray!50!black,
  fonttitle=\bfseries,
  title=Slogan,
  boxrule=0.4pt,
  arc=2pt,
  left=5pt,
  right=5pt,
  top=4pt,
  bottom=4pt,
}

\theoremstyle{plain}

\theoremstyle{definition}

\begin{document}

\title{Distilling Knowledge from Search:
  Decoder Model Factory for Quantum LDPC Codes}
\date{}
\author[edencode]{Anil Bilgin}
\author[edencode]{Wanda Hou}
\author[edencode]{Pranav Srinivas}
\author[edencode,ucsd]{Yi-Zhuang You}
\affiliation[edencode]{EdenCode Inc., San Jose, CA 95134, USA}
\affiliation[ucsd]{Department of Physics, University of California, San Diego, CA 92093, USA}

\maketitle

\begin{abstract}
\noindent
We introduce GraphMP, a learning-based message-passing decoder for quantum low-density parity-check codes that transfers without fine-tuning across code families, code sizes, physical error rates, and noise models. GraphMP uses local message passing to keep computation scalable as the decoding graph grows. Guided by its predictions, beam search progressively refines candidate error patterns, trading additional decoding time for lower logical error rates. We develop the decoder model factory as a paradigm for training successive models on experience generated by earlier models during decoding. Model-guided search supplies verified trajectories for knowledge distillation, folding computation spent on decoding into reusable training data. Repeated collection and distillation improve the decoder across generations and adapt its strategy as the noise changes. These results motivate a coupled quantum--classical architecture with two interacting cycles: rapid decoding protects quantum information, while slower learning renews the models from operating experience. The long-term vision is a scalable decoder model factory that produces high-performance, low-latency decoders and continually adapts them to evolving quantum hardware.
\end{abstract}

\bibnamespace{main}
\begin{bibunit}[unsrturl]

\section{Introduction}
\label{sec:introduction}

\begin{figure}[!htbp]
  \centering
  \includegraphics[width=0.89\textwidth]{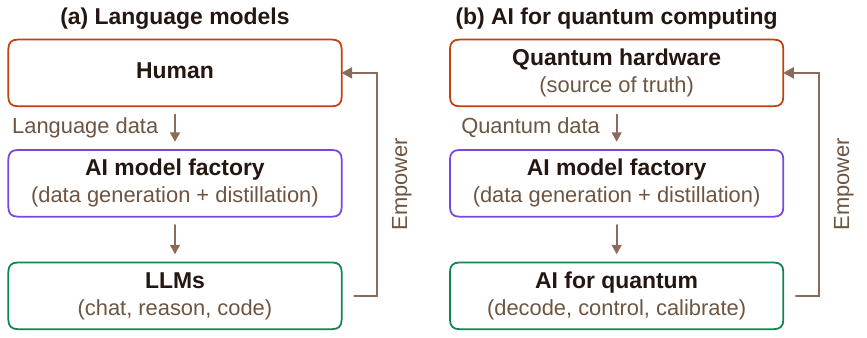}
  \caption{AI model factories connect a source of training data
    to new models and their applications. The language-model
    analogy motivates a corresponding process for quantum computing,
    with quantum hardware providing experimentally testable outcomes.
    The return arrows represent model assistance to humans or quantum hardware.
    In the illustrated approach, data generation supplies training examples, and distillation
    trains later models on these examples.
    This paper develops the decoding instance
    through supervised distillation from search trajectories, with
    outcome verification supplied by simulation; hardware learning,
    control, and calibration are prospective applications.}
  \label{fig:main-paradigm}
\end{figure}

Advances in quantum hardware allow repeated experiments to
generate large volumes of measurement data~\cite{google2025belowthreshold}.
How can artificial intelligence (AI) learn from this quantum big data
to help quantum devices operate more reliably and efficiently? The
training of large language models
(LLMs) offers inspiration for how to make use of big data.
LLMs are first trained on collections of language data and can then
be refined using human feedback~\cite{ouyang2022}. An evaluator can
also check text generated by the model, with verified outputs added
to the training data~\cite{zelikman2022,singh2023restem,guo2025}.
Learning from these additions can help the model produce higher-quality
data for the next round of training.
This \emph{recursive self-improvement} cycle~\cite{zelikman2022,singh2023restem} suggests that existing models can help generate data to train their successors, enabling an \emph{AI model factory} that continually produces new generations of models. For quantum computing, AI models
could learn from experimental measurements to support
error-correction decoding~\cite{bausch2024}, quantum control~\cite{reuer2023},
and device calibration
\cite{moon2020} (Figure~\ref{fig:main-paradigm}). As models perform
these tasks, they may also generate data that can train later
models and support continued improvement.

To make the discussion concrete, we focus on quantum error correction,
where a decoder infers an error pattern from the measured syndrome
and determines a recovery operation.
The optimal decoding strategy depends on the code structure and on
the probability distribution of error patterns. In operating quantum devices, noise
can be difficult to characterize accurately, and error probabilities
may drift over time~\cite{huo2017noise,spitz2018adaptive,bhardwaj2026drift}.
Learning from data accumulated during decoding offers a way to adapt
the strategy to changing device noise.

Motivated by the above ideas, we propose a \emph{decoder model factory}:
a systematic process that
trains new decoder models on data accumulated as the current decoder
performs decoding tasks. In our implementation, a search algorithm
lets the decoder refine its prediction over several steps.
An independent evaluator uses experimental outcomes (or simulation
feedback, as demonstrated here) to verify the decoder's prediction
~\cite{derks2025dem}. We train the next model on verified search
trajectories. The updated model is then redeployed for decoding,
generating new trajectories for further learning. This
cycle provides a route to decoding strategies that improve through
repeated training rounds and adapt as the noise encountered during
operation changes.

Our design also draws on several related ideas from the literature. AlphaQubit and recurrent-transformer decoders infer logical errors from observed syndromes~\cite{bausch2024,blue2026}, while neural pre-decoders assist conventional decoders through first-pass corrections~\cite{chamberland2026predecoder,miao2026qadapt}. Transfer learning has been demonstrated by Astra, which shows a frozen model can decode codes of different sizes within a code family~\cite{astra2024}, and explored through pretrained initialization at larger distances by foundation decoders~\cite{yan2026foundation}. AlphaQubit and QAdapt further address the adaptivity of device noise through processor-based fine-tuning and training across changing noise conditions~\cite{bausch2024,miao2026qadapt}. These developments motivate us to build a decoder that infers error patterns from syndromes, transfers across decoding tasks without fine-tuning, and adapts its strategy as device noise changes.

In our approach, the decoder is embedded in a search algorithm and repeatedly called to refine candidate error predictions. An independent
evaluator checks each final prediction against the logical outcome;
the successful search trajectory then supply
training examples for the next model. This connects to expert iteration,
where search results are used to train improved models~\cite{anthony2017}, and
to GNN-assisted BP, which uses its decoder to collect harder training
examples~\cite{gong2024}. Our approach uses these verified trajectories for knowledge distillation, aiming to improve the decoder's ability to
guide later searches.

\begin{figure}[!htbp]
  \centering
  \includegraphics[width=0.90\textwidth]{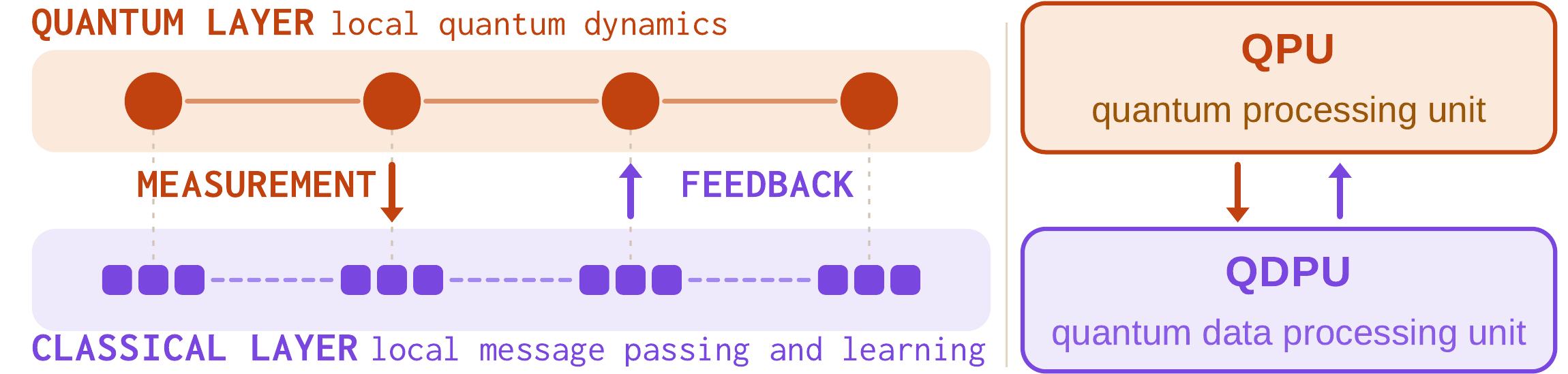}
  \caption{Proposed coupled quantum--classical deployment of a local learned
    decoder. Circles represent qubits and square groups represent local
    classical memory. Solid and dashed links indicate quantum couplings
    and classical message passing; vertical arrows carry measurement data
    and recovery feedback between the layers.}
  \label{fig:main-architecture}
\end{figure}

Inspired by local active error correction~\cite{lake2025}, we envision the coupled quantum and classical architecture in Figure~\ref{fig:main-architecture}. Local measurement and feedback couple qubits to classical bits, while neighboring classical processors exchange messages to guide error correction. The architecture supports two interacting cycles on different time scales~\cite{gong2026selfcalibration}: a fast decoding cycle uses the learned decoder model to perform active error correction, while a slower training cycle learns from verified decoding outcomes and supplies updated models for deployment. This arrangement resembles an immune system that learns through its fight against disease: rapid decoding actively combats errors, while experience gained in that fight trains future models to defend quantum information more effectively.

A hardware realization would pair a quantum processing unit (QPU), which executes quantum operations, with a quantum data processing unit (QDPU), which processes measurement data and returns feedback for decoding and control. A companion paper develops a QDPU architecture for neutral-atom quantum computers, with inference on a field-programmable gate array (FPGA) and a slower graphics processing unit (GPU) loop for model updates~\cite{bilginQDPU}. GPU-based decoding offers another deployment path, with FPGAs handling quantum control and readout. NVIDIA NVQLink provides an architecture for coupling GPUs to quantum system controllers, with CUDA-Q coordinating computation and data exchange~\cite{caldwell2025nvqlink}. Together, these components suggest a hardware blueprint for the decoder model factory: a quantum--classical system that couples real-time error correction with ongoing learning from verified outcomes.

The remainder of the paper is organized as follows.
Section~\ref{sec:decoder} introduces GraphMP and its message-passing
decoder. Section~\ref{sec:search} explains how beam search refines its
predictions, and Section~\ref{sec:factory} describes how search
trajectories are verified and distilled into updated models.
Section~\ref{sec:results} evaluates decoding performance, transfer
across tasks, and noise adaptation. Section~\ref{sec:outlook} concludes
with the implications and next steps for device-based learning and
deployment.

\section{GraphMP: A message-passing decoder}
\label{sec:decoder}

The goal of the research is to build a foundational decoder: a single learned decoding strategy
that can be applied to different codes and noise models without
retraining for every task. We focus on quantum low density parity check (qLDPC) codes~\cite{ninkovic2024}.
These codes comprise several families with different code structures, including
hypergraph-product~\cite{tillich2014}, generalized-bicycle~\cite{panteleev2021},
bivariate-bicycle~\cite{bravyi2024}, and
lifted-product codes~\cite{panteleev2022,huang2026},
including the mitten family~\cite{bhardwaj2026}.
Different block sizes, noise models, and measurement schedules
generate further decoding tasks. The code families and instances are
summarized in Appendix~\ref{app:task} (Table~\ref{tab:codes}).

\begin{figure}[!htbp]
  \centering
  \includegraphics[width=0.76\textwidth]{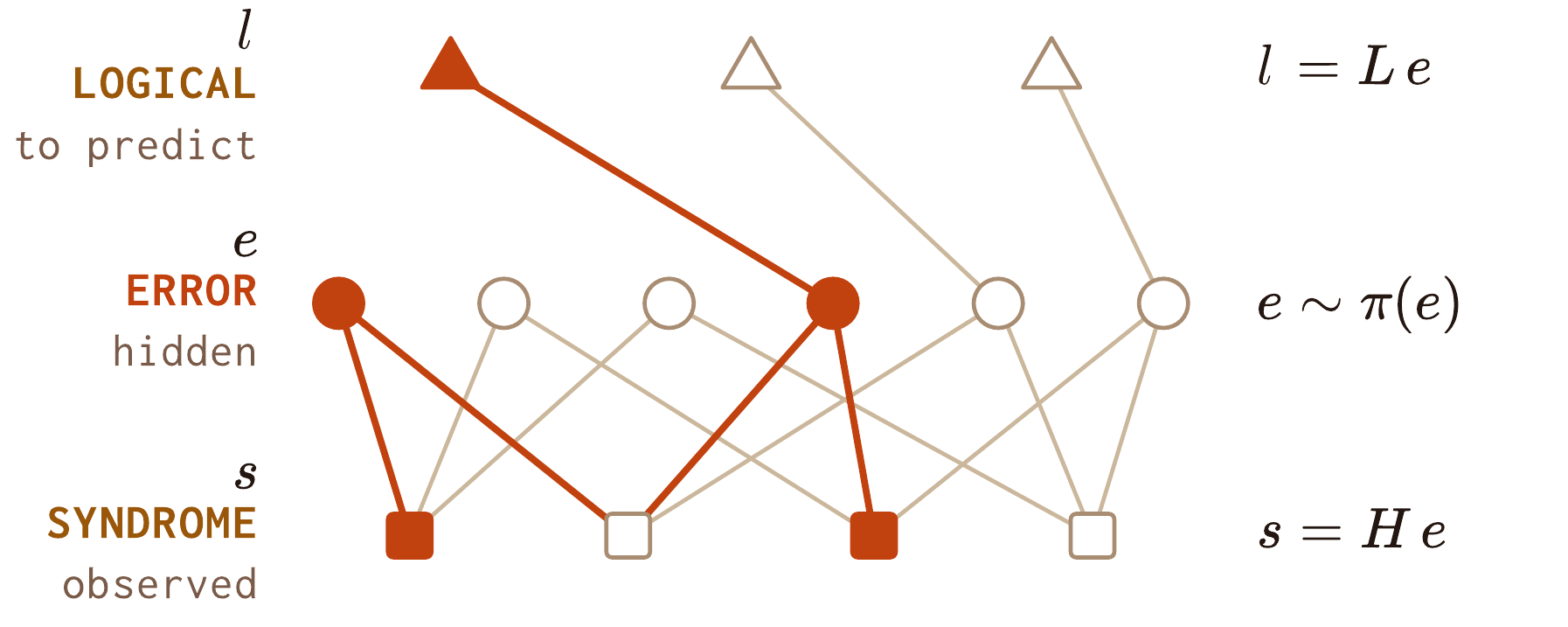}
  \caption{Detector error model. Hidden error events connect to the
    syndrome bits and logical observables they flip. The decoder sees
    the syndrome and predicts an error pattern. Highlighted nodes
    illustrate one realization and its syndrome and logical outcome.}
  \label{fig:main-dem}
\end{figure}

The detector error model (DEM) describes how hidden fault events produce observed detector responses and affect logical observables~\cite{gidney2021,derks2025dem}. The detector matrix $H$ records which detectors each event flips, the logical matrix $L$ records which logical observables it flips, and the prior $\pi(e)$ specifies the probability of each error pattern. For a binary error vector $e$ indicating which events occurred, the observed syndrome is $s=He$ and the logical error is $l=Le$, with arithmetic modulo $2$ (Figure~\ref{fig:main-dem}). The framework also accommodates noisy syndrome extraction: detectors formed from repeated measurements capture both data and measurement errors. This common representation provides an interface for training and applying a foundational decoder across codes and noise models. Appendix~\ref{app:task} details the DEM constructions used here.

The essential task for a decoder is to predict the most likely error pattern $\hat e$ from the syndrome $s$.
Different error patterns may explain the same syndrome while having
different logical effects. Matching the observations, $H\hat e=s$,
therefore does not by itself preserve the logical information: the
prediction must also satisfy $L\hat e=Le$. We evaluate
decoding performance using the logical error rate
$\LER=\Pr(L\hat e\ne Le)$, the probability that the prediction and
the actual error differ in their logical effects.

We aim to build a decoder that transfers its learned strategy across
tasks by applying shared local update rules to different decoding graphs.
To do so, we use local message passing over the sparse bipartite graph
of detector and error nodes defined by the matrix $H$. This is a natural
strategy for sparse qLDPC decoding: belief propagation (BP) is an
established approach that exchanges scalar beliefs along graph edges,
combining incoming messages with local constraints and priors
~\cite{liu2019,miao2024,kschischang2001}. GraphMP follows the same
local message-passing principle, but replaces BP's scalar beliefs with
learned multi-channel states and messages.
These richer representations support more flexible rules for processing
and combining information across the graph.

\begin{figure}[!htbp]
  \centering
  \includegraphics[width=0.74\textwidth]{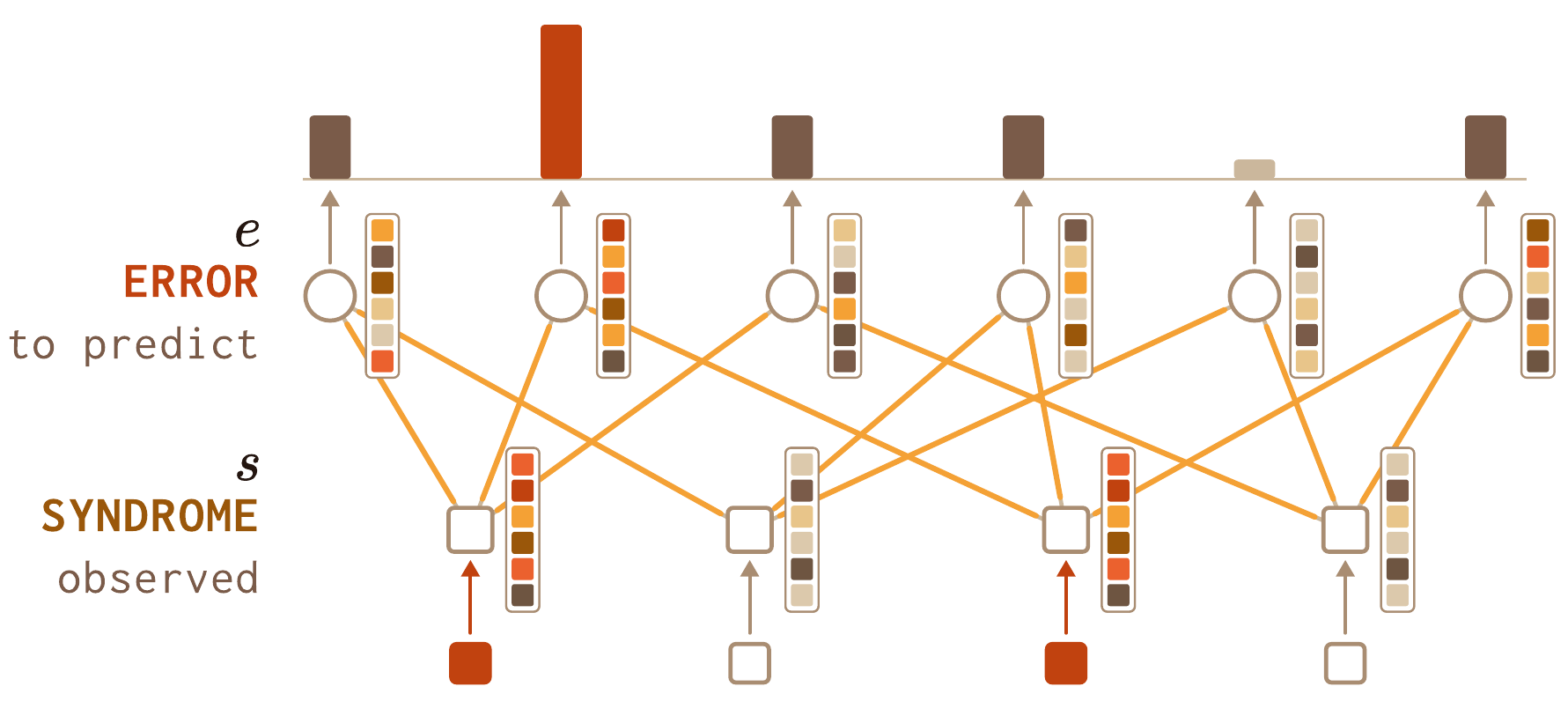}
  \caption{GraphMP predicts errors through local message passing.
    The syndrome enters at detector nodes; shared learned updates
    update node states and messages, and error nodes read out probabilities.
    The same learned functions operate on different decoding graphs. Hidden states
    and logits are illustrative.}
  \label{fig:main-decoder}
\end{figure}

Figure~\ref{fig:main-decoder} illustrates the GraphMP architecture, which applies learned message-passing rules on the Tanner graph connecting check and error nodes. Each syndrome variable in the input $s$ is embedded into a feature vector at the corresponding check node. Messages then pass in both directions along each edge, and each node aggregates its incoming messages to update its state. One forward pass performs several rounds of these updates, followed by a shared readout that maps each error node's final state to an estimated error probability $p(e_i=1\mid s)$. Together, these probabilities define a factorized approximation to the conditional error distribution, $p(e\mid s)\approx\prod_i p(e_i\mid s)$.

With a fixed number of message-passing rounds, syndrome information propagates only over a bounded distance on the decoding graph, so GraphMP remains a local decoder. Node updates within each round can run in parallel, and total computation grows linearly with the number of graph edges. Related architectures have been explored in neural BP and graph-neural-network decoders~\cite{nachmani2016,liu2019,astra2024,ninkovic2024,yan2026foundation}.

The same trained model can then be applied to different decoding tasks just by providing the corresponding decoding graphs, as specified by the DEM. This is the architectural
basis for transferring the trained decoder across code families and noise models.
Its performance on untrained decoding tasks is an empirical question.

To train an initial decoder, we choose code-capacity decoding tasks on several qLDPC codes. For each task, we sample error realizations from its DEM and compute the corresponding syndromes $s$. A BP teacher then provides per-error posterior probability estimates $p^{\mathrm{tch}}(e_i\mid s)$, which serve as soft targets for the GraphMP student. We train GraphMP by minimizing the KL divergence from the teacher's predicted distributions to the student's model predictions. Matching those predictions transfers an existing
inference procedure into learned parameters. This teacher--student
training is \emph{knowledge distillation}~\cite{hinton2015}.

\begin{figure}[!htbp]
  \centering
  \includegraphics[width=\textwidth]{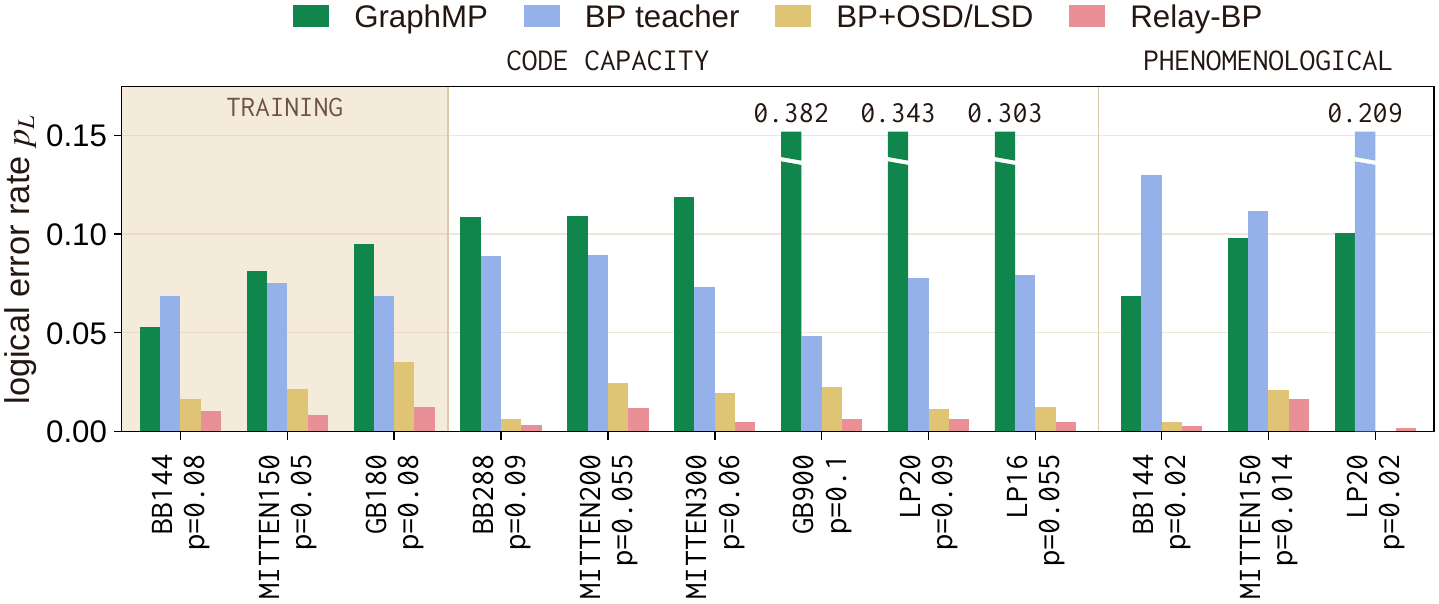}
  \caption{BP distillation supplies a transferable starting decoder,
    while leaving a gap to stronger classical methods. The three
    shaded code-capacity tasks supply training data; the other nine
    evaluate the same model without fine-tuning. Logical error rates
    use $2048$ realizations per task. Bars above $0.15$ are truncated.
    Training and comparison settings are in Appendix~\ref{app:sl}.}
  \label{fig:main-bootstrap}
\end{figure}

We bootstrap GraphMP on three code-capacity tasks, one each from the
bivariate-bicycle, generalized-bicycle, and mitten families, and
apply the trained model to nine unseen tasks without fine-tuning
(Figure~\ref{fig:main-bootstrap}; Appendix~\ref{app:sl}).
GraphMP largely matches its BP teacher, although transfer weakens on
the largest unseen tasks. Matching the teacher fulfills the bootstrap
objective of learning a reusable local inference rule, but
Figure~\ref{fig:main-bootstrap} shows a clear performance gap to
stronger classical decoders such as BP+OSD~\cite{roffe2020},
BP+LSD~\cite{hillmann2025}, and Relay-BP~\cite{muller2025}.

A structural source of this performance gap is the lack of coordination
in single-pass decoding. Both GraphMP and BP assemble their predictions
from separate per-variable probabilities, using the factorized
approximation $p(e\mid s)\approx\prod_i p(e_i\mid s)$.
Although every marginal conditions on the same syndrome, their product
omits the correlations needed to enforce the joint parity constraints,
so predicting each error variable separately does not guarantee
$H\hat e=s$. Further imitation of the BP teacher alone does not address
this limitation. We therefore place GraphMP inside beam search,
using successive model calls conditioned on earlier choices to refine
candidate error patterns. The learned local decoder supplies the
predictions, while search provides the additional coordination.

\section{GraphMP-guided beam search}
\label{sec:search}

We adopt the beam-search decoding framework of
Ref.~\cite{ye2025}, replacing its BP soft decoder with GraphMP.
At each search state, the error variables are divided into two subsets:
pinned variables, whose values are fixed along that path, and free
variables, whose values remain to be determined. Given the observed
syndrome $s$ and the pinned values, GraphMP predicts the free variables
to complete the error pattern $\hat e$.

\begin{figure}[!htbp]
  \centering
  \includegraphics[width=\textwidth]{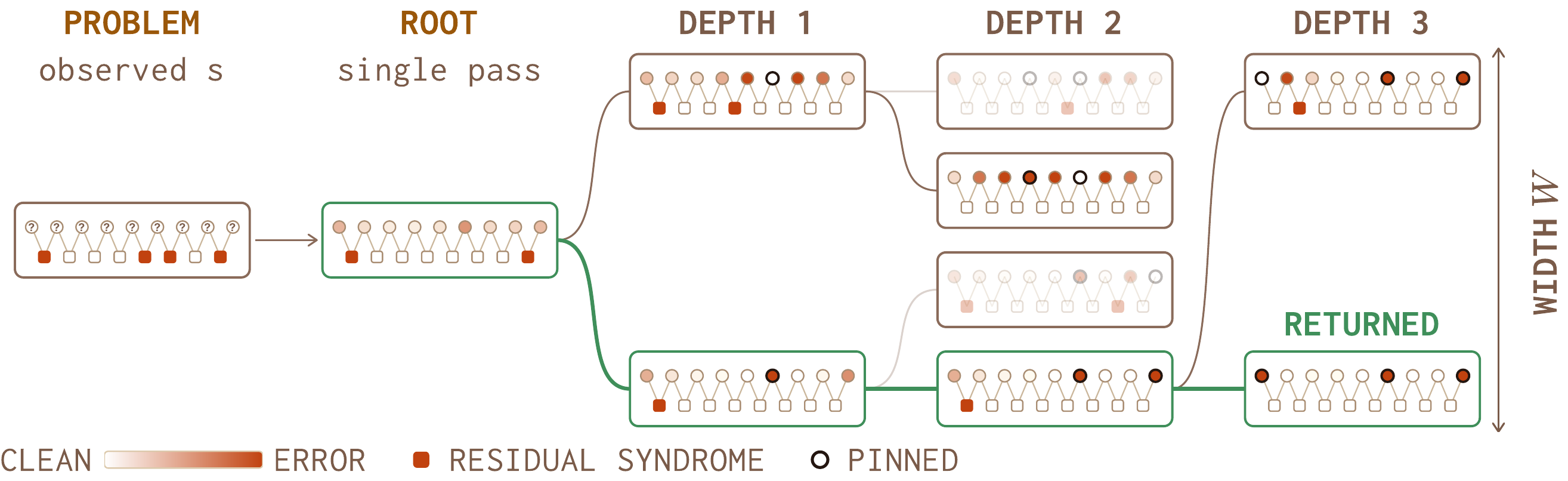}
  \caption{Search turns successive conditional predictions into a
    coordinated error pattern. Each branch fixes an uncertain variable
    and predicts the rest again; pruning bounds the computation.
    The green frame marks the returned completion. This illustrative
    BP-guided tree shows the mechanism; the GraphMP implementation
    is specified in Appendix~\ref{app:beam}.}
  \label{fig:main-search}
\end{figure}

The search begins with all variables free. If the residual syndrome
$s_{\mathrm{res}}:=s\oplus H\hat e$ is nonzero, the algorithm branches
on a least-confident free variable, whose predicted probability is
closest to $1/2$, pinning it to $0$ in one branch and $1$ in the other.
Each branch thus enlarges its pinned subset and removes those
variables' syndrome contribution from the original syndrome $s$.
GraphMP uses this updated syndrome and the pinned values to predict
the remaining free variables. These predictions and the pins together
form a full candidate error pattern, so a syndrome-consistent completion
can be found while some variables remain free. The algorithm retains
promising alternatives and repeats the process
(Figure~\ref{fig:main-search}).

Two budgets control how extensively the algorithm explores these
choices. \emph{Search width} $W$ bounds the number of competing
unfinished histories retained at each step; a wider search preserves
alternatives that an early commitment might discard.
\emph{Search depth} $R$ bounds the number of successive conditional
decisions along a history; a deeper search lets later predictions
respond to more earlier choices. Syndrome-consistent candidates,
$H\hat e=s$, are retained separately and ranked by their error prior. The returned
prediction is selected using syndrome consistency and these rankings, with logical
correctness assessed during evaluation. Appendix~\ref{app:beam}
gives the complete pruning and termination rules.

\begin{figure}[!htbp]
  \centering
  \includegraphics[width=0.86\textwidth]{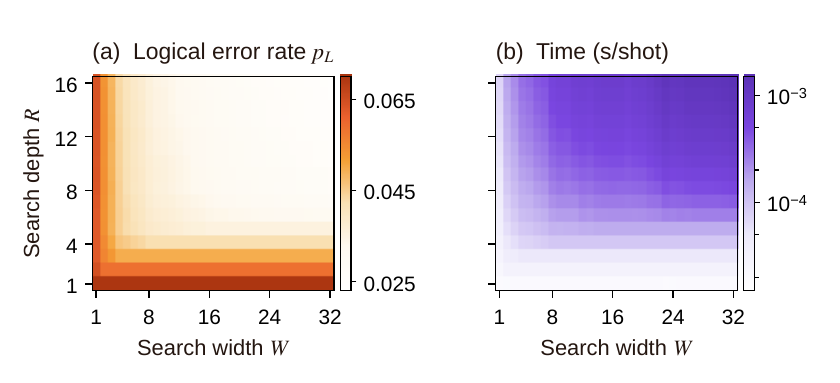}
  \caption{Inference-time search trades computation for accuracy.
    With the starting GraphMP model fixed, broader and deeper search
    lowers logical error while increasing decoding time. The surfaces
    aggregate the three training tasks; measurements and budgets
    are given in Appendix~\ref{app:search-numbers}.}
  \label{fig:main-tradeoff}
\end{figure}

Broader and deeper beam search can further suppress logical errors
without additional model training, at the cost of more decoding time
(Figure~\ref{fig:main-tradeoff}). For the BP-distilled GraphMP model
on the three code-capacity training tasks, a modest search budget
more than halves the pooled logical error rate; larger budgets yield
further gains with diminishing returns
(Appendix~\ref{app:search-numbers}). The same trained model thus
supports fast single-pass decoding or more accurate search-enhanced
decoding when additional computation is available.

Beam search serves two roles: improving the current decoding prediction
and generating training data for future models. Each search trajectory records the
intermediate problems leading to a final predicted error pattern.
During data collection, a syndrome-consistent final prediction that
passes independent logical verification supplies a valid completion
for every problem along its path. These examples can train a successor
model to guide later searches, folding thinking into learning.
When the updated model collects new trajectories for its successors,
this process becomes a decoder model factory: a recurring cycle of
collection, verification, and distillation.

\section{Decoder model factory}
\label{sec:factory}

BP-distilled GraphMP is trained to reproduce its teacher's predictions
from original syndromes in a single pass, with all error variables
free. When deployed inside beam search, it encounters updated syndromes
and variables pinned by earlier decisions, which shape its subsequent
inputs~\cite{ross2011}.
The training and deployment settings therefore differ in both the
inputs presented to the model and the role of its predictions.
A model that performs best in single-pass decoding need not be the
best guide for search: BP distillation does not train it for this
new objective. To address this training--deployment misalignment,
we train GraphMP on intermediate search states as well as original
syndromes, teaching it to complete partially pinned problems.

\begin{figure}[!htbp]
  \centering
  \includegraphics[width=\textwidth]{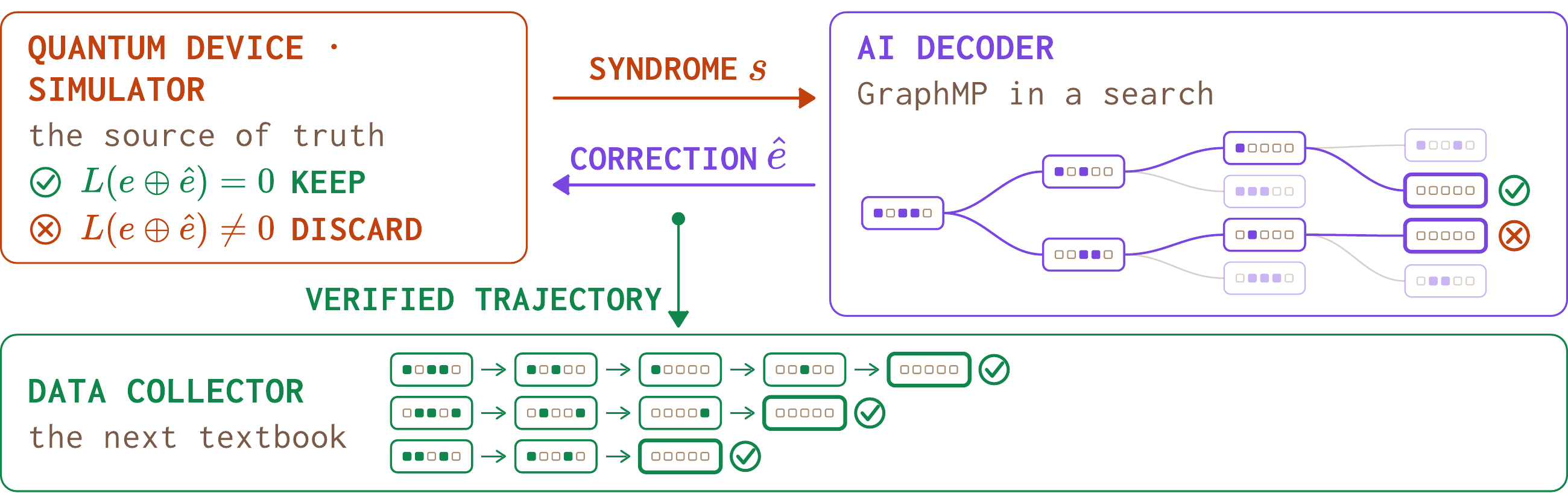}
  \caption{Outcome feedback selects search trajectories for training.
    The collector retains syndrome-consistent, logically correct final
    predictions together with the intermediate states along their search paths. This work obtains
    logical feedback in simulation,
    while dedicated quantum-memory experiments provide a route to
    device-based verification.}
  \label{fig:main-collection}
\end{figure}

The training data for these intermediate states come from the
decoder's own beam-search trajectories. We record the updated syndromes
and pinned values along the path leading to the returned error
prediction $\hat e$. We then select trajectories according to the
correctness of their final decoding outcomes. In our simulations,
the sampled error $e$ supplies an independent reference: a path is
retained only when its final prediction explains the observed syndrome,
$H\hat e=s$, and preserves the logical outcome, $L(e\oplus\hat e)=0$.
On hardware, dedicated quantum-memory experiments could supply
logical feedback by preparing a known logical state and measuring
the relevant logical observable in the prepared basis after recovery
(Figure~\ref{fig:main-collection}). Logical feedback is used for
trajectory collection and evaluation; ordinary decoding selects
predictions by syndrome consistency and candidate score.

Each retained trajectory supplies examples of this completion task.
At every recorded state, the final predicted error pattern agrees with
the values already pinned along the path. Its entries on the free
subset therefore give target values that complete the pattern
consistently with the observed syndrome. We train GraphMP to predict
these free-variable values from the recorded pinned values and the
updated syndrome, obtained by removing the pins' contribution from
the original observation. The model thus learns how to complete a
prediction under earlier search decisions. In our controlled
comparisons, this supervised distillation reduces decoding failures
during beam search relative to training on original syndromes and
final predictions alone (Appendix~\ref{app:alignment-result}).

\begin{figure}[!htbp]
  \centering
  \includegraphics[width=\textwidth]{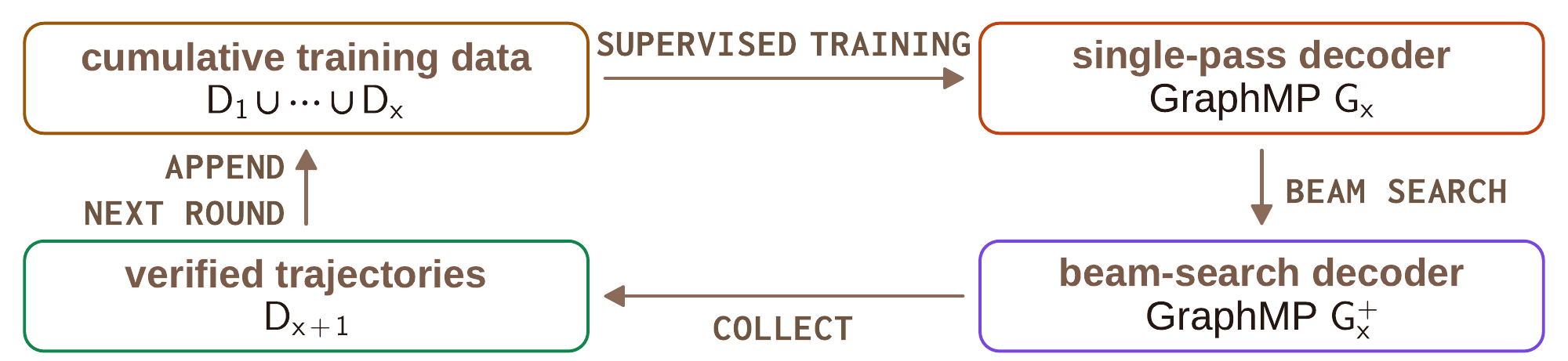}
  \caption{The GraphMP implementation alternates trajectory collection and
    model training. The current model guides search, selected trajectories are
    added to the accumulated data, and distillation produces its
    successor. \Gmodel{x} and \Gsearch{x} are the single-pass and
    search-enhanced decoders, and \Ddata{x} labels each trajectory dataset.}
  \label{fig:main-cycle}
\end{figure}

After distillation, the trained model guides a new round of beam
search and trajectory collection. The selected records are added to
the accumulated training data and used to train a successor, which
in turn guides further collection. Repeating these steps closes the
decoder model factory's collection-and-training cycle
(Figure~\ref{fig:main-cycle}). The message-passing architecture remains
fixed, while the model parameters and training data evolve.

To distinguish successive generations, we denote a single-pass model
by \Gmodel{x}, the same model inside beam search by \Gsearch{x}, and
each new trajectory dataset by \Ddata{x}. The BP-distilled model is
\Gmodel{0}; its search-enhanced decoder collects \Ddata{1}, which
trains \Gmodel{1}. This model collects \Ddata{2}, and the combined
$\Ddata{1}\cup\Ddata{2}$ trains \Gmodel{2}. A further round produces
\Ddata{3} and trains \Gmodel{3} on all three datasets.

This cycle gives \emph{recursive self-improvement} a concrete form:
improved decoder models can guide searches that produce more useful
training experience, which can in turn improve subsequent models.
For a fixed decoding task, this improvement is bounded by the
performance of an optimal decoder. The gains in our experiments
diminish across generations, and further iterations need not produce
additional improvement. Building on this feedback, an \emph{AI model factory} can produce
successive generations of decoders through iterative training on
accumulated experience. Related ideas appear in search-based model
improvement~\cite{anthony2017}, learning from outcome-filtered model
outputs~\cite{singh2023restem}, and decoder-assisted collection of
harder examples~\cite{gong2024}. Our implementation brings these ideas
together through verified conditional search trajectories that can
train later decoder models.

The factory therefore produces both trained models and a growing
resource of reusable decoding experience. A stored trajectory can
supervise a new student after its collector has been replaced, allowing
computation spent on one generation to benefit later generations.
Accumulated records preserve earlier experience, while fresh
trajectories can redirect learning as the noise changes. The model
factory thus provides a framework for maintaining and adapting decoding
strategies through experience acquired during decoding.

\section{Decoder performance and noise adaptation}
\label{sec:results}

\begin{figure}[!htbp]
  \centering
  \includegraphics[width=\textwidth]{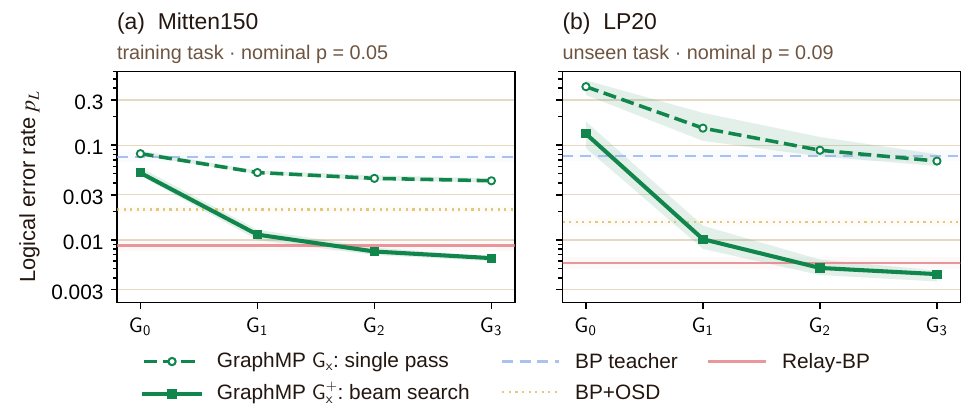}
  \caption{Successive models improve single-pass decoding and transfer
    that improvement to an unseen task. Dashed green curves show
    single-pass decoding; solid green curves show beam-search decoding.
    Horizontal lines show reference decoders. Curves are means over
    five independent model factory runs; bands are the range across those
    runs. Full generation results are in
    Appendix~\ref{app:generations}.}
  \label{fig:main-progress}
\end{figure}

We first test whether the model factory improves decoding over
successive generations and whether these gains transfer beyond its
training tasks. Figure~\ref{fig:main-progress} tracks logical error
rate on $\M_{150}$, a training task, and \LPtwenty, an unseen task.
Averaged over five independent factory runs, both single-pass and
search-enhanced decoding improve at every generation on these two
tasks. The largest gain occurs from \Gmodel{0} to \Gmodel{1}; later
updates yield progressively smaller reductions in logical error,
demonstrating iterative improvement with diminishing returns.
The gains on \LPtwenty\ occur without any training on that task,
showing that the learned improvements transfer beyond the training
portfolio. Because the message-passing architecture remains fixed,
better single-pass predictions come without increased inference cost,
while beam search continues to provide additional accuracy.

\begin{figure}[!htbp]
  \centering
  \includegraphics[width=\textwidth]{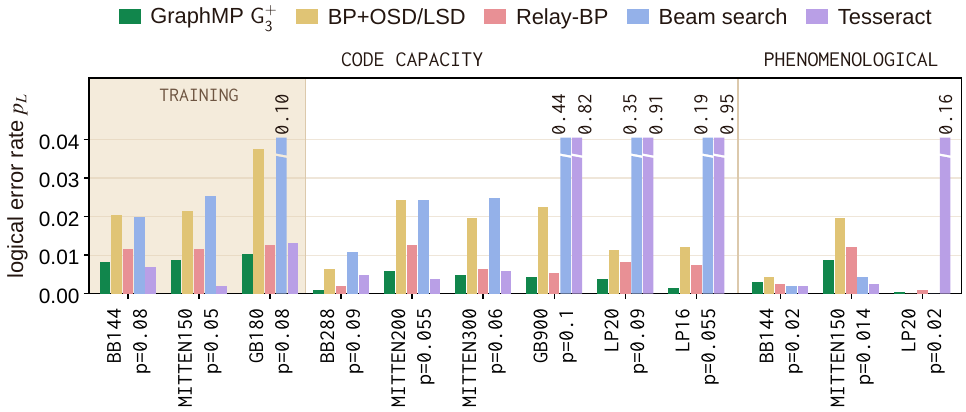}
  \caption{The final search-enhanced model \Gsearch{3} compared with
    established decoders on $12$ tasks. The three shaded tasks supply
    training data; the other nine test the same model without further
    training. Comparisons use the same $2048$ error realizations per
    task, except for shorter Tesseract runs on the largest tasks.
    Bars above $0.04$ are truncated. Decoder settings and evaluation
    procedures are specified in Appendix~\ref{app:factory-classical}.}
  \label{fig:main-products}
\end{figure}

We next assess the decoding performance reached after three rounds of
model factory training. Figure~\ref{fig:main-products} compares the
final model combined with beam search, \Gsearch{3}, with
BP+OSD~\cite{roffe2020}, BP+LSD~\cite{hillmann2025},
Relay-BP~\cite{muller2025}, BP-based beam search~\cite{ye2025},
and Tesseract~\cite{beni2025tesseract} on $12$ tasks.
We train GraphMP on only three code-capacity tasks. We then apply the
same model without further training to nine unseen tasks spanning
different code sizes, code families, and noise models. Under the reported decoder
configurations, \Gsearch{3} achieves lower measured logical error rates
than BP+OSD on all nine code-capacity tasks and is broadly competitive
with Relay-BP across the portfolio. Its advantage over BP-based beam
search and Tesseract is particularly pronounced on the larger
generalized-bicycle and lifted-product tasks. On phenomenological
$\M_{150}$, however, BP-based beam search and Tesseract remain more
accurate. The model factory and beam search thus yield a decoder that
competes with strong classical methods beyond its training tasks,
while leaving task-dependent performance gaps.

\begin{figure}[!htbp]
  \centering
  \includegraphics[width=\textwidth]{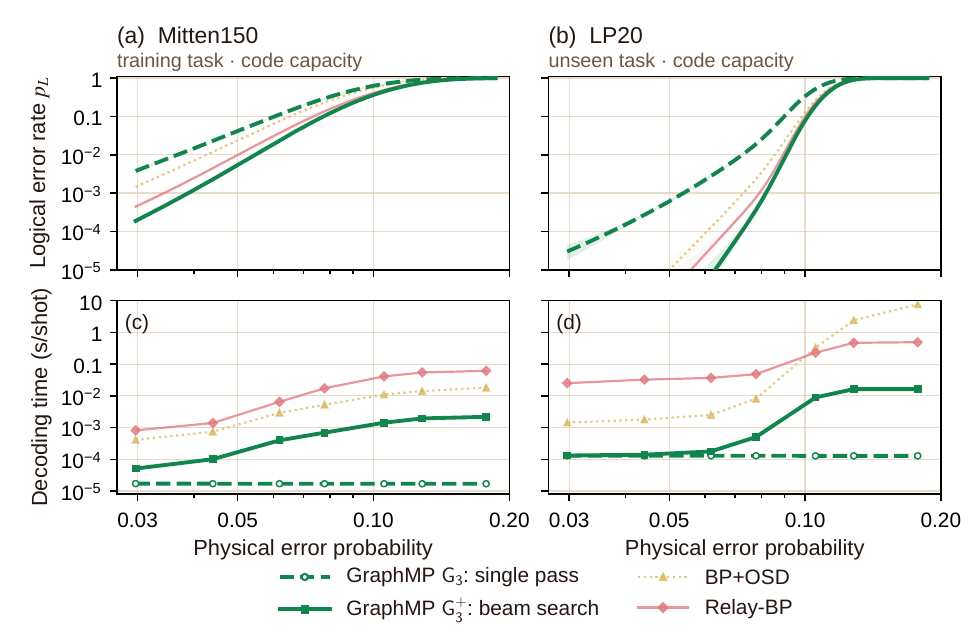}
  \caption{Accuracy and decoding time on a training task and an unseen
    task. Logical-error curves are fitted to measured data; bands are
    $90\%$ bootstrap intervals conditional on that model. Times compare
    GPU GraphMP with single-core
    CPU reference decoders. They measure time per shot in a stream of
    syndromes. Fit diagnostics and timing procedures are in
    Appendices~\ref{app:factory-curves} and~\ref{app:factory-time}.}
  \label{fig:main-cost-curves}
\end{figure}

The practical value of these accuracy gains also depends on decoding
time. Figure~\ref{fig:main-cost-curves} examines this tradeoff on the
training task $\M_{150}$ and the unseen task \LPtwenty.
Single-pass decoding uses a fixed number of message-passing rounds,
so its decoding time remains nearly independent of physical error
rate. Beam search lowers logical error rates across the measured
range, but requires additional model calls. Its cost rises as more
frequent errors cause more realizations to require search. On both
tasks, search-enhanced GraphMP achieves logical error rates close to
Relay-BP with less measured time per shot. These timings measure time
per shot in a stream of syndromes, with GraphMP running on one NVIDIA
B300 GPU and the reference decoders running single-threaded on one CPU
core. The measurements primarily characterize how decoding time varies
with physical error rate and search workload. Absolute times depend
on hardware and implementation, and the performance attainable through
further engineering optimization remains an open question.

\begin{figure}[!htbp]
  \centering
  \includegraphics[width=\textwidth]{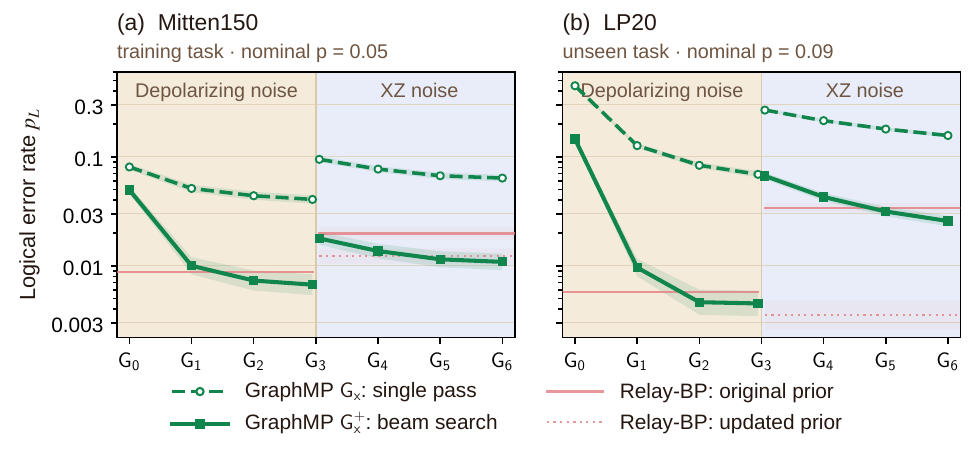}
  \caption{Adaptation after the noise changes at \Gmodel{3}.
    The same model is evaluated immediately before and after the
    switch from depolarizing to XZ noise; \Gmodel{4}--\Gmodel{6} learn
    from newly collected trajectories. (a) $\M_{150}$ supplies training
    data; (b) \LPtwenty\ tests adaptation without further training on
    that task. Dashed green curves
    show single-pass decoding; solid green curves show beam-search
    decoding. Relay-BP is evaluated with its original and updated noise
    priors; GraphMP's beam-search candidate selector retains the
    original prior. Bands are $90\%$ Wilson intervals. The full
    experiment is described in Appendix~\ref{app:factory-adapt}.}
  \label{fig:main-adaptation}
\end{figure}

Finally, we test whether a continuously running model factory can
adapt when its decoder encounters a different noise environment. After
three rounds of improvement under depolarizing noise, we switch to XZ
noise, replacing independent $X$, $Y$, and $Z$ error sources of
probability $p/3$ each with independent $X$ and $Z$ sources of probability
$p/2$ each. This abrupt switch is a toy-model test of adaptation, not a
model of realistic device drift. The decoding graphs and nominal
error-rate settings remain fixed. Figure~\ref{fig:main-adaptation}
follows adaptation on $\M_{150}$, a training task, and tests its
transfer to the unseen task \LPtwenty.

The logical error rate jumps when \Gmodel{3} encounters XZ noise
(Figure~\ref{fig:main-adaptation}): a decoding strategy learned under
one noise model can lose accuracy when the error statistics change.
The factory continues collecting trajectories on the three training
tasks, now under XZ noise, and distilling them into
\Gmodel{4}--\Gmodel{6}. These successive models adapt to the new noise,
lowering logical error rates in both single-pass and beam-search
decoding. Most of the recovery occurs at \Gmodel{4}, with smaller gains
in later generations. Relay-BP with its original prior also loses
accuracy after the switch, while updating its prior improves its
performance, illustrating the importance of adjusting the decoding
strategy to the new error statistics.

The learned adaptation also transfers to \LPtwenty, which undergoes
the same noise change but is excluded from adaptation training
(Figure~\ref{fig:main-adaptation}(b)). Continued factory
training lowers its logical error rate, with \Gsearch{6} outperforming
Relay-BP retaining the original noise prior.
All three independent adaptation runs show this recovery, with the
largest gain in the first update. Relay-BP with the updated prior
remains more accurate on this task. Thus, in this toy-model test,
continued factory training restores performance as successive models
learn from the changed noise, and the resulting adaptation extends
beyond the tasks used for training.

These results suggest that a decoder model factory could help sustain
logical protection as device noise changes. Related approaches to
adaptive decoding update noise estimates or decoder weights from
syndrome statistics~\cite{spitz2018adaptive,bhardwaj2026drift}, while
hardware calibration uses detector events to adjust physical control
parameters~\cite{sivak2026control,gong2026selfcalibration}. Model factory
training could complement these efforts by using verified decoding
outcomes to train models for the noise currently encountered. Extending
this approach to hardware requires logical feedback from dedicated
memory or benchmark experiments, with collection and training fast
enough to follow the relevant changes in device noise.

\section{Conclusion}
\label{sec:outlook}

We have introduced a decoder model factory that produces successive
models from experience accumulated during decoding. Our implementation
for quantum LDPC codes combines GraphMP's local message passing with
beam search and distillation from verified trajectories. In simulation,
this process improves single-pass decoding across generations,
transfers those improvements to unseen tasks, and supports adaptation
after a change in the noise model. Beam search supplies additional
accuracy at inference time, while distillation improves the decoder
without changing its message-passing architecture. These results
provide a concrete way to fold thinking into learning: search improves
the current prediction and creates experience that can train better
decoders.

The broader contribution is an AI model factory built around reusable
decoding experience. Recursive self-improvement in this setting is
bounded by optimal decoding performance for a fixed task, and our
experiments show diminishing gains across generations. Retained
trajectories preserve experience beyond any one model, while fresh
trajectories can teach later models about changed noise. The factory
thus provides a framework for both improving decoding under fixed
conditions and adapting as the conditions for protecting logical
information evolve.

The next step is to collect search trajectories from quantum-memory
experiments, with measured syndromes as inputs and dedicated logical
benchmarks supplying outcome feedback. Circuit-level noise, sustained
drift, and repeated model updates will test the reach of this learning
cycle. Real-time deployment also requires engineering the decoder to
meet error-correction deadlines, with streaming neural decoding and
parallel-window decoding offering relevant approaches~\cite{senior2025alphaqubit2,skoric2023windows}.
The fast decoding cycle could then operate alongside asynchronous
model training. The long-term vision is a quantum immune system for
fault-tolerant computing: a classical decoding layer that combats
errors as they arise and learns from verified experience to strengthen
future logical protection.

\section*{Author contributions}
Wanda Hou and Yi-Zhuang You conceived and organized the research. Yi-Zhuang You, Anil Bilgin and Pranav Srinivas carried out the research, developed the methods and software, and designed, performed, and analyzed the experiments. All authors contributed to writing, revising, and reviewing the manuscript.

The authors used Codex, Claude Code, and Cursor for assistance with code development and manuscript editing. The authors take full responsibility for the code and the scientific content of this work.

\section*{Acknowledgments}
We thank Ethan Lake, Nathaniel Selub, Aditya Bhardwaj, Shengtao Wang,
Hossein Dehghani, Hong-Ye Hu, Chen Zhao, and Jin-Sung Kim for helpful
discussions. 
This work was partially supported by the U.S. Department of Energy under Award
No.~DE-SC0026807, and by grant NSF PHY-2309135 to the Kavli Institute for Theoretical
Physics (KITP). 

{\raggedright
\putbib[refs]
}

\end{bibunit}

\clearpage
\appendix
\bibnamespace{appendix}
\renewcommand{\citenumfont}[1]{A#1}
\renewcommand{\bibnumfmt}[1]{[A#1]}
\renewcommand{\refname}{Appendix References}
\begin{bibunit}[unsrturl]

\section{Decoding Problem}
\label{app:architecture}
\subsection{Decoding Task}
\label{app:task}

A quantum error-correcting code protects logical information by
encoding it redundantly into a larger collection of physical qubits.
After noise acts, a decoder infers from the measured syndrome a
correction that preserves the encoded logical state. Here we briefly
review the decoding tasks defined by the code families and noise
models studied in this paper.

\paragraph{Quantum error-correcting codes.}
A quantum error-correcting code with parameters $[[n,k,d]]$ encodes $k$ logical qubits
into $n$ physical qubits, with distance $d$ being the minimum weight
of an undetectable operation that acts nontrivially on the encoded
logical information.  We consider stabilizer codes~\cite{gottesman1997}, whose
codespace is the simultaneous $+1$ eigenspace of a set of commuting
Pauli stabilizers, and focus on the Calderbank--Shor--Steane (CSS)
subclass~\cite{calderbank1996,steane1996}.  In a CSS code, the stabilizers split into $X$-type and $Z$-type specified by
binary check matrices $H_X$ and $H_Z$ respectively, satisfying
$H_XH_Z^{\mathsf T}=0$.  We further restrict our scope to quantum low-density parity-check (qLDPC) codes,
for which the row and column weights of $H_X$ and $H_Z$ remain bounded
as the code size grows.
This low-density property introduces a notion of locality in the code
that is particularly well suited to graph-based message-passing
decoders, which admit local computation whose complexity grows linearly
with the code size.
The qLDPC codes have attracted increasing attention for their favorable rate and distance scalings~\cite{tillich2014,panteleev2021}.

Table~\ref{tab:codes} lists the code families and instances studied
throughout the paper.
Generalized bicycle (GB) and bivariate-bicycle (BB) codes both take
the two-block CSS form $H_X=[A\mid B]$, $H_Z=[B^{\mathsf T}\mid A^{\mathsf T}]$;
GB is constructed from a pair of circulants~\cite{panteleev2021}, BB from a pair of
bivariate polynomials over a torus~\cite{bravyi2024}.
Lifted-product (LP) codes are products of two base matrices over a
group ring~\cite{panteleev2022,huang2026}; mitten codes are the
non-abelian case~\cite{bhardwaj2026}.
We select three codes to train our decoder model: \BBgross,
\GBeighteen, and $\M_{150}$, one from each of the BB, GB, and
mitten families. The remaining codes are unseen during training.

\begin{table}[!htbp]
  \centering
  \small
  \begin{tabular}{lllll}
    \toprule
    Family & Instance & $[[n,k,d]]$ & Training & Ref. \\
    \midrule
    Bivariate bicycle
      & \BBgross
      & $[[144,12,12]]$
      & yes
      & \cite{bravyi2024} \\
    Bivariate bicycle
      & \BBtwogross
      & $[[288,12,18]]$
      & no
      & \cite{bravyi2024} \\
    Generalized bicycle
      & \GBeighteen
      & $[[180,10,\le 18]]$
      & yes
      & \cite{panteleev2021} \\
    Generalized bicycle
      & \GBnine
      & $[[900,50,15]]$
      & no
      & \cite{panteleev2021} \\
    Mitten
      & $\M_{150}$
      & $[[150,30,10]]$
      & yes
      & \cite{bhardwaj2026} \\
    Mitten
      & $\M_{200}$
      & $[[200,40,12]]$
      & no
      & \cite{bhardwaj2026} \\
    Mitten
      & $\M_{300}$
      & $[[300,60,14]]$
      & no
      & \cite{bhardwaj2026} \\
    Lifted product
      & \LPtwenty
      & $[[1122,148,\le 20]]$
      & no
      & \cite{huang2026} \\
    Lifted product
      & \LPsixteen
      & $[[2610,744,\le 16]]$
      & no
      & \cite{huang2026} \\
    \bottomrule
  \end{tabular}
  \caption{Code instances studied in the experiments.
    A yes marks a code used to train the decoder model.
    Distances written $\le d$ are the published upper bounds.}
  \label{tab:codes}
\end{table}

\paragraph{Detector error model.}
We consider decoding a quantum error-correcting code under a specified
noise model. The code supplies the check matrices
\(H_X\) and \(H_Z\); the noise model specifies error mechanisms and
their probabilities. A detector error model (DEM) compiler represents
these mechanisms as error variables and constructs the decoding graph,
represented by the check matrix \(H\). This graph communicates how
errors affect the measured syndrome. Supplying the corresponding
decoding graph and error probabilities allows the same decoding
algorithm to operate across different codes and noise models.
The literature describes three standard constructions~\cite{dennis2002,fowler2012,gidney2021}.
\begin{itemize}[leftmargin=*]
\item \emph{Code-capacity noise.}
  Data-qubit Paulis; syndrome measurements are perfect.
  At nominal error-rate setting \(p\), our depolarizing-noise
  construction has independent \(X\), \(Y\), and \(Z\) sources
  on each data qubit, each with prior \(p/3\).
  Their composition gives a nonidentity Pauli with probability
  \(p-p^2/3\), the physical error probability.
  A \(Y\) fault is incident to both the \(X\)-type and the \(Z\)-type
  checks that contain that qubit~\cite{dennis2002,roffe2020}.
  With detectors stacked as \(Z\)-then \(X\)-stabilizers and columns
  ordered as all-\(X\), all-\(Y\), all-\(Z\),
  \begin{equation}
    H
    =
    \begin{pmatrix}
      H_Z & H_Z & 0 \\
      0 & H_X & H_X
    \end{pmatrix},
    \qquad
    \pi_i = p/3.
    \label{eq:codecap-H}
  \end{equation}
  The logical matrix \(L\) is built the same way from the \(X\)- and
  \(Z\)-type logicals: a \(Y\) source is incident to both.
\item \emph{Phenomenological noise.}
  Besides data errors, measurement errors can flip the recorded check
  outcomes, making a single round of syndrome extraction unreliable.
  We therefore repeat syndrome extraction and construct detectors
  that compare the outcomes of the same check in successive rounds.
  These comparisons turn decoding into a space-time problem:
  the decoding graph extends across successive time layers, with
  data errors affecting detectors within a layer and measurement
  errors connecting detectors in adjacent layers~\cite{dennis2002,fowler2012}.
  Throughout this paper, phenomenological evaluations refer to a
  memory experiment with 5 noisy syndrome-extraction rounds followed
  by a noisy final readout of the data qubits, from which the final
  syndrome is inferred. Each noisy round includes independent
  $X$ and $Z$ data errors and measurement-outcome flips at the
  task's stated error rate; the final readout also has errors at
  that rate.
\item \emph{Circuit-level noise.}
  Every gate, idle, and measurement in a noisy circuit can fail.
  The Stim compiler~\cite{gidney2021} extracts \((H,L,\pi)\) from
  the noisy circuit. We construct the decoding graph from $H$ and
  use $\pi$ to specify the error distribution.
\end{itemize}

\noindent
All three constructions express the decoding task through a
DEM~\cite{gidney2021,derks2025dem}.

\begin{figure}[!htbp]
  \centering
  \begin{tikzpicture}[
    >=Stealth,
    node distance=0.5cm and 2.6cm,
    box/.style={
      draw=black!55,
      rounded corners=2pt,
      minimum width=1.15cm,
      minimum height=0.72cm,
      align=center,
      fill=black!2,
    },
    lab/.style={font=\small},
  ]
    \node[box] (e) {$e$};
    \node[lab, below=0.2cm of e] {$e\sim\pi(e)$};
    \node[box, above left=1.15cm and 0.15cm of e] (s) {$s$};
    \node[box, above right=1.15cm and 0.15cm of e] (l) {$l$};
    \draw[->, thick] (e) -- node[lab, left, pos=0.55] {$s=He$} (s);
    \draw[->, thick] (e) -- node[lab, right, pos=0.55] {$l=Le$} (l);
  \end{tikzpicture}

  \caption{Detector error model.
    The fault $e$ is drawn from the error-variable prior.
    The syndrome $s=He$ is observed; the logical error $l=Le$ is not.}
  \label{fig:dem}
\end{figure}
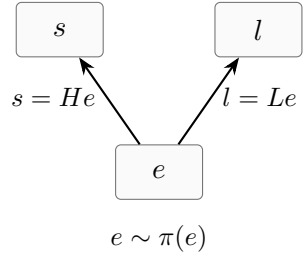

A DEM records two kinds of object.
A \emph{detector} is an additional parity constraint among measurement
outcomes that is satisfied when the circuit is noiseless~\cite{delfosse2023spacetime,derks2025dem}.
An \emph{error variable} records an independent fault event---a data
flip, a measurement flip, or a two-qubit gate fault---and is indexed by
$i=1,\ldots,n$. Error variable $i$ flips a subset of the $m$ detectors
and a subset of the
$k$ logical observables (the logical information to be protected).
Those incidences are the check (detector) matrix
$H\in\GF^{m\times n}$ and the logical (observable) matrix
$L\in\GF^{k\times n}$:
\begin{equation}
  s \;=\; H e,
  \qquad
  l \;=\; L e
  \pmod{2}.
  \label{eq:dem}
\end{equation}
Here $e\in\GF^{n}$ is the \emph{error vector} ($e_i=1$ if error variable
$i$ occurred), $s=(s_1,\ldots,s_m)\in\GF^{m}$ is the observed
\emph{syndrome}, a binary string with one bit $s_c$ for each detector,
and
$l\in\GF^{k}$ is the \emph{logical error}
(Figure~\ref{fig:dem}).
Addition in $\GF$ is denoted by $\oplus$.
The nonzero entries of this detector matrix $H$ define a bipartite
Tanner graph~\cite{kschischang2001} between detector nodes and
error-variable nodes.
By passing messages along these edges, a graph-based local decoder
predicts the error variables conditioned on the detector responses of
a given decoding task.

In the simple setting, the prior on $e$ can be chosen as a product of
independent Bernoulli random variables,
\begin{equation}
  \pi(e)
  \;=\;
  \prod_{i=1}^{n}
  \mathrm{Bernoulli}(e_i;\,\pi_i),
  \qquad
  \mathrm{Bernoulli}(e_i;\,\pi_i)
  \;=\;
  \pi_i^{e_i}\,(1-\pi_i)^{1-e_i},
  \label{eq:bernoulli}
\end{equation}
for bits $e_i\in\{0,1\}$.
The DEM supplies the prior vector $(\pi_1,\ldots,\pi_n)$.
Code capacity, phenomenological, and circuit-level noise all compile
to some $(H,L,\pi)$.
A \emph{shot} is one independent draw $e\sim\pi(e)$ together with the
induced syndrome $s=He$ and logical error $l=Le$.
The decoder sees only $(H,L,\pi)$ and $s$.
The Tanner graph of $H$ is the \emph{decoding graph} of the instance:
write $\mathcal{G}=(C\cup V,E)$ for it, with check nodes the detectors
and variable nodes the error variables.
We work in that language for the rest of the paper.

Event probabilities can also be estimated from syndrome data.
Early work extracted decoding error models from detection-event
statistics~\cite{fowler2014extraction}; identifiability analyses establish
conditions for recovering Pauli-noise parameters from syndrome
measurements~\cite{wagner2021noise,wagner2022pauli}.
DEM-focused methods estimate individual or aggregated event probabilities
and reconstruct decoding graphs and hypergraphs~\cite{blumekohout2025dem,takou2025dem,ye2026dem}.
These methods provide a way to calibrate the decoder's probabilistic inputs.
In the simulations reported here, the configured noise model supplies
$(H,L,\pi)$.

\paragraph{Decoding task.}
The decoder is given $s$ and returns a predicted error
$\hat e\in\GF^{n}$.
Write $r=e\oplus\hat e$ for the \emph{residual}.
Success requires both syndrome consistency and logical correctness,
\begin{equation}
  H r \;=\; 0
  \qquad\text{and}\qquad
  L r \;=\; 0,
  \label{eq:success}
\end{equation}
i.e.\ $r\in\ker H\cap\ker L$: the residual is a stabilizer of the DEM,
and $e$ and $\hat e$ lie in the same logical coset.
The residual can be classified along two axes: syndrome consistency
($Hr=0$ versus $Hr\neq 0$) and logical outcome ($Lr=0$ versus
$Lr\neq 0$).
The resulting four cells are named in Table~\ref{tab:residual} as a
working vocabulary for later discussion.
Many distinct $e$ share the same $s$ (they differ by an element of
$\ker H$), so $e\mapsto s$ is
many-to-one~\cite{dennis2002,panteleev2021}.
A \emph{miscorrection}---syndrome-consistent, wrong coset---still
counts as a logical error: the decoder has accepted an answer that
looks valid, but the logical coset is wrong.
When the syndrome has not been cleared ($Hr\neq 0$) but $Lr=0$, we
call the outcome an \emph{escape}: the prediction passes the specified
logical check while leaving a residual syndrome. Successful recovery
requires the residual to lie in the code space as well as preserve the
logical information, as in \eqnref{eq:success}.

\begin{table}[!htbp]
  \centering
  \small

  \begin{tabular}{l|cc}
      & logical success ($Lr=0$)
      & logical error ($Lr\neq 0$) \\
    \hline
    syndrome consistent ($Hr=0$)
      & success
      & miscorrection \\
    syndrome inconsistent ($Hr\neq 0$)
      & escape
      & failure \\
  \end{tabular}
  \caption{Classification of the residual $r=e\oplus\hat e$ along
    syndrome consistency and logical outcome.}
  \label{tab:residual}
\end{table}

A key metric to evaluate decoder performance is the word-level (block)
logical error rate
\begin{equation}
  \LER
  \;=\;
  \Pr\bigl(L r\neq 0\bigr),
  \label{eq:ler}
\end{equation}
that is the probability that any logical observable flips, under
$e\sim\pi(e)$ and $s\mapsto\hat e$.
A trial fails if the decoded logical word is wrong, including both
\emph{miscorrection} and \emph{failure}, regardless of syndrome
consistency: $Lr\neq 0$ is a logical error whether or not $Hr=0$.
An escape ($Hr\neq 0$ and $Lr=0$) therefore counts as a logical
success under $\LER$. To assess whether the predicted error also
explains the syndrome, we record the syndrome-inconsistency rate
$\Pr(Hr\neq 0)$ and the joint failure rate
$\Pr(Hr\neq 0\;\text{or}\;Lr\neq 0)$. The latter equals the logical
error rate plus the escape rate. Both checks determine which
trajectories the model factory accepts for training.

\subsection{Training Objective}
\label{app:objective}

Recovering the encoded logical information still leaves a choice of
what the decoder should predict. The training objective can be the
logical error $l$, the error pattern $e$, or the individual
error variables $e_i$ (bit-wise), and each objective comes with its
own probability model. We refer to these objectives as
maximum-likelihood decoding (MLD), most-likely-error decoding (MLE),
and bit-wise maximum a posteriori (MAP)
decoding, respectively~\cite{dennis2002,iyer2015,hsieh2011,kschischang2001}
(Table~\ref{tab:estimands}).

\begin{table}[!htbp]
  \centering
  \small
  \begin{tabular}{llll}
    \toprule
    Objective & Probability model & Decision & \\
    \midrule
    $l$   & $p(l\mid s)$           & $\hat l=\arg\max_{l} p(l\mid s)$ & MLD \\
    $e$   & $p(e\mid s)$           & $\hat e=\arg\max_{e:\,He=s} p(e\mid s)$, $\hat l=L\hat e$ & MLE \\
    $e_i$ & $\prod_i p(e_i\mid s)$ & $\hat e_i=\ind[\,p(e_i\mid s)\ge\tfrac12\,]$, $\hat l=L\hat e$ & bit-wise MAP \\
    \bottomrule
  \end{tabular}
  \caption{Three training objectives and their probability models.
    MLD is optimal~\cite{dennis2002,iyer2015}; MLE and bit-wise MAP
    are successive approximations.}
  \label{tab:estimands}
\end{table}

The three models form a chain of approximations. MLD is exact: the
logical error $l=Le$ is a parity of the error variables, so its
posterior is a sum over the coset. MLE replaces the sum by its
largest term,
\begin{equation}
  p(l\mid s)
  \;=\;
  \sum_{e:\,Le=l} p(e\mid s)
  \;\approx\;
  \max_{e:\,Le=l} p(e\mid s),
  \label{eq:coset-sum}
\end{equation}
valid when each coset is dominated by its most likely
configuration~\cite{dennis2002,hsieh2011}. In turn, $p(e\mid s)$
factorizes exactly in autoregressive form, which bit-wise MAP
replaces by a product of single-variable marginals,
\begin{equation}
  p(e\mid s)
  \;=\;
  \prod_{i=1}^{n} p(e_i\mid e_{<i},s)
  \;\approx\;
  \prod_{i=1}^{n} p(e_i\mid s),
  \label{eq:factorized-posterior}
\end{equation}
which treats the error variables as conditionally independent given
the syndrome and so ignores the correlations the syndrome induces
among them.

\begin{figure}[!htbp]
  \centering
  \includegraphics[width=0.72\textwidth]{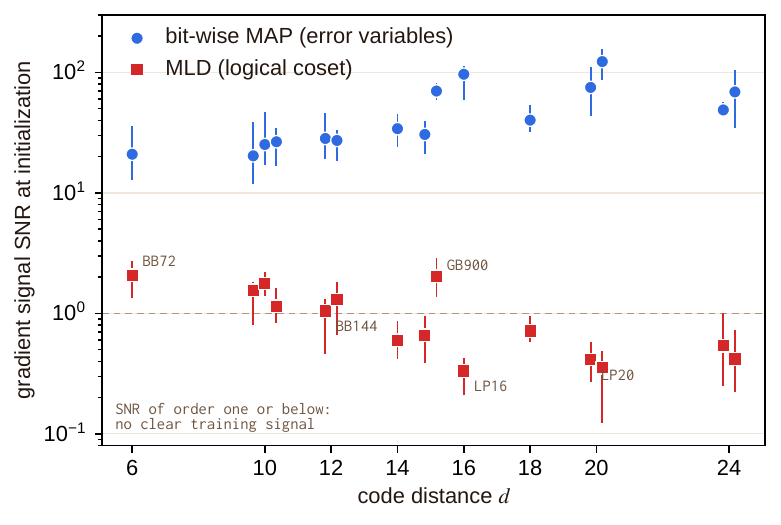}
  \caption{Gradient signal of the two objectives at random
    initialization: SNR~\eqref{eq:snr} of the gradient signal on the
    shared model parameters, estimated over $8$
    minibatches of $64$ syndromes (mean over $5$ initializations,
    bars the range), for $15$ qLDPC codes at code-capacity
    depolarizing noise $p=0.05$. The dashed line marks
    $\mathrm{SNR}=1$. Codes sharing a distance are offset
    horizontally.}
  \label{fig:grad-snr}
\end{figure}

Each step is a further approximation, so on exactness alone the
logical error $l$ would be the training objective of choice and the
individual error variables $e_i$ the last resort. For a learned
decoder, however, exactness is not the only criterion: the objective
also has to be learnable, and the two objectives turn out to differ
sharply in this respect. To compare them we built one decoder model
with two outputs. The decoder model processes the decoding graph
and produces a feature vector per error variable; one
output predicts $e_i$ by mapping each feature vector linearly to the
logit of $p(e_i\mid s)$, the other predicts $l$ by pooling the
feature vectors on the support of each logical operator and mapping
them to the logit of $p(l\mid s)$. The comparison holds the decoder
model, syndromes, and random initialization fixed while changing the
prediction objective.
Figure~\ref{fig:grad-snr} reports, for $15$ qLDPC codes, the
signal-to-noise ratio (SNR) of the gradient signal that each
objective sends back into the shared model parameters
$\theta$ at initialization; the objective is $e_i$ for one output
and $l$ for the other. Let $g=\nabla_\theta\mathcal{L}$ be the
gradient of an objective on a minibatch of shots, each shot a
syndrome paired with the true value of the objective, and let
$\tilde g$ be the same gradient after those true values are shuffled
among the shots, so that every syndrome is paired with the value
from another shot. The difference $\Delta g=g-\tilde g$ is the part
of the gradient tied to the correspondence between syndrome and
objective, the part a decoder can learn from. We define
\begin{equation}
  \mathrm{SNR}
  \;=\;
  \frac{\bigl\lVert\,\mathbb{E}[\Delta g]\,\bigr\rVert}
       {\sqrt{\mathbb{E}\,\bigl\lVert\Delta g-\mathbb{E}[\Delta g]\bigr\rVert^{2}}},
  \label{eq:snr}
\end{equation}
with the expectation over independent minibatches. An SNR far above
one means a clear gradient signal: different minibatches agree on
the direction of optimization. An SNR of order one or below means
the direction fluctuates from minibatch to minibatch and there is no
clear gradient signal.

The SNR of the two objectives differs by more than an order of
magnitude on every code, and the separation grows with the code
distance $d$. The
gradient signal for $e_i$ is strong on every code (SNR $20$ to
$120$) and strengthens with code size; the signal for $l$ starts at
an SNR of about $2$ on the smallest code and drops below one for
$d\gtrsim 14$, where the gradients from different minibatches are
nearly orthogonal. The two prediction heads aggregate supervision
differently: error-variable prediction supplies a target at every
variable node, while logical-error prediction pools features over the
support of each logical operator. This difference offers a possible
explanation for the trend. At initialization, error-variable prediction
provides a stronger gradient signal than logical-error prediction in
this comparison.

To build a machine-learning decoder, we therefore train on $e_i$:
the decoder learns $\prod_i p(e_i\mid s)$ and decides each bit from
its own marginal, the same choice belief propagation
makes~\cite{kschischang2001,roffe2020,liu2019}, and the predicted
logical error is $L\hat e$.

\subsection{Universal Decoder Interface}
\label{app:interface}

We implement the bit-wise predictions of
\eqnref{eq:factorized-posterior} through \emph{message passing} on the
decoding graph (Section~\ref{app:task}): each node holds a state, each
edge carries a message from the states at its endpoints, and each node
updates from the messages on its incident edges.
Belief propagation, neural BP~\cite{nachmani2016,liu2019}, and other
local rules all run on this graph.
These algorithms use different message rules on the same graph.

Our decoder receives
the decoding graph $H$ and the observed syndrome $s$, and returns the
per-variable probabilities
$p(e_i\mid s)$ in \eqnref{eq:factorized-posterior}.  The complete decoding
system also reads $\pi$ when search ranks completed candidates.
After decoding, we compute the predicted logical error as $L\hat e$.
A Stim DEM exposes these objects~\cite{gidney2021}, so the same model
class applies to the decoding graphs considered here.

The BP family can instantiate the same inference algorithm on a new
decoding graph with its event priors and observed syndrome. We adopt
this task specification, $(H,\pi,s)$, for the complete decoding system.
GraphMP learns its per-error predictions from training data, while
the search algorithm reads the prior when ranking candidates. Graph
dimensions and connections can change without changing GraphMP's
parameterized message, update, and readout functions. This is
architectural compatibility; useful decoding with a frozen checkpoint
on a new task must be established by evaluation.

\paragraph{Input representations and prediction targets.}
Learned decoders make different choices about the information supplied
to the model and the quantity it predicts
(Table~\ref{tab:extras}). Coordinates, check types, soft readouts, and
logical-node annotations can encode relevant structure or additional
observations. Their availability defines which tasks a particular
input representation supports. A logical-output decoder predicts the
protected observables directly, whereas an error-output decoder
predicts the variables from which their logical effect can be computed.
This output choice is separate from the choice of input features.
\begin{table}[!htbp]
  \centering
  \small

  \begin{tabularx}{\textwidth}{@{}l>{\raggedright\arraybackslash}X>{\raggedright\arraybackslash}X@{}}
    \toprule
    Decoder & Model input representation & Prediction target \\
    \midrule
    AlphaQubit~\cite{bausch2024}
      & Roundwise stabilizer observations and positional information;
        optional soft readout and leakage features
      & Logical-observable errors \\
    Astra~\cite{astra2024}
      & Code Tanner graph and syndrome-node features
      & Per-error probabilities \\
    GraphQEC~\cite{hu2025}
      & Syndrome history on an extended code Tanner graph with
        data, check, and logical nodes
      & Logical flips \\
    Foundation decoders~\cite{yan2026foundation}
      & Transformer: detector histories and relation embeddings;
        Neural-BP: DEM graph, priors, and syndrome
      & Transformer: logical flips; Neural-BP: error mechanisms \\
    GraphMP (this work)
      & DEM detector--error graph, syndrome, and optional pin mask
      & Per-error probabilities \\
    \bottomrule
  \end{tabularx}
  \caption{Representative model interfaces. Input representations and
    output targets are separate design choices. The demonstrated range
    of parameter reuse depends on the training and evaluation protocol.}
  \label{tab:extras}
\end{table}

Parameter reuse has several experimentally distinct meanings.
AlphaQubit demonstrates one jointly trained model across surface-code
distances $3$--$11$ in simulation with relative positional embeddings,
and separately
fine-tunes models pretrained on simulation using processor
data~\cite{bausch2024}. Astra demonstrates frozen-model decoding across
sizes within surface and bivariate-bicycle families, as well as
larger-size training initialized from earlier
checkpoints~\cite{astra2024}. Its reported surface and
bivariate-bicycle models have different feature dimensions.
GraphQEC applies a common architecture to multiple code families with
task-specific training; transfer across different decoding graphs is
left open~\cite{hu2025}. The foundation decoders of
Ref.~\cite{yan2026foundation} reuse
relation-tied weights within scalable families and fine-tune them for
the target distance. These are different
forms of reuse: a compatible architecture, a frozen model evaluated
on new tasks, and a checkpoint that initializes further training.

Other architectures illustrate further choices of representation and
inference. HyperNQ alternates qubit-to-stabilizer and
stabilizer-to-qubit aggregation on a hypergraph~\cite{bhave2025hypernq}.
Recurrent transformers process circuit-level syndrome histories for
bivariate-bicycle codes and generate logical-error bits
conditionally~\cite{blue2026}. Autoregressive generative decoding learns
a joint distribution over syndromes and logical variables, then predicts
logical variables sequentially~\cite{cao2025generative}.
Masked diffusion instead learns to complete partially masked logical-error
vectors conditioned on the syndrome~\cite{liu2025diffusion}.
DiffQEC conditions discrete denoising on multi-round syndrome histories
and trains against logical outcomes~\cite{xu2026diffqec}.
These prediction targets and conditioning choices determine which
intermediate decoding problems the models can answer.

GraphMP's error-node readout scales with the number of DEM variables,
and its graph operations require no lattice coordinates or typed
Pauli-error labels. A DEM variable may describe a correlated fault or
a measurement event as well as a single-qubit error. This interface
supports the conditional error predictions used inside our beam search
algorithm. The range and accuracy of cross-task parameter reuse are
reported in the evaluations below. For logical-output models, a change
in the protected observables may require changing their representation
or readout, while shared inference components can still be reusable.

\paragraph{Optional mask for pinning error variables.}
The standard decoder interface $(H,\pi,s)$ is sufficient for
single-pass decoding. To embed a decoder within a search procedure, we
extend the model with an optional mask $\pin$ that pins selected error
variables to specified values. For each error variable, $\pin_i=\pinf$
leaves the variable free, whereas $\pin_i\in\{0,1\}$ pins it to the
indicated value. The mask supplies additional conditioning information
about pinned error variables within the beam-search algorithm
(Section~\ref{app:search}). Incorporating it into
\eqnref{eq:factorized-posterior} gives
\begin{equation}
  p(e\mid s,\pin)
  \;\approx\;
  \prod_{i=1}^{n}p(e_i\mid s,\pin).
  \label{eq:pinned-factorized-posterior}
\end{equation}
Setting every mask variable $\pin_i$ to $\pinf$ recovers the single-pass
model. Models intended only for single-pass decoding may omit this
optional input; models embedded in search are trained on examples whose
masks specify pinned error variables.

\subsection{GraphMP Decoder}
\label{app:graphmp}

We introduce GraphMP, a learned graph message-passing decoder, to model
the factorized conditional distribution in
\eqnref{eq:pinned-factorized-posterior}. GraphMP assigns a hidden
state to every detector node and error-variable node of the decoding
graph and exchanges learned messages along its edges, following the
structure of belief propagation~\cite{nachmani2016,liu2019,roffe2020}. A forward
pass contains $T$ message-passing rounds, repeatedly applying learned
message and recurrent-update functions along the graph edges.

\begin{figure}[!htbp]
  \centering
  \begin{tikzpicture}[
    >=Stealth,
    node distance=0.48cm and 0.55cm,
    box/.style={
      draw=black!55,
      rounded corners=2pt,
      minimum height=0.64cm,
      align=center,
      font=\small,
      fill=black!2,
      inner sep=4pt,
    },
    nnbox/.style={
      box,
      font=\small\sffamily,
      fill=black!5,
    },
    stage/.style={
      draw=black!35,
      dashed,
      rounded corners=3pt,
      inner sep=9pt,
      fill=none,
    },
    innerbox/.style={
      draw=black!60,
      rounded corners=2pt,
      inner sep=8pt,
      fill=black!5,
    },
    arr/.style={->, thick, black!65},
    wire/.style={thick, black!65},
    stagelab/.style={font=\small\bfseries, text=black!72, anchor=east},
  ]
    \def\stagevgap{0.82cm}

    \node[box, minimum width=1.55cm] (syn) {syndrome $s$};
    \node[box, minimum width=1.70cm, right=0.50cm of syn] (lam)
      {mask $\pin$};
    \node[box, minimum width=2.15cm, right=0.50cm of lam] (graph)
      {decoding graph $H$};
    \begin{scope}[on background layer]
      \node[stage, fit=(syn)(lam)(graph)] (inputbox) {};
    \end{scope}

    \node[nnbox, minimum width=1.55cm, below=\stagevgap of syn] (encC) {Enc$_C$};
    \node[nnbox, minimum width=1.70cm, below=\stagevgap of lam] (encV) {Enc$_V$};

    \coordinate (mpaxis) at ($(encC.south)!0.5!(encV.south)+(0,-\stagevgap)$);
    \node[box, minimum width=4.15cm, minimum height=0.68cm, anchor=north]
      (h0) at (mpaxis) {initial states $h^{(0)}$};

    \node[innerbox, minimum width=4.15cm, minimum height=1.15cm,
          below=0.40cm of h0] (mpinner) {};
    \node[box, font=\scriptsize, minimum width=1.20cm, minimum height=0.52cm]
      (chk) at ($(mpinner.center)+(-1.20,0)$) {checks};
    \node[box, font=\scriptsize, minimum width=1.35cm, minimum height=0.52cm]
      (bitn) at ($(mpinner.center)+(1.20,0)$) {variables};
    \draw[arr] ([yshift=3pt]chk.east) to[bend left=16]
      ([yshift=3pt]bitn.west);
    \draw[arr] ([yshift=-3pt]bitn.west) to[bend left=16]
      ([yshift=-3pt]chk.east);

    \node[box, minimum width=4.15cm, minimum height=0.68cm, below=0.40cm of mpinner]
      (hT) {final states $h^{(T)}$};
    \node[font=\normalsize, anchor=west] (ttimes)
      at ([xshift=0.28cm]mpinner.east) {$\times T$};

    \node[nnbox, minimum width=1.70cm, below=\stagevgap of hT] (ro) {Readout};
    \node[box, minimum width=1.70cm, below=0.42cm of ro] (z) {logits $z$};
    \node[box, minimum width=2.05cm, below=0.42cm of z] (pe)
      {$p_\theta(e_i=1\mid s,\pin)=\sigma(z_i)$};

    \draw[arr] (syn.south) -- (encC.north);
    \draw[arr] (lam.south) -- (encV.north);
    \draw[dashed, thick, black!45]
      (graph.south) |- ([yshift=12pt]mpinner.east);
    \coordinate (embbus) at ($(h0.north)+(0,0.4)$);
    \draw[wire] (encC.south) -- (encC.south |- embbus);
    \draw[wire] (encV.south) -- (encV.south |- embbus);
    \draw[wire] (encC.south |- embbus) -- (encV.south |- embbus);
    \draw[arr] (embbus) -- (h0.north);
    \draw[arr] (h0.south) -- (h0.south |- mpinner.north);
    \draw[arr] (mpinner.south) -- (hT.north);
    \draw[arr] (hT.south) -- (ro.north);
    \draw[arr] (ro.south) -- (z.north);
    \draw[arr] (z.south) -- (pe.north);

    \begin{scope}[on background layer]
      \node[stage, fit=(encC)(encV)] (encbox) {};
      \node[stage, fit=(h0)(hT)(mpinner)(ttimes)] (mpbox) {};
      \node[stage, fit=(ro)(z)(pe)] (robox) {};
    \end{scope}

    \coordinate (labcol) at ([xshift=-0.55cm]mpbox.west);
    \node[stagelab] at (labcol |- inputbox) {Input};
    \node[stagelab] at (labcol |- encbox) {Encode};
    \node[stagelab] at (labcol |- mpbox) {Message passing};
    \node[stagelab] at (labcol |- robox) {Readout};
  \end{tikzpicture}

  \caption{GraphMP forward pass.
    The syndrome $s$ and the mask $\pin$ are encoded.
    The decoding graph $H$ specifies the bipartite support of the $T$
    check$\leftrightarrow$variable message-passing rounds (dashed).
    Readout emits logits $z$; per-variable probabilities are
    $\sigma(z_i)$, with
    $\sigma$ the logistic sigmoid~\eqnref{eq:sigmoid}.}
  \label{fig:graphmp-flow}
\end{figure}
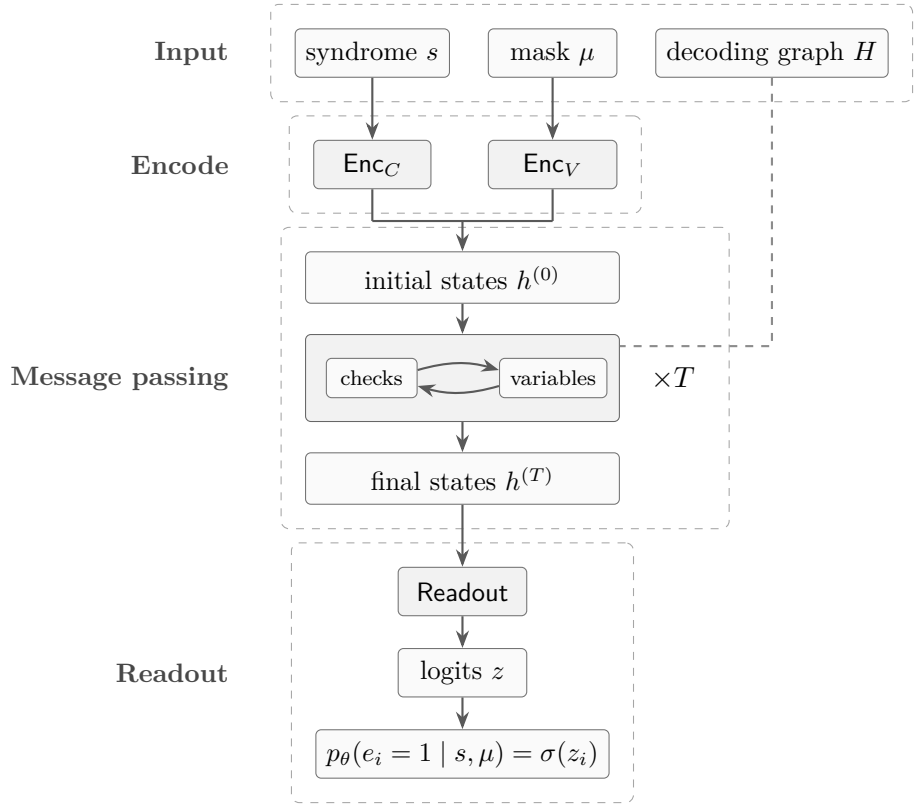

We denote the learned model parameters by $\theta$. A GraphMP forward pass
outputs one logit $z_i\in\mathbb{R}$ per error variable and parameterizes
the conditional product model as
\begin{equation}
  p_\theta(e\mid s,\pin)
  \;=\;
  \prod_{i=1}^{n}
  \mathrm{Bernoulli}\!\left(e_i;\sigma(z_i)\right),
  \qquad
  \sigma(z)=\frac{1}{1+e^{-z}}.
  \label{eq:sigmoid}
\end{equation}
These logits form the soft output of GraphMP. For a given decoding graph
$H$, they are a function of the syndrome and the mask,
\begin{equation}
  z \;=\; z_\theta(s,\pin\mid H)
  \;\in\;
  \mathbb{R}^{n},
  \label{eq:logit-map}
\end{equation}
computed by GraphMP.
The decoding graph $H$ supplies the computational structure of this
mapping, while the observed syndrome $s$ and the optional mask $\pin$
are the data supplied to each forward pass.
Message passing evaluates this mapping on the decoding graph. The
forward pass has three stages
(Figure~\ref{fig:graphmp-flow}): \emph{encode}, recurrent
\emph{message passing}, and \emph{readout}.

\paragraph{Encoding.}
The observed syndrome $s\in\GF^{m}$ and the optional mask $\pin$ are
encoded into the initial states
\begin{equation}
  h_c^{(0)}
  \;=\;
  \nn{Enc}_{C}(s_c),
  \qquad
  h_i^{(0)}
  \;=\;
  \nn{Enc}_{V}(\pin_i),
  \label{eq:gmp-init}
\end{equation}
Both encoders are learned maps shared across nodes of their type. The
mask encoder distinguishes a free error variable from one pinned to
either binary value and incorporates this information into the initial
variable states.

\paragraph{Message passing.}
Let $\mathcal G=(C\cup V,E)$ denote the decoding graph, where $C$ and
$V$ are the detector and error-variable nodes and
$E=\{c\leftrightarrow i\mid c\in C,\ i\in V,\ H_{ci}=1\}$
is its bidirectional edge set.
Each graph edge supports messages in both directions. For
$u\to v\in E$, the message at step $t$ is
\begin{equation}
  \phi_{u\to v}^{(t)}
  \;=\;
  \nn{MLP}_\theta\!\left(
    h_u^{(t-1)},h_v^{(t-1)},d_{u\to v}
  \right).
  \label{eq:gmp-message}
\end{equation}
The destination sums its incoming messages and incorporates the result
through a recurrent update,
\begin{equation}
  h_v^{(t)}
  \;=\;
  \nn{GRU}_\theta\!\left(
    h_v^{(t-1)},
    \sum_{u:\,u\to v\in E}\phi_{u\to v}^{(t)}
  \right).
  \label{eq:gmp-aggregate-update}
\end{equation}
Here $\nn{GRU}_\theta(h,x)$ is a standard GRUCell~\cite{cho2014}; $h$
is the previous node state and $x$ is the newly aggregated message.

The direction embedding $d_{u\to v}$ distinguishes check-to-variable
from variable-to-check messages. The message function is a two-layer
MLP with ReLU between its linear maps. Each round updates variable
nodes first, then check nodes; the check-directed messages in
\eqnref{eq:gmp-message} therefore take the newly updated variable
state $h_i^{(t)}$ as input. After their updates, check states receive
the syndrome encoding again, and pin-conditioned variable states
receive their mask encoding again. The GRU parameters are shared
between the two node types, and all message and update parameters are
shared across rounds. The models evaluated here have hidden dimension
$32$ and $24$ message-passing rounds. The BP-distilled starting model
has zero initial variable states; trajectory-trained models learn the
mask encoder so that the same update rule can condition on pinned
values.

\paragraph{Readout.}
A shared learned head maps each final variable state to the logit of
\eqnref{eq:logit-map},
\begin{equation}
  z_i \;=\; \nn{Readout}_\theta\bigl(h_i^{(T)}\bigr),
  \qquad
  p_\theta(e_i=1\mid s,\pin)
  \;=\;
  \sigma(z_i),
  \label{eq:gmp-readout}
\end{equation}
Training applies the same head after every round (deep supervision;
Appendix~\ref{app:sl-loss}); decoding uses the final-round logits
$z^{(T)}$.
The decoder forms the predicted error $\hat e$ by thresholding free
logits and copying pinned values,
\begin{equation}
  \hat e_i
  \;=\;
  \begin{cases}
    \ind[z_i>0] & \text{if }\pin_i=\pinf,\\
    \pin_i & \text{if }\pin_i\in\{0,1\}.
  \end{cases}
  \label{eq:hard-decode}
\end{equation}
Together, \eqnref{eq:logit-map} and \eqnref{eq:hard-decode} define the
GraphMP soft-decoder interface
\begin{equation}
  \mathsf D_\theta(s,\pin\mid H)=(z,\hat e).
  \label{eq:graphmp-soft-decoder}
\end{equation}
We call this a soft decoder because it returns the logits $z$ in
addition to the predicted error $\hat e$. Once $\hat e$ is available, the
decoding system computes the residual syndrome $s\oplus H\hat e$ from
$s$ and $H$, and the predicted logical error $L\hat e$ from $L$.

As a universal decoder in the sense of Section~\ref{app:interface},
GraphMP requires no additional node-type annotations, lattice
coordinates, or other graph-specific metadata, allowing a single
trained model to be applied across decoding tasks.
Each forward pass costs $O(\lvert E\rvert T)$ and updates nodes in
parallel within each round. Single-pass decoding consists of one
forward pass with every pin free; beam-search decoding reuses the same
model parameters to guide candidate selection and improve decoding
quality (Appendix~\ref{app:beam}).

\subsection{Classical Decoders}
\label{app:bp-baselines}

BP refinements combine soft inference with several forms of structured
postprocessing. Belief-matching and belief-find convert BP posteriors
into matching or weighted union-find costs~\cite{higgott2023belief}.
Ambiguity Clustering uses BP information to divide the problem into
clusters for independent analysis~\cite{wolanski2024ambiguity}.
Ordered Tanner forests remove cycles before a further BP
pass~\cite{demarti2024otf}; SymBreak introduces syndrome-split
constraints to distinguish competing error patterns~\cite{yin2024symbreak}.
Automorphism ensembles decode transformed syndromes and select a
high-probability result~\cite{koutsioumpas2025autdec}.
Learned sequential BP schedules and their list-search extension provide
another route to coordinating variable updates~\cite{moradi2026rls,moradi2026rlls}.

The experimental comparisons in this paper use the following classical decoders.
Each reads the detector matrix $H$, the mechanism priors
$\pi$, and the measured syndrome $s$ of Section~\ref{app:interface}.
Their exact implementations and parameters are specified below.

\begin{itemize}[leftmargin=*,nosep]
\item \emph{BP teacher.}
  The BP teacher is our independently maintained PyTorch
  implementation of layered normalized min-sum BP. Given an input
  syndrome, the decoder computes a logit for every error variable using
  8 layers, at most 30 iterations, and scaling $\alpha=0.625$.
  Within one iteration the checks are partitioned into 8 contiguous
  groups; each group updates from the current messages, then the
  variables they touch refresh before the next group.
  The teacher processes each syndrome with one BP invocation and
  supplies its per-variable logits as soft labels.
\item \emph{BP+OSD/LSD.}
  In this work, BP+OSD and BP+LSD are implemented by the third-party
  \texttt{ldpc} library~\cite{roffe2022ldpc,roffe2020}. They run normalized
  min-sum belief propagation with a serial
  check-update schedule, 200 iterations, and the same scaling
  $\alpha=0.625$.
  Serial updates visit checks in sequence, so later checks in a sweep
  already see earlier updates.
  After BP, combination-sweep ordered-statistics decoding of order 7
  (OSD-CS-7)~\cite{roffe2020,panteleev2021} post-processes
  code-capacity tasks, and combination-sweep localized-statistics
  decoding of order 7 (LSD-CS-7)~\cite{hillmann2025} post-processes phenomenological
  tasks.
\item \emph{Relay-BP.}
  Relay-BP runs version 0.2.2 of the authors' upstream \texttt{relay-bp} reference
  implementation~\cite{muller2025,relaybpsoftware}. The
  package is copyright IBM 2025 and is released under Apache-2.0. We
  use 60 pre-relay iterations, then
  60 relay legs of 60 iterations each, with disordered
  $\gamma$-memory on the bit-node updates.
  Min-sum scaling is $\alpha=0.625$.
  The first leg uses uniform memory $\gamma_0=0.1$; later legs draw
  per-bit memory strengths independently from $(-0.24,0.66)$.
  We stop after 5 syndrome-consistent legs.
\item \emph{Beam search decoder.}
  The beam search decoder runs the authors' reference
  implementation~\cite{ye2025}, released for non-commercial research
  use under CC BY-NC-SA 4.0, which guides beam search with belief
  propagation. We use search width $W=8$ and search depth $R=10$,
  with 30 belief-propagation iterations before the search and 20 per
  search step; the remaining parameters follow the authors' published
  defaults.
\item \emph{Tesseract.}
  Tesseract runs the authors' \texttt{tesseract-decoder}
  implementation~\cite{beni2025tesseract}, an A$^{*}$ search for the
  most likely error pattern with beam pruning. We keep the upstream
  search parameters, a detector beam of 5 and a priority-queue limit of
  $2\times10^{5}$, and add an ensemble of 50 detector orderings
  generated by breadth-first search.
\end{itemize}
\noindent Whenever the paper writes BP teacher, BP+OSD/LSD, Relay-BP,
the beam search decoder, or Tesseract, it refers to these
configurations.

We use the layered min-sum BP decoder defined above as the bootstrap
teacher because its per-variable logits come directly from one
invocation of local message passing, matching GraphMP's computational
form (Appendix~\ref{app:graphmp}). Section~\ref{app:sl-results} shows
that GraphMP largely reproduces this teacher. BP+OSD/LSD instead adds
global ordered-statistics post-processing to BP, while Relay-BP
aggregates repeated BP legs with re-drawn memory. Their final
predictions combine these successive stages. The bootstrap teacher's
soft logits provide an architecture-aligned message-passing target
for GraphMP.

In this work, GraphMP is initially distilled from the BP teacher
defined above. Subsequent self-distillation trains
GraphMP on verified trajectories supplied by the
model-guided beam-search decoder, with collection criteria specified
in Appendix~\ref{app:factory-setup}.
The third-party \texttt{ldpc} and \texttt{relay-bp} implementations
serve as evaluation baselines.

\section{Bootstrapping by Supervised Learning}
\label{app:sl}

GraphMP learns to decode from syndrome--error training pairs or
from the outputs of another decoder. We first describe these data
formats and the training losses, then give the training and
evaluation procedure used in our experiments.

\subsection{Supervised Data}
\label{app:sl-data}

We generate training data after choosing a decoding task, which
specifies the check matrix $H$ and the error distribution $\pi$.
For each example, we independently sample an error realization and
compute the observed syndrome $s$. We train GraphMP to map the input
$s$ to a label supplied by a simulator or a decoder acting as the
teacher. The label is either a hard binary error vector
$e^{\mathrm{tch}}\in\GF^{n}$ or a soft logit vector
$z^{\mathrm{tch}}\in\mathbb{R}^{n}$ on the same $n$ error variables.

\begin{itemize}[leftmargin=*]
  \item \emph{Hard labels.}
    To construct a hard training pair $(s,e^{\mathrm{tch}})$, a
    simulator draws $e^{\mathrm{tch}}\sim\pi$ and computes
    $s=He^{\mathrm{tch}}$ over $\GF$.
    Alternatively, we pass the sampled syndrome $s$ to a decoder and
    take its predicted error pattern as $e^{\mathrm{tch}}$.
  \item \emph{Soft labels.}
    To construct a soft training pair $(s,z^{\mathrm{tch}})$, we use
    a teacher decoder's per-variable posterior estimates
    $p^{\mathrm{tch}}(e_i=1\mid s)$. We express each estimate as
    log odds (a logit),
    \begin{equation}
      z_i^{\mathrm{tch}}
      \;=\;
      \log
      \frac{p^{\mathrm{tch}}(e_i=1\mid s)}{p^{\mathrm{tch}}(e_i=0\mid s)}.
      \label{eq:teacher-logit}
    \end{equation}
    A decoder that estimates these posterior probabilities, such as belief
    propagation, can supply those logits.
\end{itemize}

\begin{table}[!htbp]
  \centering
  \small
  \begin{tabular}{llll}
    \toprule
    Label type & Input & Label & Teacher \\
    \midrule
    Hard & $s$ & $e^{\mathrm{tch}}\in\GF^{n}$
      & simulator or a decoder \\
    Soft & $s$ & $z^{\mathrm{tch}}\in\mathbb{R}^{n}$
      & a decoder \\
    \bottomrule
  \end{tabular}
  \caption{Supervised formats.
    GraphMP distills from a simulator or a decoder teacher.
    A simulator supplies a hard label; a decoder may supply a hard
    label, a soft label, or both.}
  \label{tab:sl-data}
\end{table}

A simulator teacher only supplies hard labels.
A decoder teacher may supply hard, soft, or both labels.
We represent all three cases by a target probability
$q$.
\begin{equation}
  q_i =
  \begin{cases}
    e_i^{\mathrm{tch}}, & \text{hard},\\
    \sigma(z_i^{\mathrm{tch}}), & \text{soft},\\
    \eta e_i^{\mathrm{tch}}
      +(1-\eta)\sigma(z_i^{\mathrm{tch}}), & \text{mixed},
  \end{cases}
  \qquad i=1,\ldots,n,\quad 0\le\eta\le 1.
  \label{eq:supervision-label}
\end{equation}
The mixed case interpolates the hard bit and the teacher's soft
probability with weight $\eta$.
The experiments in Section~\ref{app:sl-results} use soft labels
from a belief-propagation (BP) decoder.

\subsection{Losses}
\label{app:sl-loss}

Write $p_i=\sigma(z_i)$ for the student's predicted posterior
probability for error variable $i$ in
Appendix~\ref{app:graphmp}, and
\begin{equation}
  \ell(q,p)
  \;=\;
  -q\log p-(1-q)\log(1-p)
  \label{eq:bce}
\end{equation}
for the binary cross-entropy between the target probability $q$
and the predicted probability $p$.
Building on error-variable and parity supervision~\cite{nachmani2016,liu2019,astra2024},
we use binary cross-entropies for error-variable predictions,
logical parities, and syndrome parities.

\begin{itemize}[leftmargin=*]
  \item \emph{Error-variable loss.}
    Per-variable binary cross-entropy on the training pairs of
    Section~\ref{app:sl-data}.
    The student's predicted posterior for error variable $i$ is $p_i$.
    The target probability $q_i$ is defined in
    \eqnref{eq:supervision-label}.
    \begin{equation}
      \mathcal{L}_{\mathrm{var}}
      \;=\;
      \frac{1}{n}\sum_{i=1}^{n}\ell(q_i,p_i).
      \label{eq:loss-mech}
    \end{equation}
    An optional extra weight on the rare $q_i=1$ terms accounts for
    sparse errors.
  \item \emph{Logical loss.}
    The logical bits are parities of error variables.
    A matrix $A$ with one column per error variable defines a
    \emph{soft parity} on the student's predicted posterior
    probabilities $p$.
    \begin{equation}
      \Pi_j(p;A)
      \;=\;
      \frac12\left(1-\prod_{i:A_{ji}=1}(1-2p_i)\right).
      \label{eq:soft-parity}
    \end{equation}
    The product runs over the error variables in the support of row
    $j$ of $A$; $\Pi_j$ gives the probability of odd parity under the
    factorized prediction.
    When every $p_i$ is $0$ or $1$, $1-2p_i=\pm 1$ and $\Pi_j$
    recovers the hard parity $(Ae)_j$ over $\GF$.
    When some $p_i$ lie in $(0,1)$, $\Pi_j$ lies in $(0,1)$ and
    moves continuously between the two parity values.
    Applying this soft-parity operation to the student's predicted
    posteriors with the logical matrix $L$, and matching the teacher's
    logical label $l=Le^{\mathrm{tch}}$, gives
    \begin{equation}
      \mathcal{L}_{\mathrm{log}}
      \;=\;
      \frac{1}{k}\sum_{j=1}^{k}
      \ell\bigl(l_j,\,\Pi_j(p;L)\bigr).
      \label{eq:loss-log}
    \end{equation}
    This term encourages the prediction to reproduce the
    logical outcome of its label.
  \item \emph{Syndrome loss.}
    The same soft parity through the check matrix $H$ matches the
    observed syndrome,
    \begin{equation}
      \mathcal{L}_{\mathrm{syn}}
      \;=\;
      \frac{1}{m}\sum_{c=1}^{m}
      \ell\bigl(s_c,\,\Pi_c(p;H)\bigr).
      \label{eq:loss-syn}
    \end{equation}
    Minimizing this differentiable loss encourages the predicted
    error pattern to satisfy $H\hat e=s$.
\end{itemize}

The three losses are combined as
\begin{equation}
  \mathcal L
  =\mathcal L_{\mathrm{var}}
  +\lambda_{\mathrm{syn}}\mathcal L_{\mathrm{syn}}
  +\lambda_{\mathrm{log}}\mathcal L_{\mathrm{log}},
  \qquad
  \lambda_{\mathrm{syn}},\lambda_{\mathrm{log}}\ge 0.
  \label{eq:sl-total}
\end{equation}
The coefficients control the balance between per-variable accuracy,
syndrome consistency, and logical correctness. A coefficient may
vanish when the corresponding supervision is unavailable, and its
value is otherwise selected as part of the training procedure.

\paragraph{Deep supervision.}
\eqnref{eq:sl-total} evaluates the output of one
message-passing round.
The same readout can be applied after every round $t=1,\ldots,T$.
Write
\begin{equation}
  z_i^{(t)}
  \;=\;
  \nn{Readout}_\theta\bigl(h_i^{(t)}\bigr),
  \qquad
  p_i^{(t)}
  \;=\;
  \sigma\bigl(z_i^{(t)}\bigr),
  \label{eq:round-readout}
\end{equation}
and write $\mathcal L_{\mathrm{var}}^{(t)}$ for the error-variable
loss evaluated on $p^{(t)}$. Following deep supervision~\cite{lee2015,astra2024},
we compare every round with the teacher using discounted weights,
while parity supervision acts on the final prediction:
\begin{equation}
  \mathcal{L}_{\mathrm{deep}}
  =
  \frac{\sum_{t=1}^{T}\gamma^{T-t}\mathcal L_{\mathrm{var}}^{(t)}}
       {\sum_{t=1}^{T}\gamma^{T-t}}
  +\lambda_{\mathrm{syn}}\mathcal L_{\mathrm{syn}}^{(T)}
  +\lambda_{\mathrm{log}}\mathcal L_{\mathrm{log}}^{(T)},
  \qquad 0<\gamma\le 1.
  \label{eq:deep-loss}
\end{equation}
The normalized weights set the relative emphasis on the rounds;
$\gamma<1$ places greater emphasis on later predictions.
At inference, the last-round logits $z^{(T)}$ determine the returned
predicted error pattern.

\subsection{Training and Evaluation}
\label{app:sl-results}

Classical knowledge can enter training through different targets.
An earlier GNN study combines logical-outcome supervision with an
auxiliary edge-probability loss against DEM priors~\cite{ikeda2025distillation}.
Foundation decoders train on classical-decoder predictions for
intermediate measurement-time prefixes~\cite{yan2026foundation}.
Here, BP provides syndrome-conditioned per-error probabilities in the
same output format as GraphMP.

We train one GraphMP model on three code-capacity tasks:
\BBgross, $\M_{150}$, and \GBeighteen, each with the noise model
specified in Appendix~\ref{app:task}.
For each task, we choose a nominal training error-rate setting, sample an error
$e\sim\pi$, and compute its syndrome $s=He$ over $\GF$.
The pair $(s,e)$ can be used directly as a training example.
Alternatively, we can decode $s$ with the BP teacher of
Section~\ref{app:bp-baselines} and use its
output as the label, as described in Section~\ref{app:sl-data}.
In this case, the BP decoder serves as the teacher, and GraphMP is
trained to match its output. This teacher--student training strategy
is called knowledge distillation.
We have found that both simulator supervision and decoder distillation
can yield effective GraphMP decoders. The model evaluated here is
distilled from a BP teacher.
Specifically, we record the teacher's per-variable logits
$z^{\mathrm{tch}}$ on each sampled syndrome and use
$q_i=\sigma(z_i^{\mathrm{tch}})$ as the soft label, following the
soft-label case of \eqnref{eq:supervision-label}.

The training error rate is chosen separately for each task,
slightly above its decoding threshold, to include difficult error
patterns on which the student can learn from the teacher's
predictions.
We minimize the training loss using the discounted deep-supervision
scheme of \eqnref{eq:deep-loss}: the error-variable loss
of \eqnref{eq:loss-mech} matches the BP soft labels at every
message-passing round, and the syndrome loss of
\eqnref{eq:loss-syn} is added at the final round.
The nominal training settings are $p=0.10$, $0.06$, and $0.10$ on
\BBgross, $\M_{150}$, and \GBeighteen, respectively. Training takes
$20\,000$ updates with minibatches of $64$ examples and deep-supervision
discount $\gamma=0.9$; the syndrome-loss coefficient is $1$ and the
logical-loss coefficient is $0$.

\begin{figure}[!htbp]
  \centering
  \includegraphics[width=\textwidth]{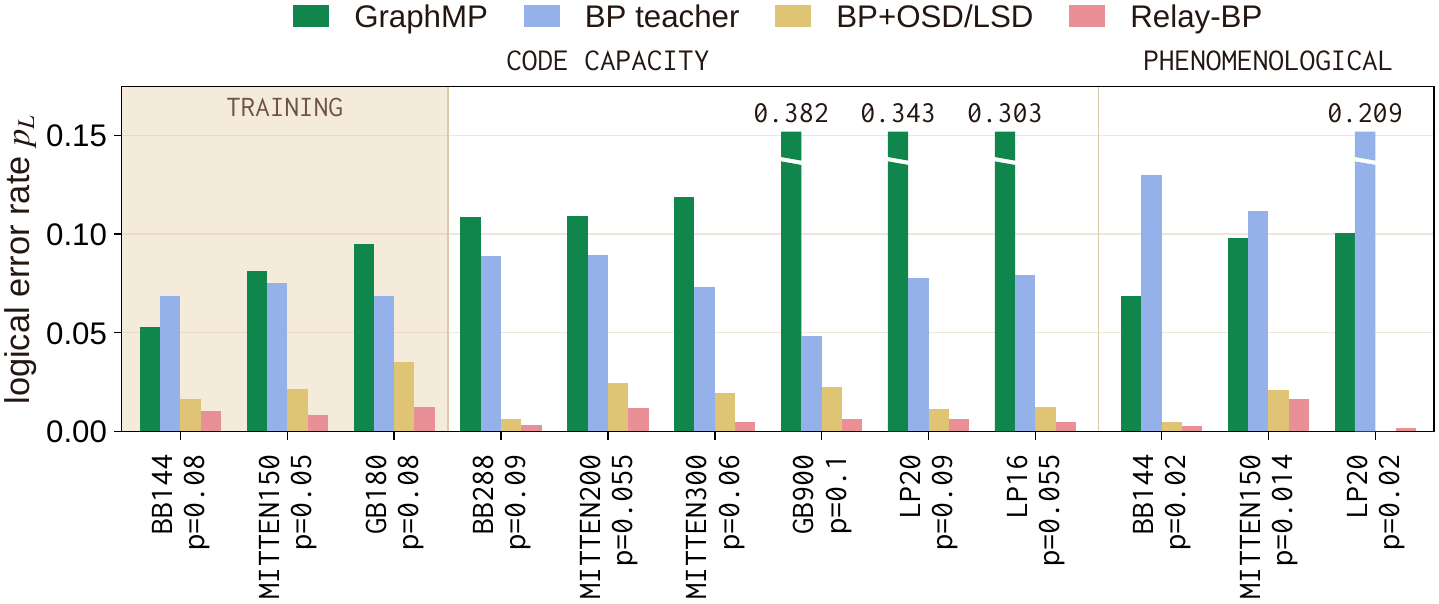}
  \caption{Logical error rate $p_L$ of GraphMP,
    its BP teacher, and reference decoders on the decoding tasks of
    Appendix~\ref{app:task}, with $2048$ samples per task.
    The shaded band marks the three code-capacity training tasks.
    The same model is applied without fine-tuning to the unshaded
    unseen tasks. The BP teacher, BP+OSD/LSD, and Relay-BP
    configurations are those of Section~\ref{app:bp-baselines}.}
  \label{fig:sl-ler}
\end{figure}

We evaluate GraphMP and the BP teacher on $2048$
sampled errors per decoding task, at the evaluation rates shown in
Figure~\ref{fig:sl-ler}. The first comparison asks how closely the
student approaches its teacher. On the three training tasks, GraphMP
achieves a comparable logical error rate: it improves on the teacher
for \BBgross\ and trails it for $\M_{150}$ and \GBeighteen.

Only the three code-capacity tasks in the shaded region are used for
training: no examples from the other nine tasks appear at any
training stage, and the model is not fine-tuned on them. We evaluate
the same model on these unseen tasks, which change the code size, the
code family, the physical error rate, or the noise model.
The comparison remains broadly similar on the smaller unseen
code-capacity tasks, while GraphMP achieves a lower logical error
rate than the teacher on each phenomenological task shown.
The model therefore transfers across the universal decoding interface
of Section~\ref{app:interface}.

The weakest code-capacity transfer occurs on
\GBnine, \LPtwenty, and \LPsixteen. GraphMP's logical error rates
range from $30.3\%$ to $38.2\%$, compared with $4.8\%$ to $7.9\%$
for the BP teacher. These codes contain 5 to 15 times as many
physical qubits as the largest training code
(Table~\ref{tab:codes}). The lifted-product code tasks also introduce
new error-variable degrees that were not seen in the training tasks.
We suspect that the transfer difficulty is associated with the change
in graph structure and in the distribution of syndrome patterns; the
specific causes and their individual effects remain an open question
for future work.

Finally, BP+OSD/LSD and Relay-BP, configured as in
Section~\ref{app:bp-baselines}, set a stronger benchmark in
Figure~\ref{fig:sl-ler} than the BP teacher, with substantially
lower logical error rates than GraphMP on every task shown.
Matching much of the teacher's performance therefore leaves a clear
gap to stronger decoders. Appendix~\ref{app:search} examines the
remaining decoding errors and develops a search algorithm to reduce
this gap.

\section{Enhancing by Search}
\label{app:search}

The supervised model represents the distribution over error variables as a
product of per-variable probabilities conditioned on the measured
syndrome.
The model's predicted error $\hat e$, assembled from those probabilities,
often leaves
unsatisfied detectors. We examine these violations and develop a
beam search algorithm that revises the predicted error $\hat e$ along
multiple candidate decoding paths.

Additional inference can also approximate probabilities of logical
classes. Tensor-network decoders estimate these probabilities by
approximate contraction, including for circuit-level surface-code
noise~\cite{piveteau2024tensor}.
The Frontier decoder processes error variables in order, sums the
probabilities of prefixes with matching residual syndromes and logical
labels, and prunes the retained list~\cite{leverrier2026frontier}.
Our beam search ranks candidate error patterns through repeated
conditional decoder queries. Its paths also provide the conditional
training examples developed in Appendix~\ref{app:search-learning}.

\subsection{Diagnosis: Syndrome Inconsistency}
\label{app:sl-diagnosis}
\label{app:factorization-problem}

\begin{figure}[!htbp]
  \centering
  \includegraphics[width=\textwidth]{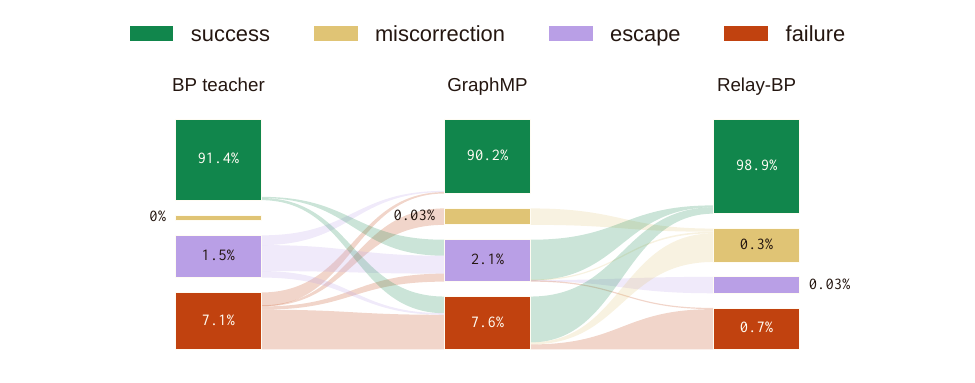}
  \caption{Decoding-outcome profiles of GraphMP in
    single-pass decoding, its BP teacher, and
    Relay-BP. Each profile classifies decoding outcomes as in
    Table~\ref{tab:residual}. Columns report the fraction of error
    realizations in each category. The logarithmic scale keeps rare
    outcomes visible. The shaded bands between columns follow groups
    of error realizations from their bucket under GraphMP to their
    bucket under each reference decoder; band width shows how many
    realizations take that path. Percentages are pooled over three
    code-capacity training tasks.}
  \label{fig:sl-residual}
\end{figure}

Figure~\ref{fig:sl-ler} shows that GraphMP approaches its BP teacher
while retaining a gap to stronger classical decoders. To locate that
gap, Figure~\ref{fig:sl-residual} compares the decoding-outcome profiles
of GraphMP, its BP teacher, and the stronger Relay-BP decoder
(Section~\ref{app:bp-baselines}). A
decoding-outcome profile is the distribution over the four residual
classes of Table~\ref{tab:residual}---success, miscorrection, escape,
and failure---as determined by the residual syndrome $Hr$ and the
logical error $Lr$.
Each column represents one decoder and is divided into buckets labeled
by the fraction of error realizations in each category. The displayed
bucket heights are plotted on a logarithmic scale so that rare outcomes remain
visible. The shaded bands between columns connect the buckets: each
band follows a group of error realizations from its bucket under
GraphMP to its bucket under the reference decoder, and the band's
width shows how many realizations take that path.

GraphMP and its teacher are comparable in every bucket, down to the
split between syndrome-consistent and syndrome-inconsistent outcomes;
the clear difference in Figure~\ref{fig:sl-residual} is between both
of them and Relay-BP. The bands show that the resemblance holds
realization by realization: about $67\%$ of the error realizations
GraphMP leaves syndrome inconsistent are also left inconsistent by
the teacher. The substantial overlap shows that both decoders struggle
with many of the same syndrome patterns and motivates extending
inference beyond single-pass teacher imitation.

The syndrome-inconsistent outcomes retain a precise, observable signal:
A nonzero residual syndrome ($H\hat e\neq s$) indicates that the
predicted error pattern still violates the measured syndrome constraints. Relay-BP turns
about $92\%$ of these outcomes into syndrome-consistent predictions and
$89\%$ into successes, demonstrating that most of this population is
recoverable by a stronger decoder. Miscorrections present a different
opportunity: they account for only $0.03\%$ of GraphMP outcomes, carry
no residual-syndrome signal, and remain miscorrections under Relay-BP in
this evaluation. The main recoverable gap is therefore the
syndrome-inconsistent population, where the decoder both knows that
work remains and has evidence that continued decoding can succeed.

GraphMP, like BP, represents the conditional error distribution with the
factorized conditional model
\(p(e\mid s)\approx\prod_i p(e_i\mid s)\)
(Appendix~\ref{app:architecture}). This representation enables local,
parallel inference, while independently predicted errors can leave a
residual syndrome. The
diagnosis above motivates a search procedure that follows the residual
syndrome and continues decoding.

\subsection{Beam Search}
\label{app:beam}

A search algorithm explores a state space by
applying actions to a current state, evaluates the resulting states
against a goal or objective, and selects which states to expand next.
Each later state is therefore conditioned on the previous state.
This conditioning allows the variables in the final state to be
correlated, even if each step performs an uncorrelated state update.
For decoding, a state contains a current syndrome and a pin mask that
records which error variables are free or pinned, which enables the
decoder to progressively improve the predicted error \(\hat e\),
building up correlations among the error variables through the search
process.

We implement this general search principle with the beam-search
decoding framework proposed in Ref.~\cite{ye2025}.
For each state, the pin mask \(\pin\) records whether each error variable
is free or pinned along the current path.
The soft decoder introduced in \eqnref{eq:graphmp-soft-decoder} maps
the current syndrome and pin mask to
\[
  \mathsf D_\theta(s,\pin\mid H)=(z,\hat e),
\]
where \(z\) contains one logit for each error variable, the magnitude
\(|z_i|\) measures the confidence of that prediction, and \(\hat e\) is
the predicted error under the current pins.
In the original work~\cite{ye2025}, the soft decoder is realized by
belief propagation.
In this work, GraphMP replaces belief propagation within the
beam-search framework through the same input--output interface. A
trained GraphMP model can therefore enter beam search without
additional training. At inference time, the beam-search algorithm
repeatedly calls the same trained GraphMP model, aiming to improve the
predicted error.

\begin{figure}[t]
  \centering
  \newcommand{\tstate}[5]{%
    \begin{scope}[shift={(#2,#3)}]
      \foreach \f [count=\j from 1] in {#4} {
        \node[chk, fill=\f] (#1c\j) at ({0.45*\j-0.225},0.55) {};
      }
      \foreach \f/\b [count=\i from 1] in {#5} {
        \node[bit, fill=\f, \b] (#1b\i) at ({0.45*\i-0.45},0) {};
      }
      \foreach \j in {1,...,4} {
        \pgfmathtruncatemacro{\k}{\j+1}
        \draw[ecLine, line width=0.4pt] (#1c\j) -- (#1b\j);
        \draw[ecLine, line width=0.4pt] (#1c\j) -- (#1b\k);
      }
      \node[frame, fit=(#1c1) (#1c4) (#1b1) (#1b5)] (#1) {};
    \end{scope}}
  \definecolor{ecInk}{HTML}{241610}
  \definecolor{ecInk2}{HTML}{6E5541}
  \definecolor{ecLine}{HTML}{8A6B5A}
  \definecolor{ecCard}{HTML}{DCC9AC}
  \definecolor{ecVerified}{HTML}{11864C}
  \begin{tikzpicture}[
      x=0.975cm, y=1cm,
      chk/.style={draw=ecLine, line width=0.5pt, minimum size=2.4mm, inner sep=0pt},
      bit/.style={circle, draw=ecLine, minimum size=3.0mm, inner sep=0pt},
      free/.style={line width=0.4pt},
      pin/.style={draw=ecInk, line width=1.1pt},
      frame/.style={draw=ecCard, line width=0.5pt, rounded corners=2pt, inner sep=3pt},
      done/.style={draw=ecLine, line width=1.0pt, rounded corners=2pt, inner sep=0pt},
      edge/.style={->, >={Stealth[length=4pt,width=3.5pt]}, draw=ecLine, line width=0.7pt},
      lbl/.style={font=\small, inner sep=1pt, text=ecInk},
      lvl/.style={font=\small, anchor=west, text=ecInk2}]
    \tstate{s}{-1.7}{0}{white,ecInk,white,ecInk}{white/free,white/free,white/free,white/free,white/free}
    \node[lvl, anchor=east] at ($(s.west)+(-0.15,0)$) {Problem};
    \node[lbl, anchor=south] at ($(s.north)+(0,0.05)$) {observed $s$};
    \node[lbl, anchor=north] at ($(s.south)+(0,-0.05)$) {unknown $e$};
    \definecolor{pOne}{HTML}{C1420F}
    \definecolor{pS0}{HTML}{FFFCF3}
    \definecolor{pS1}{HTML}{FFFCF3}
    \definecolor{pS1000000}{HTML}{C1420F}
    \definecolor{pS155172}{HTML}{F9E0D1}
    \definecolor{pS300000}{HTML}{F2C6B2}
    \definecolor{pS500000}{HTML}{E6A288}
    \definecolor{pS700000}{HTML}{D97E5E}
    \definecolor{pS999999}{HTML}{C1420F}
    \definecolor{pZero}{HTML}{FFFCF3}
    \tstate{r}{3.0}{0}{white,white,ecInk,white}{pS300000/free,pS500000/free,pS700000/free,pS500000/free,pS700000/free}
    \draw[edge] (s.east) -- node[lbl, above, align=center] {single-pass\\prediction} (r.west);
    \tstate{A}{0.6}{-2.3}{white,white,white,white}{pS999999/free,pOne/pin,pS0/free,pS1/free,pS700000/free}
    \tstate{B}{5.4}{-2.3}{white,white,white,ecInk}{pS0/free,pZero/pin,pS999999/free,pS999999/free,pS700000/free}
    \tstate{C}{2.9}{-4.6}{white,white,white,white}{pS0/free,pZero/pin,pS1000000/free,pS1000000/free,pZero/pin}
    \tstate{D}{8.0}{-4.6}{white,ecInk,white,white}{pS0/free,pZero/pin,pS300000/free,pS155172/free,pOne/pin}
    \node[done, fit=(A)] {};
    \node[done, draw=ecVerified, fit=(C)] {};
    \draw[edge] (r.south) -- node[lbl, above left] {$e_2=1$} (A.north);
    \draw[edge, draw=ecVerified, line width=1.4pt, >={Stealth[length=6pt,width=5.5pt]}] (r.south) -- node[lbl, above right] {$e_2=0$} (B.north);
    \draw[edge, draw=ecVerified, line width=1.4pt, >={Stealth[length=6pt,width=5.5pt]}] (B.south) -- node[lbl, above left] {$e_5=0$} (C.north);
    \draw[edge] (B.south) -- node[lbl, above right] {$e_5=1$} (D.north);
    \node[lbl, anchor=north] at ($(A.south)+(0,-0.05)$) {completed, weight $3$};
    \node[lbl, anchor=north, align=center] at ($(C.south)+(0,-0.05)$) {completed, weight $2$\\$\downarrow$ returned};
    \node[lbl, anchor=north, align=center] at ($(D.south)+(0,-0.05)$) {no completion\\at the final depth};
    \node[lvl] at (10.3,0) {Root};
    \node[lvl] at (10.3,-2.3) {Step 1};
    \node[lvl] at (10.3,-4.6) {Step 2};
  \end{tikzpicture}

  \caption{Beam search on a toy decoding task. Each box is a search
    state on the code's Tanner graph. Squares are checks, filled
    according to the residual syndrome left by the predicted error.
    Circles are error variables, shaded by their predicted error
    probability from cream at $0$ to orange at $1$. The soft decoder
    runs two rounds of belief propagation. A thick border marks a pinned
    variable. Each step pins the least-confident variable to $0$ and to
    $1$. A state with no residual syndrome is completed (thick frame).
    The search returns the completed state of lowest weight, and the
    green edges mark the trajectory that leads to it.}
  \label{fig:beam-tree}
\end{figure}
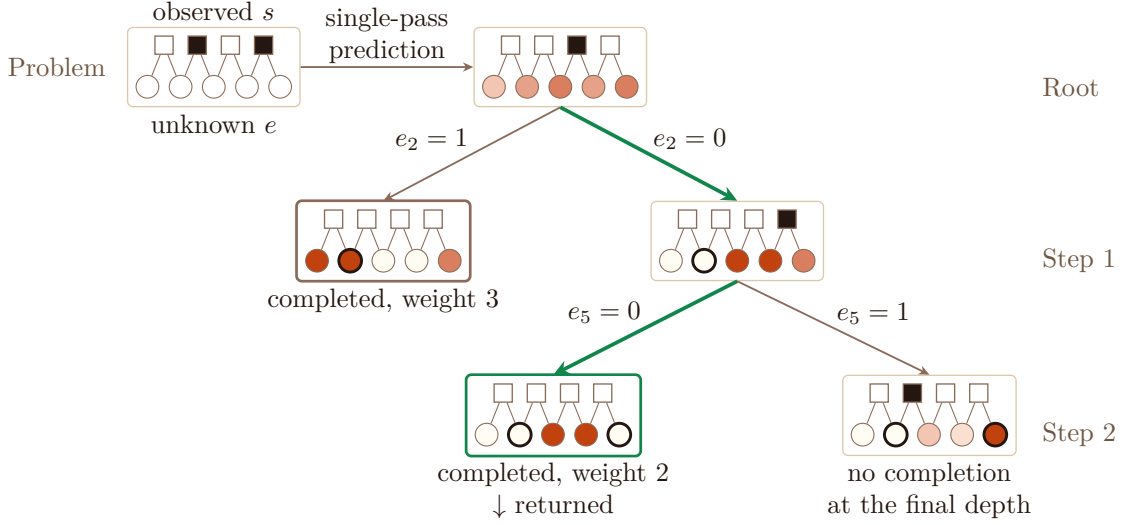

The states that the search creates form a tree
(Figure~\ref{fig:beam-tree}).
The first call of \(\mathsf D_\theta\) evaluates the \emph{root state},
in which every error variable is free.
If that predicted error already explains the measured syndrome, the
search stops and returns it.
Otherwise, from each state kept so far, the search selects the index
\(i_\star\) of least confidence among the free coordinates of the pin
mask, i.e., the coordinate that minimizes \(|z_i|\), and
\emph{branches} the state into two \emph{child states}
by pinning the corresponding error variable to \(0\) and to \(1\).
Each child updates the residual syndrome by the \(i_\star\)-th column
of the detector matrix,
\(s\leftarrow s\oplus H_{:,i_\star}e_{i_\star}\),
and the soft decoder \(\mathsf D_\theta\) evaluates the child and
fills in its logits and predicted error.
Each search state is a tuple
\((s,\pin,z,\hat e)\): the current syndrome, the pin mask, and the
logits and predicted error returned by the soft decoder.
Branching every kept state doubles the number of states at each step,
so the search needs \emph{pruning}: of all children created in one
step it keeps the \(W\) most promising, ranked by a frontier score
defined below, and discards the rest. The \emph{search width} \(W\) is
therefore the number of states kept after each step, and the
\emph{search depth} \(R\) bounds the number of steps.
A state whose predicted error explains the measured syndrome is
\emph{completed}. The sequence of states from the root state to a state whose
predicted error is returned is a \emph{search trajectory}. The state
at its end is the trajectory's \emph{final state}. When a completed
state is available, the search returns the predicted error of the
best completed state, and
Appendix~\ref{app:search-learning} reuses the trajectory that leads
to it.

Algorithm~\ref{alg:beam} maintains three sets of these states.
\(\mathcal A\) contains the active states that the search expands,
\(\mathcal B\) is a bucket that holds the child states created in the
current search step, and
\(\mathcal C\) collects completed states whose predicted error explains
the measured syndrome.
Write \(F(\pin)=\{i:\pin_i=\pinf\}\) for the indices of the free error
variables.
The frontier score \(u\) and the operator \(\operatorname{Top}_W\) are
defined below; they determine which children remain in \(\mathcal A\).

\begin{algorithm}[H]
  \caption{Decoder-guided beam search}
  \label{alg:beam}
  \small
  \begin{algorithmic}[1]
    \Require Measured syndrome $s_{\mathrm{obs}}$, DEM $(H,\pi)$,
      soft decoder $\mathsf D_\theta(\,\cdot\mid H)$,
      width $W$, depth $R$
    \Ensure Predicted error $\hat e$
    \State $s\gets s_{\mathrm{obs}}$,\quad $\pin\gets(\pinf)^n$,\quad
      $(z,\hat e)\gets\mathsf D_\theta(s,\pin\mid H)$ \Comment{evaluate the all-free root}
    \If{$H\hat e=s_{\mathrm{obs}}$} \Comment{test the full single-pass prediction}
      \State \Return $\hat e$ \Comment{return a syndrome-consistent root prediction}
    \EndIf
    \State $\mathcal A\gets\{(s,\pin,z,\hat e)\}$,\quad
      $\mathcal C\gets\varnothing$ \Comment{initialize active and completed states}
    \For{$r=1,\ldots,R$} \Comment{advance the search depth}
      \State $\mathcal B\gets\varnothing$ \Comment{start a new set of children}
      \ForAll{$(s,\pin,z,\hat e)\in\mathcal A$} \Comment{expand each active state}
        \State $i_{\star}\gets\arg\min_{i\in F(\pin)}|z_i|$
          \Comment{choose the index of least confidence among free error variables}
        \For{$e_{i_{\star}}\in\{0,1\}$} \Comment{pin the selected error variable to both values}
          \State $s'\gets s\oplus H_{:,i_{\star}}e_{i_{\star}}$
            \Comment{update the residual syndrome}
          \State $\pin'\gets\pin$,\quad
            $\pin'_{i_{\star}}\gets e_{i_{\star}}$
            \Comment{copy the mask and pin the error variable at $i_{\star}$}
          \State $(z',\hat e')\gets\mathsf D_\theta(s',\pin'\mid H)$
            \Comment{evaluate the child state}
          \State $\mathcal B\gets\mathcal B\cup
            \{(s',\pin',z',\hat e')\}$ \Comment{record the child state}
        \EndFor
      \EndFor
      \State $\mathcal C\gets\mathcal C\cup
        \{(s,\pin,z,\hat e)\in\mathcal B:H\hat e=s_{\mathrm{obs}}\}$
        \Comment{collect completed states}
      \State $\mathcal A\gets\operatorname{Top}_W
        (\mathcal B\setminus\mathcal C;u)$ \Comment{retain the strongest frontier states}
    \EndFor
    \If{$\mathcal C\neq\varnothing$} \Comment{a completed state was found}
      \State \Return the $\hat e$ from
        $\arg\min_{(s,\pin,z,\hat e)\in\mathcal C}w_\pi(\hat e)$
        \Comment{select by DEM weight}
    \EndIf
    \State \Return the $\hat e$ from
      $\arg\max_{(s,\pin,z,\hat e)\in\mathcal A}u(s,\pin,z)$
      \Comment{return the best remaining state}
  \end{algorithmic}
\end{algorithm}

\paragraph{Frontier score.}
While the beam-search algorithm explores partially pinned inputs,
it ranks them by a frontier score.
Let $s$ be the residual syndrome of the current assignment, the
detectors still unexplained by the pins, and let $z$ be the soft
decoder's logits on the free coordinates $F(\pin)$.
The frontier score
\begin{equation}
  u
  =\frac{1}{|F(\pin)|}\sum_{i\in F(\pin)}|z_i|
  -\lambda\frac{\lVert s\rVert_0}{m},
  \qquad\lambda>0,
  \label{eq:beam-frontier-score}
\end{equation}
favors assignments whose remaining predictions are decisive and whose
residual syndrome is small.
The operator $\operatorname{Top}_W$ retains the $W$ partial
assignments with the largest frontier scores.

\paragraph{Candidate selection.}
Let $\mathcal C$ collect the completed states found by the beam-search algorithm;
each full prediction, including its pinned values, satisfies
$H\hat e=s_{\mathrm{obs}}$ for the original measured syndrome.
The state syndrome $s$ is the part left unexplained by its pins.
The DEM assigns each candidate the weight
\begin{equation}
  w_\pi(\hat e)
  =\sum_{i=1}^{n}\hat e_i\log\frac{1-\pi_i}{\pi_i},
  \qquad
  \hat e^{\star}=\arg\min_{(s,\pin,z,\hat e)\in\mathcal C}w_\pi(\hat e).
  \label{eq:beam-weight}
\end{equation}
This weight is the negative log of the DEM prior up to an additive
constant.
The syndrome constraint defines the feasible candidates, and the DEM
prior ranks them.
Frontier scores rank partially pinned inputs; DEM weights rank completed
predictions.
When $\mathcal C$ is empty at the depth limit, the algorithm returns
the predicted error of the highest-ranked remaining state.

\subsection{Search Gain and Computational Cost}
\label{app:search-numbers}

We first quantify the gain from search for a search-trained GraphMP
model, \Gmodel{3} from the first of the five model factory runs (the runs are
defined in Appendix~\ref{app:factory}).
We compare single-pass decoding with two search widths on the
code-capacity tasks used for training. The later depth and budget
sweeps instead use the BP-distilled starting model \Gmodel{0} of the
same run, as does Figure~\ref{fig:main-tradeoff}.

\begin{table}[!htbp]
  \centering
  \small
  \begin{tabular}{lcccc}
    \toprule
    Code & $p$ & Single-pass & Search ($W=8$)
      & Search ($W=32$) \\
    \midrule
    \BBgross     & $0.08$ & $2.21^{+0.28}_{-0.25}\%$
      & $0.55^{+0.15}_{-0.12}\%$ & $0.46^{+0.14}_{-0.11}\%$ \\
    $\M_{150}$   & $0.05$ & $4.06^{+0.37}_{-0.34}\%$
      & $0.67^{+0.17}_{-0.13}\%$ & $0.33^{+0.12}_{-0.09}\%$ \\
    \GBeighteen  & $0.08$ & $3.91^{+0.37}_{-0.34}\%$
      & $0.73^{+0.17}_{-0.14}\%$ & $0.51^{+0.15}_{-0.11}\%$ \\
    \midrule
    Pooled       & ---    & $3.39^{+0.20}_{-0.18}\%$
      & $0.65^{+0.09}_{-0.08}\%$ & $0.44^{+0.07}_{-0.06}\%$ \\
    \bottomrule
  \end{tabular}
  \caption{Logical error rate $p_L$ of GraphMP \Gmodel{3},
    with pointwise $90\%$ Wilson confidence intervals. Each code is
    evaluated on $8192$ sampled errors. Single-pass decoding and both
    search settings process the same samples; the pooled row contains
    $24\,576$ samples. Single-pass decoding consists of one GraphMP
    forward pass. Search repeatedly queries that model with width $W$
    and maximum depth $R=10$.}
  \label{tab:search-gain}
\end{table}

Across the pooled evaluation, search reduces the logical error rate
from $3.39\%$ in single-pass decoding to $0.65\%$ at $W=8$ and
$0.44\%$ at $W=32$. Thus width-$8$ search lowers the rate by a factor
of $5.2$ with the same GraphMP model, and increasing
the width to $32$ lowers it by a factor of $7.8$ relative to
single-pass decoding. The same ordering appears on all three tasks,
showing that search turns the local predictions of a fixed GraphMP
model into substantially more accurate error-pattern predictions.

\begin{figure}[!htbp]
  \centering
  \includegraphics[width=0.90\textwidth]{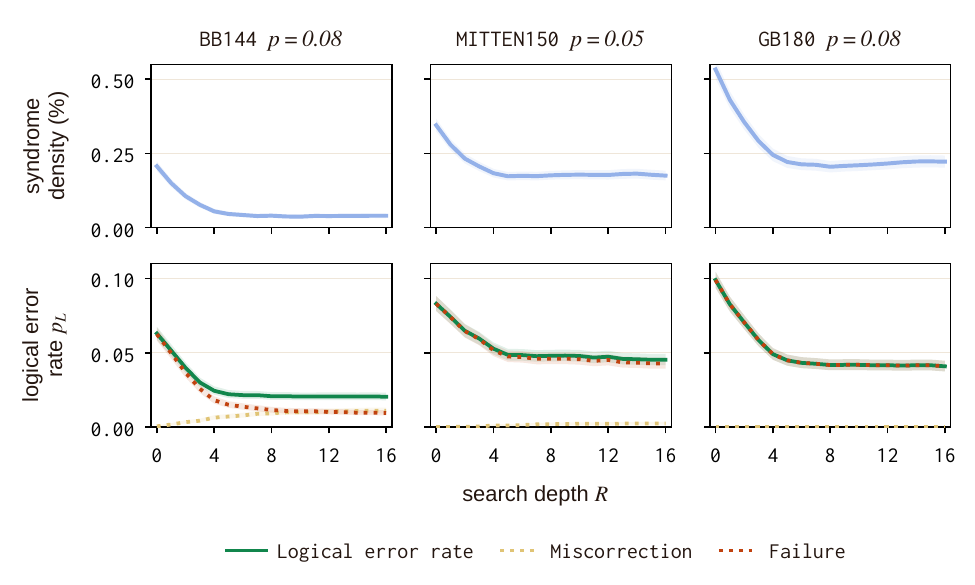}
  \caption{The BP-distilled starting model \Gmodel{0} at fixed search width $W=8$ on the
    three code-capacity training tasks.
    The left endpoint is single-pass decoding; the remaining points
    increase the search depth.
    Residual-syndrome density of the predicted error pattern is shown
    above logical error rate, which is decomposed into failures and
    miscorrections.
    Shaded bands are $90\%$ confidence intervals; $8192$ shots per
    task, the same shots at every search depth.}
  \label{fig:search-depth}
\end{figure}

Table~\ref{tab:search-gain} establishes the gain available from
search after trajectory training. To examine the starting model's
dependence on inference-time computation,
Figure~\ref{fig:search-depth} holds the search width fixed at $W=8$
and shows how decoding performance changes as the search depth
increases, starting from single-pass decoding.
Logical error rate is the evaluation objective, whereas syndrome
density diagnoses the parity inconsistency that remains along the
search trajectory.
Across all three decoding tasks, the logical error rate falls rapidly
at small search depths and then saturates; further depth yields
diminishing returns.
Additional depth shifts outcomes away from failure; on
\BBgross\ and $\M_{150}$, a growing contribution from miscorrections
offsets part of that improvement. Syndrome density on $\M_{150}$ and \GBeighteen\ rises
slightly after reaching its minimum while logical error remains near,
or continues toward, its minimum. Search depth is therefore selected
from the tradeoff between logical error rate and decoding cost, with
syndrome density serving as a diagnostic of the search trajectory.

\begin{figure}[!htbp]
  \centering
  \includegraphics[width=\textwidth]{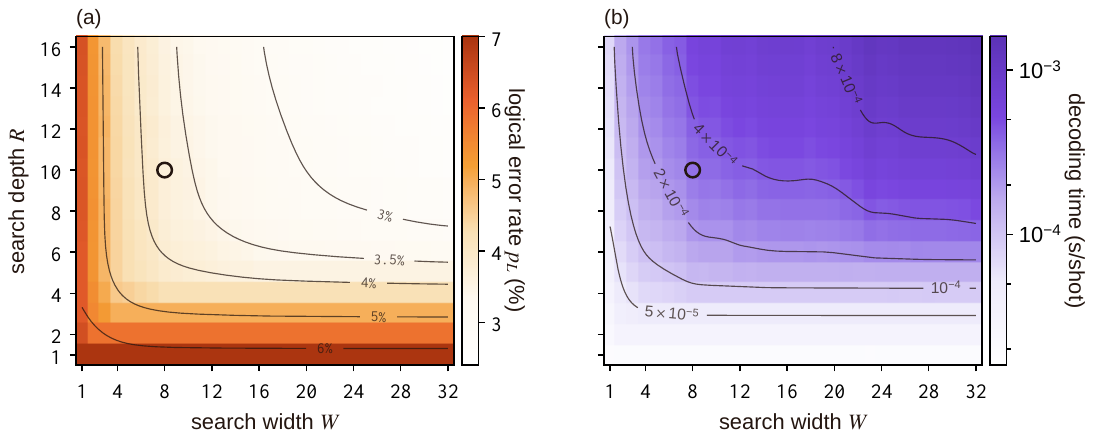}
  \caption{Accuracy--cost tradeoff of beam search with the
    BP-distilled starting model \Gmodel{0}.
    (a)~Pooled logical error rate \(p_L\) over search width \(W\) and
    depth \(R\) on the three training tasks at nominal evaluation
    error-rate settings \(0.08\) on \BBgross, \(0.05\) on
    \(\M_{150}\), and \(0.08\) on \GBeighteen; \(24\,576\)
    shots per cell, with the same sampled errors in every cell.
    (b)~Throughput-level decoding time
    (s/shot), averaged over the same three tasks.
    Circles mark the primary evaluation setting \(W=8\), \(R=10\).}
  \label{fig:search-budget}
\end{figure}

Depth and width provide complementary forms of coordination.
Increasing the depth extends each conditional assignment sequence,
while increasing the width preserves alternative pinning histories.
Figure~\ref{fig:search-budget} expands the starting-model
evaluation from the fixed-width slice of
Figure~\ref{fig:search-depth} to the full width--depth plane.
Figure~\ref{fig:search-budget}(a) reports the logical error rate over the three
training tasks. The pooled single-pass logical error rate is
$8.17\%$; search lowers it to $3.67\%$ at $W=8$, $R=10$ and
$2.64\%$ at $W=32$, $R=10$. The error rate falls rapidly over the smaller search
budgets and then gradually flattens, making the tradeoff between a
compact search and the remaining improvement directly visible.
Figure~\ref{fig:search-budget}(b) reports the throughput-level decoding time,
the incremental wall time associated with each
shot.
Decoding time grows with search width and search depth; the far
corner \(W=32\), \(R=16\) spends the most time to improve on
single-pass accuracy.
The circle marks the primary evaluation setting \(W=8\), \(R=10\).

\section{Distilling from Search}
\label{app:search-learning}

Appendix~\ref{app:sl} trains GraphMP to predict the error from the
measured syndrome in a single pass. Appendix~\ref{app:search} then
places the trained model inside beam search, which pins error
variables one at a time and calls the model again at every search
state along the way. Here we study the misalignment between training
and inference that results, show how to address the mismatch in model inputs by
distilling a new GraphMP model from the search states themselves, and
find that the distilled model is both a better single-pass decoder
and a better guide for beam search.
The student minimizes supervised losses on the free error variables at the recorded
search states.

\subsection{Training Objective versus Decoding inside Beam Search}
\label{app:misalignment}

Beam search starts from a root state, the measured syndrome $s$ with
every error variable free, and after each pinned variable calls the
model again at a new search state consisting of a residual syndrome
and a pin mask. Inside the search, each model call proposes how to continue from the current state, and only
the predicted error at the end of the trajectory is judged. The
supervised training of Appendix~\ref{app:sl} optimizes each call in
isolation. Its loss asks the prediction at the measured syndrome to be
the complete answer in one step, and the search procedure that will
call the model plays no part in the training. The mismatch affects both inputs and objectives.
\begin{itemize}[leftmargin=*]
  \item \emph{Inputs.} The states that the search visits, with pinned
    variables and residual syndromes, are absent from the training
    data.
  \item \emph{Objective.} The best one-step answer at a state and the
    best proposal for continuing the search from that state can
    differ, because the search pins one variable and calls the model
    again instead of committing to the whole prediction.
\end{itemize}
We call this mismatch the \emph{misalignment} between training and
inference.

\subsection{Learning from Search Trajectories}
\label{app:trajectory-data}

Appendix~\ref{app:beam} defines the trajectory of a beam search as
the sequence of search states from the root state to the final
state. We call a trajectory \emph{successful} when the predicted error
at its final state reproduces both the measured syndrome and the
logical outcome of the decoding experiment. Every state on a successful
trajectory is an input the model meets again when it guides a search,
and the outcome of the search tells how each of those states should
have been completed. Search trajectories can therefore serve as
supervision: each state of a successful trajectory becomes one
training record, constructed as follows.

\paragraph{Supervised data.}
For a search state with pin mask $\pin$, write $F=F(\pin)$ for the
indices of the free error variables, as in Appendix~\ref{app:beam},
and $P$ for the indices of the pinned ones, whose values $e_P$ the pin
mask records. The residual syndrome is the part of the measured
syndrome that the pinned values leave unexplained,
\begin{equation}
  s_{\mathrm{res}} = s \oplus H_P e_P .
\end{equation}
For a trajectory $\tau$, index its states by $t=0,\dots,T_\tau$,
with $t=0$ the root state and $t=T_\tau$ the final state, and mark quantities of state $t$
by the superscript $(t)$; in particular $F^{(t)}=F(\pin^{(t)})$ are
the free indices at state $t$. Two quantities from the search label
each state. The first is the \emph{final predicted error}, the
predicted error at the final state, which as a label we denote
$e^{\mathrm{tch}}$. Each pinned value along the trajectory agrees with
the final predicted error,
because the search reached the final state through those pins, so at
every state the free part of the final predicted error satisfies
$H_{F^{(t)}}e^{\mathrm{tch}}_{F^{(t)}} = s^{(t)}_{\mathrm{res}}$
and is a valid completion. The second is the logits
$z^{\mathrm{tch}}(x^{(t)})$ that the guiding model returned when the
search called it at state $t$. The beam-search decoder thereby supplies
the labels of Appendix~\ref{app:sl-data}, with the final predicted error as
its hard label and the logits as its soft label, and the training
record of state $t$ is
\begin{equation}
  x^{(t)}=\bigl(s^{(t)}_{\mathrm{res}},\pin^{(t)}\bigr),\qquad
  y^{(t)}=\Bigl(z^{\mathrm{tch}}_{F^{(t)}}\bigl(x^{(t)}\bigr),\;
               e^{\mathrm{tch}}_{F^{(t)}}\Bigr),
  \label{eq:trajectory-record}
\end{equation}
in the same order as the input and output of the soft decoder in
Appendix~\ref{app:beam}. The two labels depend on $t$ differently. The
hard label is one vector for the whole trajectory, restricted to the
variables still free at state $t$; the soft label is a function of the
state, because the guiding model's logits change as variables are
pinned. These logits describe the guide's predictions at each visited
state.
At the root state every variable is free, so
$x^{(0)}=(s,\pin^{(0)})$ is the measured syndrome with an all-free pin
mask and $y^{(0)}$ carries the whole final predicted error; the root record
therefore keeps single-pass decoding in the training set.

\paragraph{Two training schemes.}
The same set $\mathcal T$ of successful trajectories can supervise a
model in two ways, which differ only in which records of each
trajectory $\tau\in\mathcal T$ enter the training set.
\begin{itemize}[leftmargin=*]
  \item \emph{Root-state training} keeps only the root record of each
    trajectory,
    \begin{equation}
      \mathcal D_{\mathrm{root}}
        = \bigl\{(x^{(0)},y^{(0)}):\tau\in\mathcal T\bigr\}.
    \end{equation}
    The input is the measured syndrome with every variable free and
    the label is the final predicted error, so the model is asked to map the
    search's input directly to the search's output.
  \item \emph{Trajectory training} keeps every record,
    \begin{equation}
      \mathcal D_{\mathrm{traj}}
        = \bigl\{(x^{(t)},y^{(t)}):\tau\in\mathcal T,\;t=0,\dots,T_\tau\bigr\}.
    \end{equation}
    The same final predicted error is presented at the root state and at
    every pinned state along the way, so the model is also asked to
    complete it from partially pinned inputs.
\end{itemize}
Both training sets come from the same trajectories and hence the same
error realizations and final predicted errors, and
$\mathcal D_{\mathrm{root}}\subset\mathcal D_{\mathrm{traj}}$; the
difference is exactly the records of pinned states. We call the
resulting models the \emph{root-state-trained} and the
\emph{trajectory-trained} model; root-state training serves as the
control in Section~\ref{app:alignment-result}.
Sampling root and pinned records can also change how often each
original problem is presented. We examine the combined effect in the
root-versus-trajectory comparison, then match the ordered source
problems and final labels to isolate conditional-state training.

\paragraph{Loss.}
The pin-conditioned model of \eqnref{eq:pinned-factorized-posterior}
predicts the free variables of each recorded state. At message-passing
round $t$, let $p_i^{(t)}$ be its predicted probability for free
variable $i$, and write $q_i^{\mathrm{soft}}=\sigma(z_i^{\mathrm{tch}})$
for the guiding model's probability at that state. The hard and soft
error-variable losses are
\begin{equation}
  \mathcal L_{\mathrm{hard}}^{(t)}
  =\frac{1}{|F|}\sum_{i\in F}\ell(e_i^{\mathrm{tch}},p_i^{(t)}),
  \qquad
  \mathcal L_{\mathrm{soft}}^{(t)}
  =\frac{1}{|F|}\sum_{i\in F}\ell(q_i^{\mathrm{soft}},p_i^{(t)}),
\end{equation}
where $\ell$ is the binary cross-entropy of \eqnref{eq:bce}.
Pinned values remain fixed and receive no prediction loss. Model factory
training combines the verified final label at the last message-passing
round with the guide's soft targets throughout message passing:
\begin{equation}
  \mathcal L_{\mathrm{traj}}
  =\mathcal L_{\mathrm{hard}}^{(T)}
   +\frac{\sum_{t=1}^{T}w_t\mathcal L_{\mathrm{soft}}^{(t)}}
         {\sum_{t=1}^{T}w_t},
  \qquad w_T=1,\quad w_t=0.5\quad(t<T).
  \label{eq:trajectory-loss}
\end{equation}
The same soft target supervises every message-passing round at a
given search state. The comparisons in Section~\ref{app:alignment-result}
minimize $\mathcal L_{\mathrm{hard}}^{(T)}$ alone, so both students learn
from the same verified final predictions and differ in their exposure
to conditional inputs and the corresponding free-variable loss.

\paragraph{How records are sampled.}
Training minibatches mix root-state and pinned-state records.
Root records train the model to predict the complete error from the
measured syndrome; pinned records train it to complete the remaining
prediction given a residual syndrome and partially fixed error variables.
The mixture thus trains the model for both single-pass decoding and
its calls inside beam search.

\subsection{Collecting Training Trajectories}
\label{app:trajectory-collection}

Training on search states requires successful trajectories that provide
both valid labels and examples with pinned variables. Collection must
therefore verify the search outcome and select problems that require
search but remain solvable within its budget.

\paragraph{Logical verification of trajectories.}
The collector retains a final prediction and its path when the prediction is
syndrome-consistent and logically correct. Simulation computes these checks from the
sampled error. The accepted final prediction explains the measured
syndrome and has the same logical effect as the sampled error.
On a quantum processor, dedicated memory or benchmark experiments can
provide feedback on an accessible logical observable: prepare a known
logical state, extract the syndrome, run the beam-search decoder,
incorporate its recovery physically or into the Pauli frame, and
measure the encoded logical observable in the prepared basis at the end
of the experiment. Separate preparations and measurement bases can test
the corresponding logical sectors. Our collection experiments run in
simulation, which checks logical correctness for all observables in
the DEM. Dedicated device experiments are a prospective extension.

This logical verification belongs to data collection. During ordinary
decoding of protected logical information, the decoder observes the
syndrome and ranks completed candidates by the DEM weight of
\eqnref{eq:beam-weight}; it receives no logical outcome. Dedicated
benchmark shots provide the successful trajectories used for later
training, while ordinary decoding uses the deployed model without that
feedback.

\paragraph{Collecting verified trajectories efficiently.}
To learn how to guide beam search, the model needs training examples
from the pinned states that search visits. These states arise only
when search proceeds beyond the root: if the single-pass prediction
already satisfies the measured syndrome, the search returns that
prediction without pinning any variables. Otherwise, search branches
by pinning variables and asks the model to complete the remaining
prediction. When the resulting trajectory passes joint syndrome and logical verification,
its final predicted error supplies a valid completion for every
intermediate state. Training on these records teaches the model how
to complete a prediction partway through search, including cases it
could not solve at the root.

The physical error rate controls how often collection reaches these
states and how often search can resolve them. At too low an error
rate, single-pass decoding usually satisfies the syndrome, so collection
provides few examples with pinned variables. At too high an error
rate, more realizations require search, but the available search width
and depth may be insufficient to find a logically correct prediction;
unsuccessful trajectories provide no accepted training records.
We therefore select and adjust the physical error rate for the decoding
task, the model's capability, and the search budget to collect successful
trajectories with pinned states efficiently. This means choosing error
rates that produce problems that challenge the model in a single pass
but remain solvable through search.

\subsection{Measuring and Resolving Misalignment}
\label{app:alignment-result}

To address the mismatch in model inputs, we train on data
that the search itself generates. Section~\ref{app:trajectory-data}
showed that a set of successful searches yields two training sets,
the root-state dataset $\mathcal D_{\mathrm{root}}$ of root records and the
trajectory dataset $\mathcal D_{\mathrm{traj}}$ of all records. We train one
model on each under identical conditions, including the
initialization, the collected trajectories, the final predicted errors used
as labels, and the number of updates. The sampling rules differ in
how frequently they present each source problem as well as in the
conditional states supplied to the model. Each model trains for $5000$ updates with minibatches of $256$ records;
the final checkpoint is evaluated for each of $3$ independent training
seeds. We compare the two models in two settings: at the states a search
visits, where the misalignment was defined, and as decoders, where it
matters.

\paragraph{Where misalignment shows.}
We measure the misalignment at the states a search visits. The
evaluation set consists of fresh $\M_{150}$ realizations that
single-pass decoding fails and search resolves, together with the
recorded trajectory of each successful search. Each recorded state is
a decoding problem in its own right: the model receives the input
$x=(s_{\mathrm{res}},\pin)$ and returns a predicted error $\hat e$,
and the state counts as solved when the decoding is a success in the
sense of Table~\ref{tab:residual}, with the pinned variables held at
their pinned values. On a successful trajectory every pinned value
agrees with the final predicted error, so moving from the root state
toward the final state hands the model more and more of a valid
answer: the root state is the original problem, and each later state
is the problem with one more of the least-confident variables already
resolved. At the final state the guiding model completed the
remaining variables correctly in a single call. A model that makes
use of the pin mask should therefore fail less and less often along
the trajectory.

\begin{table}[!htbp]
  \centering
  \small
  \begin{tabular}{lrrr}
    \toprule
    & \multicolumn{3}{c}{Failure rate when evaluated at}\\
    \cmidrule(lr){2-4}
    Model & root state & pinned states & final state\\
    \midrule
    Root-state-trained  & $95.7\pm2.8\%$ & $76.4\pm6.3\%$ & $47.5\pm11.6\%$\\
    Trajectory-trained  & $88.0\pm1.0\%$ & $48.3\pm3.6\%$ & $15.5\pm2.9\%$\\
    \bottomrule
  \end{tabular}
  \caption{The misalignment measured at the states a search visits.
    Two GraphMP models trained on the same trajectories, one on their
    root records only and one on every record, are evaluated at the
    recorded states of $622$ held-out successful searches on $\M_{150}$
    (nominal error-rate setting $0.06$, search width $W=32$). The root-state-trained
    model fails on most pinned states even though the pinned values
    are correct; the trajectory-trained model turns the pinned values
    into completions. The pinned-state column averages over the states
    of a trajectory and then over trajectories. Mean $\pm$ standard
    deviation over $3$ training seeds. A prediction fails when it is
    syndrome-inconsistent or logically incorrect. The probe evaluates
    students on selected teacher states that supply correct partial
    information.}
  \label{tab:conditional-probe}
\end{table}

Table~\ref{tab:conditional-probe} reports how often each model fails
to decode the recorded states. The evaluation set contains only
realizations that single-pass decoding fails, so both models fail on
most root states, as expected. The pinned states are
where the two models part ways. The root-state-trained model still
fails on $76\%$ of the pinned states and on $48\%$ of the final
states, although at a final state the variables the search found
most difficult have already been pinned and the guiding model
completed the
rest in one call. The trajectory-trained
model fails on $48\%$ of the pinned states and $16\%$ of the final
states. The root-state-trained model uses the pinned values less
effectively than the trajectory-trained model on these selected teacher states.

Root-state training and trajectory training differ in what they show
the model. Root-state training presents only the final predicted
error that the search arrived at. It supplies no examples of how
to complete that prediction from the partially pinned inputs visited
inside beam search. The conditional-state comparison above shows
that the root-state-trained model makes less effective use of those
inputs. Trajectory training instead
supervises the model at every step of the search, where the pinned
values are part of the input and only the remaining variables have to
be predicted, so the model learns to complete the remaining prediction
step by step.

\paragraph{Decoding inside beam search.}
We also evaluate the same training intervention in end-to-end decoding.
Table~\ref{tab:trajectory-vs-root} deploys the same two models as
decoders at the evaluation error rate. Inside beam search with width
$W=8$ and depth $R=10$ the trajectory-trained model lowers the joint
decoding-failure rate by a factor of $6$, while its single-pass failure
rate is equal or lower.

\begin{table}[!htbp]
  \centering
  \small
  \begin{tabular}{lcccc}
    \toprule
    & \multicolumn{2}{c}{Single-pass} & \multicolumn{2}{c}{Search ($W=8$, $R=10$)}\\
    \cmidrule(lr){2-3}\cmidrule(lr){4-5}
    & Root-state & Trajectory & Root-state & Trajectory\\
    \midrule
    Joint failure rate & $3.62\pm1.22\%$ & $1.94\pm0.16\%$ & $0.68\pm0.26\%$ & $0.11\pm0.02\%$\\
    \bottomrule
  \end{tabular}
  \caption{The same two models as Table~\ref{tab:conditional-probe},
    deployed as decoders on $\M_{150}$ at nominal error-rate setting $0.03$.
    Trajectory training improves both modes, with the larger relative
    gain inside beam search. A prediction fails when it is
    syndrome-inconsistent or logically incorrect. Every model is scored
    on the same $32\,768$ error realizations; mean $\pm$ standard
    deviation over $3$ training seeds. The comparison includes
    differences in source sampling as well as conditional-state training.}
  \label{tab:trajectory-vs-root}
\end{table}

The benefit of conditional training is larger inside beam search
than in single-pass decoding: between the two models, the relative single-pass gain
is smaller than the search-enhanced gain, whereas failure at search states
differs by tens of percentage points and joint failure inside beam
search by up to a factor of $7$. The misalignment of
Section~\ref{app:misalignment} came from deploying a decoder trained
for single-pass decoding inside a search, and the largest gain from
trajectory training appears precisely there, in decoding inside beam
search. Trajectory training exposes the decoder to conditional inputs
encountered inside the search and improves its performance in that setting.
The state probe measures completion on selected teacher inputs;
end-to-end evaluation measures the student's deployed decoder.

To separate conditional training from source sampling, we pair every
sampled trajectory record with a root record carrying the same final
label, restoring its original syndrome and all-free mask. Both students
then receive the same ordered source problems and labels at every
training step. The trajectory-trained student learns from residual
syndromes and pinned values, with the loss restricted to free variables.
Its search-enhanced joint failure rate is lower in all $3$ independent
training seeds. On this $\M_{150}$ task, conditional-state training
therefore contributes to the improvement beyond changes in how often
individual problems are presented.

\section{Decoder Model Factory}
\label{app:factory}

Appendices~\ref{app:sl}, \ref{app:search}, and \ref{app:search-learning}
establish the components of the decoder model factory implementation
evaluated here.
Supervised learning from a BP teacher produces an initial decoder.
Beam search improves the decoder's performance, and knowledge distillation
from verified search trajectories trains the next model.
The updated model guides the next round of search and data collection,
closing a recurring collection-and-training loop. For the target
decoding tasks, we repeat this loop to produce successive models and
measure how model performance and training-data value evolve.
In this implementation, the message-passing architecture remains fixed
while the training data accumulate.

\subsection{Model Factory Setup}
\label{app:factory-setup}

We train one shared GraphMP model for three code-capacity target tasks
on \BBgross, $\M_{150}$, and \GBeighteen. Starting from the BP-distilled
model \Gmodel{0} of Appendix~\ref{app:sl}, the model factory produces \Gmodel{1}--\Gmodel{3}.
G labels model generations and D labels new trajectory datasets:
\Ddata{1} is collected by \Gmodel{0} and first contributes to training \Gmodel{1}, and so on.
\Gmodel{x} denotes the single-pass decoder, and \Gsearch{x} denotes the same model
inside beam search, in either collection or decoding mode.

The collector accepts the selected final prediction when it is
syndrome-consistent and logically correct. Its final predicted error
supplies the hard label for the recorded states on the selected
path; the guiding model also supplies soft labels at those states,
following Appendix~\ref{app:search-learning}. The sampled error and
accepted prediction are retained so that both checks can be repeated.
We repeat the complete collection-and-training procedure five times,
starting from independent initializations and generating a new set of
trajectory data in each run.

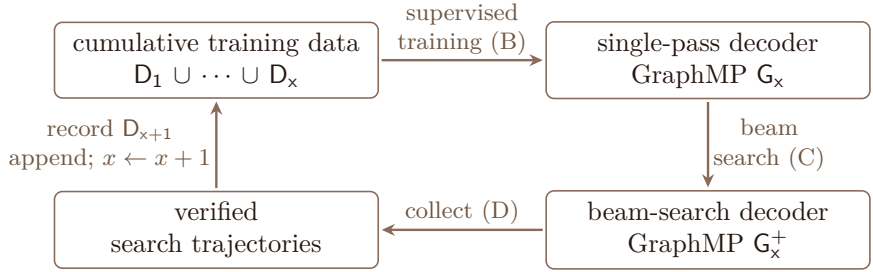
\begin{figure}[htbp]
  \centering
  \definecolor{ecInk}{HTML}{241610}
  \definecolor{ecInk2}{HTML}{6E5541}
  \definecolor{ecLine}{HTML}{8A6B5A}
  \begin{tikzpicture}[
      node distance=12mm and 22mm,
      box/.style={draw=ecLine, line width=0.6pt, rounded corners=3pt,
                  inner sep=4pt, align=center, font=\small, text=ecInk,
                  text width=4.0cm, minimum height=10mm},
      arr/.style={->, >={Stealth[length=5pt,width=4.5pt]}, draw=ecLine,
                  line width=0.9pt, shorten >=1pt, shorten <=1pt},
      lbl/.style={font=\footnotesize, text=ecInk2, inner sep=0.6pt, align=center}]
    \node[box] (data) {cumulative training data\\
      $\Ddata{1}\cup\cdots\cup\Ddata{x}$};
    \node[box, right=of data] (model) {single-pass decoder\\GraphMP \Gmodel{x}};
    \node[box, below=of model] (plus) {beam-search decoder\\GraphMP \Gsearch{x}};
    \node[box] (traj) at (data |- plus) {verified\\search trajectories};
    \draw[arr] (data) -- node[lbl, above=1pt] {supervised\\training (B)} (model);
    \draw[arr] (model) -- node[lbl, right=1pt] {beam\\search (C)} (plus);
    \draw[arr] (plus.west) -- node[lbl, above=1pt] {collect (D)} (traj.east);
    \draw[arr] (traj) -- node[lbl, left=1pt]
      {record \Ddata{x+1}\\append; $x\leftarrow x+1$} (data);
  \end{tikzpicture}

  \caption{The GraphMP collection-and-training loop for $x\geq1$. Accumulated data train \Gmodel{x};
    \Gsearch{x} collects verified trajectories that supply records \Ddata{x+1}.
    Appending those records and advancing $x$ starts the next round.
    The labels B, C, and D refer to the preceding appendices.}
  \label{fig:factory-loop}
\end{figure}

Within the model factory, beam search operates in \emph{collection mode}:
the GraphMP model guides a wider search ($W=32$, $R=10$) to generate training
data for its successors, with the acceptance rule above.
When the trained model is used for inference, beam search switches to
\emph{decoding mode}, with a narrower search ($W=8$, $R=10$). The decoder
selects its final prediction by the DEM-based candidate score of
\eqnref{eq:beam-weight}, without access to logical feedback.
We evaluate decoder performance in decoding mode.

\begin{table}[!htbp]
\centering
\small
\begin{tabular}{llll}
\toprule
Generation & Initialization & Training data & Data collection\\
\midrule
\Gmodel{0} & random & BP teacher & \Gsearch{0} collects \Ddata{1}\\
\Gmodel{1} & random & \Ddata{1} & \Gsearch{1} collects \Ddata{2}\\
\Gmodel{2} & \Gmodel{1} & \Ddata{1} $\cup$ \Ddata{2} & \Gsearch{2} collects \Ddata{3}\\
\Gmodel{3} & \Gmodel{2} & \Ddata{1} $\cup$ \Ddata{2} $\cup$ \Ddata{3} & ---\\
\bottomrule
\end{tabular}
\caption{Model initialization, training data, and subsequent data collection.
  Every model trains for $20\,000$ updates on the three
  target tasks. The model factory run ends at \Gmodel{3}.}
\label{tab:factory-lineage}
\end{table}

Table~\ref{tab:factory-lineage} specifies each model's initialization,
training data, and subsequent data collection. \Gmodel{1} starts from a fresh initialization because it
is the first model trained on pinned search states. 
The trajectory-trained models minimize cross entropy against the hard
and soft labels on free variables, as in Appendix~\ref{app:search-learning}.
All models have hidden dimension $32$ and $24$ message-passing rounds.
Each trajectory-training update trains the shared model on all three target tasks, with
$256$ records per task. We use AdamW with weight decay $0.01$ and gradient
clipping at $1$. The learning rate is $10^{-3}$ for \Gmodel{1} and
$3\times10^{-4}$ for later generations, with $1000$ warm-up updates
followed by cosine decay. Evaluation uses the final checkpoint of each
$20\,000$-update training run.
At each generation, we choose the collection error rates for the
model's current capability, following the balance between search
participation and successful completion described in
Section~\ref{app:trajectory-collection}.

\subsection{Model Improvement Across Generations}
\label{app:generations}

\begin{figure}[!htbp]
  \centering
  \includegraphics[width=\textwidth]{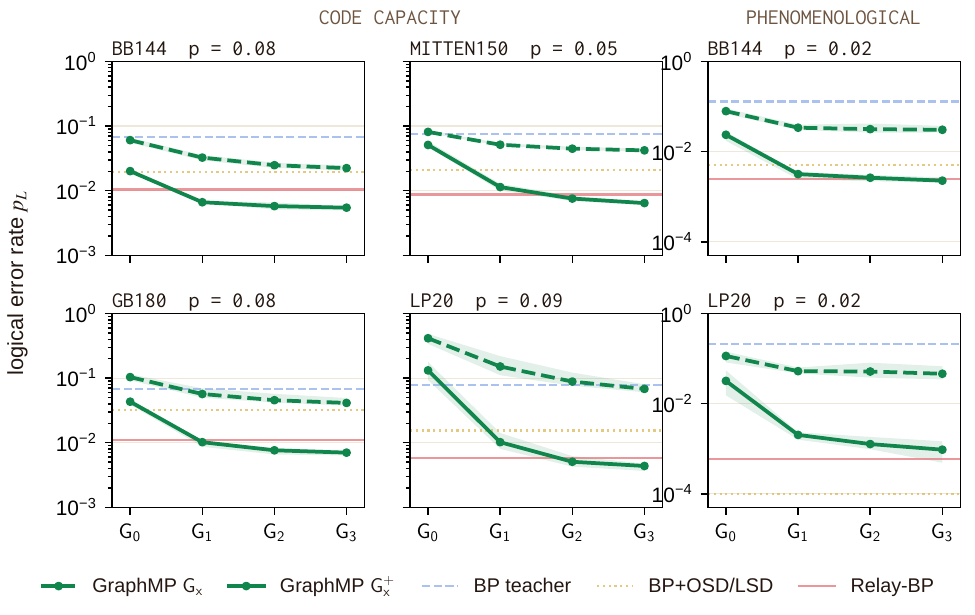}
  \caption{Logical error rate of GraphMP versus generation. Dashed:
    single-pass decoder GraphMP \Gmodel{x}; solid: search-enhanced decoder
    GraphMP \Gsearch{x}.
    Horizontal lines: BP teacher, BP+OSD (BP+LSD on the phenomenological tasks),
    and Relay-BP. Curves are means over five independent model factory runs;
    bands are the range across those runs. The
    phenomenological panels have their own vertical axis.}
  \label{fig:gen-progress}
\end{figure}

Figure~\ref{fig:gen-progress} shows the logical error rate of GraphMP
over the four generations \Gmodel{0}--\Gmodel{3}. Each generation is evaluated in two
modes: the single-pass decoder \Gmodel{x} and the search-enhanced decoder
\Gsearch{x}. These two modes appear as two curves of logical error rate
versus generation.
Single-pass logical error falls generation by generation, and so does
the search-enhanced logical error rate. The largest step is from
\Gmodel{0} to \Gmodel{1}; later steps are smaller in both modes.
The single-pass decoder falls below the BP teacher. The
search-enhanced decoder reaches Relay-BP and BP+OSD
(BP+LSD on the phenomenological tasks). Beam search together with
distillation therefore closes the performance gap between the BP-distilled
student and the stronger classical decoders shown in Figure~\ref{fig:sl-ler}.

Table~\ref{tab:generation-gains} reports the reduction in logical error
rate at each transition, averaged equally over the three target tasks.
Every transition improves this portfolio mean in all five independent
runs. The largest gains occur when the model first learns from verified
trajectories; later generations refine the decoder further. For
search-enhanced decoding, the last transition improves on $\M_{150}$ in
all five runs and on \BBgross\ and \GBeighteen\ in four of the five runs.

\begin{table}[!htbp]
\centering
\small
\begin{tabular}{lcc}
\toprule
Transition & Single-pass reduction & Search-enhanced reduction\\
\midrule
\Gmodel{0}$\to$\Gmodel{1} & $3.495\;[3.072,3.975]$ & $2.872\;[2.612,3.182]$\\
\Gmodel{1}$\to$\Gmodel{2} & $0.858\;[0.663,1.050]$ & $0.242\;[0.199,0.293]$\\
\Gmodel{2}$\to$\Gmodel{3} & $0.309\;[0.224,0.480]$ & $0.068\;[0.037,0.110]$\\
\bottomrule
\end{tabular}
\caption{Reduction in the equal-task mean logical error rate, in
  percentage points, on the three code-capacity target tasks,
  \BBgross, $\M_{150}$, and \GBeighteen. Each entry is the mean
  reduction over five independent model factory runs, followed by the
  range across those runs. Every model is evaluated on the same
  $8192$ realizations per task.}
\label{tab:generation-gains}
\end{table}

\paragraph{Logical correctness and syndrome consistency.}
\label{app:factory-recovery}
A valid recovery requires both the correct logical outcome and a
prediction consistent with the measured syndrome, as defined in
Appendix~\ref{app:task}. Table~\ref{tab:factory-recovery} reports
these two properties for the final-generation decoder on fresh
realizations. Beam search sharply reduces syndrome inconsistency,
so the joint failure rate is close to the logical error rate on the
three target tasks and on the unseen code-capacity task. The distinction
is more visible on phenomenological \LPtwenty, where beam search gives
a logical error rate of $0.095\%$ and a joint failure rate of $0.178\%$.

\begin{table}[!htbp]
\centering
\small

\begin{tabular}{lcccccc}
\toprule
 & \multicolumn{3}{c}{Single-pass \Gmodel{3}}
 & \multicolumn{3}{c}{Search-enhanced \Gsearch{3}}\\
\cmidrule(lr){2-4}\cmidrule(lr){5-7}
Task & Logical & Syndrome & Joint & Logical & Syndrome & Joint\\
\midrule
Code-capacity \BBgross & $0.02239$ & $0.02346$ & $0.02439$ & $0.005469$ & $0.000073$ & $0.005469$\\
Code-capacity $\M_{150}$ & $0.04216$ & $0.06333$ & $0.06343$ & $0.006445$ & $0.004028$ & $0.006543$\\
Code-capacity \GBeighteen & $0.04131$ & $0.04438$ & $0.04438$ & $0.007031$ & $0.007080$ & $0.007080$\\
Code-capacity \LPtwenty & $0.06807$ & $0.07605$ & $0.07605$ & $0.004370$ & $0.004419$ & $0.004419$\\
Phenomenological \BBgross & $0.03054$ & $0.04644$ & $0.04712$ & $0.002271$ & $0.000220$ & $0.002368$\\
Phenomenological \LPtwenty & $0.04607$ & $0.10203$ & $0.10203$ & $0.000952$ & $0.001782$ & $0.001782$\\
\bottomrule
\end{tabular}
\caption{Final-generation logical error rate, syndrome-inconsistency
  rate, and joint failure rate. Entries are mean probabilities over five
  independently trained models, each evaluated on the same $8192$
  realizations per task at the settings of Figure~\ref{fig:gen-progress}.
  Joint failure means that the predicted error has a wrong logical
  outcome, a nonzero residual syndrome, or both.}
\label{tab:factory-recovery}
\end{table}

\subsection{Data Improvement Across Generations}
\label{app:factory-data}

The model factory produces decoder models \Gmodel{0}--\Gmodel{3} and trajectory
datasets \Ddata{1}--\Ddata{3} through alternating collection and training.
Our hypothesis is that better models collect better training data,
which in turn train better models. We first examine how the difficulty
of the collected records and the models' ability to decode them evolve
together, then measure the training value of the records through
controlled comparisons.
The training-data and noise-adaptation controls below report the
logical error rate of \eqnref{eq:ler}. Recovery also requires syndrome
consistency, examined on fresh decoding realizations in
Section~\ref{app:factory-recovery}. The completion diagnostic below
examines the archived training states.

\paragraph{Models improve while datasets become more challenging.}
\begin{figure}[!htbp]
  \centering
  \includegraphics[width=\textwidth]{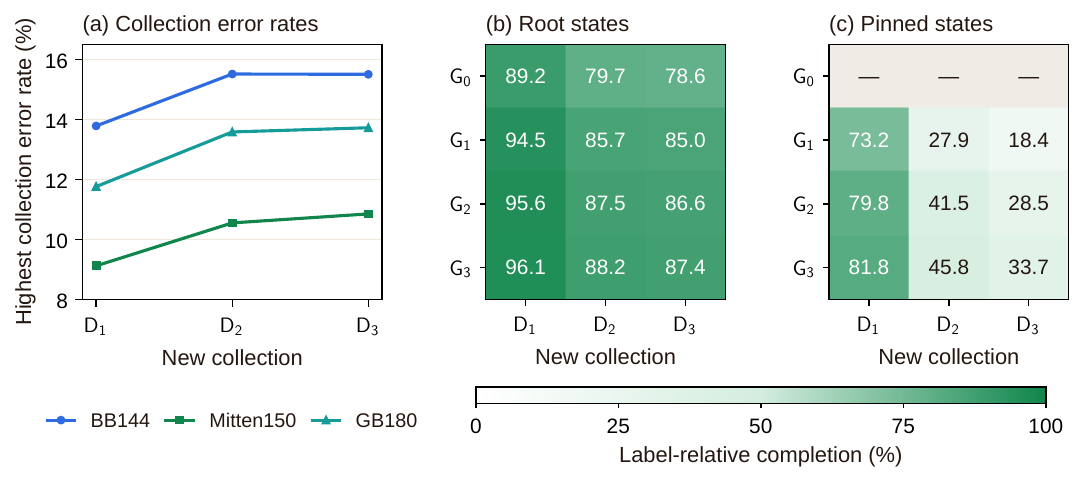}
  \caption{Later datasets challenge a fixed decoder, while later
    models achieve higher label-relative completion on fixed datasets.
    Figure~\ref{fig:factory-data-frontier}(a): the highest nominal
    error-rate setting used for collection.
    Figure~\ref{fig:factory-data-frontier}(b) and
    Figure~\ref{fig:factory-data-frontier}(c): completion relative to the recorded final label,
    the fraction of states on which the prediction matches that label's
    syndrome and logical outcome.
    Each cell pools up to $8192$ sampled records per target
    task; root and pinned states are sampled separately.
    Columns \Ddata{1}--\Ddata{3} are the new datasets defined in
    Appendix~\ref{app:factory-setup}. These are diagnostics on the archived
    training buffers. \Gmodel{0} has no pin-conditioned input and is examined
    only at root states.}
  \label{fig:factory-data-frontier}
\end{figure}

Figure~\ref{fig:factory-data-frontier} compares data difficulty and
decoder capability. In the heatmaps, each row denotes a decoder model
and each column a trajectory dataset. Each cell reports the fraction of
records from that dataset for which a single-pass prediction agrees
with the recorded label's syndrome and logical outcome. We call this
fraction the \emph{label-relative completion rate}. Root and pinned states are
evaluated separately. Reading across a row holds the model fixed and
compares data difficulty: along the \Gmodel{0} row, completion falls from \Ddata{1}
to \Ddata{3}, showing that later datasets challenge the initial model more.
Reading down a column holds the records fixed and compares decoder
capability: from \Gmodel{1} to \Gmodel{3}, later models complete more of the same
records, with the largest gains on pinned states. The models therefore
match more of the archived error-pattern labels, particularly from
partially pinned inputs.

Figure~\ref{fig:factory-data-frontier}(a) shows how collection adapts to the improving models.
The highest nominal error-rate setting used for collection is higher in \Ddata{3}
than in \Ddata{1} on all three tasks. As a model becomes stronger, raising
the collection error rate helps keep the sampled problems challenging
enough to require search, following Section~\ref{app:trajectory-collection}.
Figure~\ref{fig:factory-data-frontier}(a) thus shows the adjustment of collection difficulty,
while the \Gmodel{0} row directly confirms that the later records are harder
for a fixed decoder.

\paragraph{Updated collection procedures produce more valuable training data.}
We measure data value by how well a student trained on those data
decodes fresh errors. Think of the datasets as textbooks of worked
examples and the collection procedure as the process that compiles
them. The three comparisons examine the contributions of new data,
an updated collector, and increased collection difficulty. Each
comparison is repeated with three independent training seeds; students
in a pair start from the same \Gmodel{1} and take $20\,000$ updates
with the same optimizer and loss.

For paired comparisons, let $n_{+}$ count realizations decoded
incorrectly by student $A$ and correctly by student $B$, and let
$n_{-}$ count the reverse. We summarize these discordant outcomes by
\begin{equation}
  Z_{A\to B}=\frac{n_{+}-n_{-}}{\sqrt{n_{+}+n_{-}}}.
  \label{eq:matched-z}
\end{equation}
A positive score favors $B$. When outcomes are combined across training
runs, the score describes the paired evaluations of those models;
individual-seed results show how the improvement varies with training.

\begin{figure}[!htbp]
  \centering
  \includegraphics[width=\textwidth]{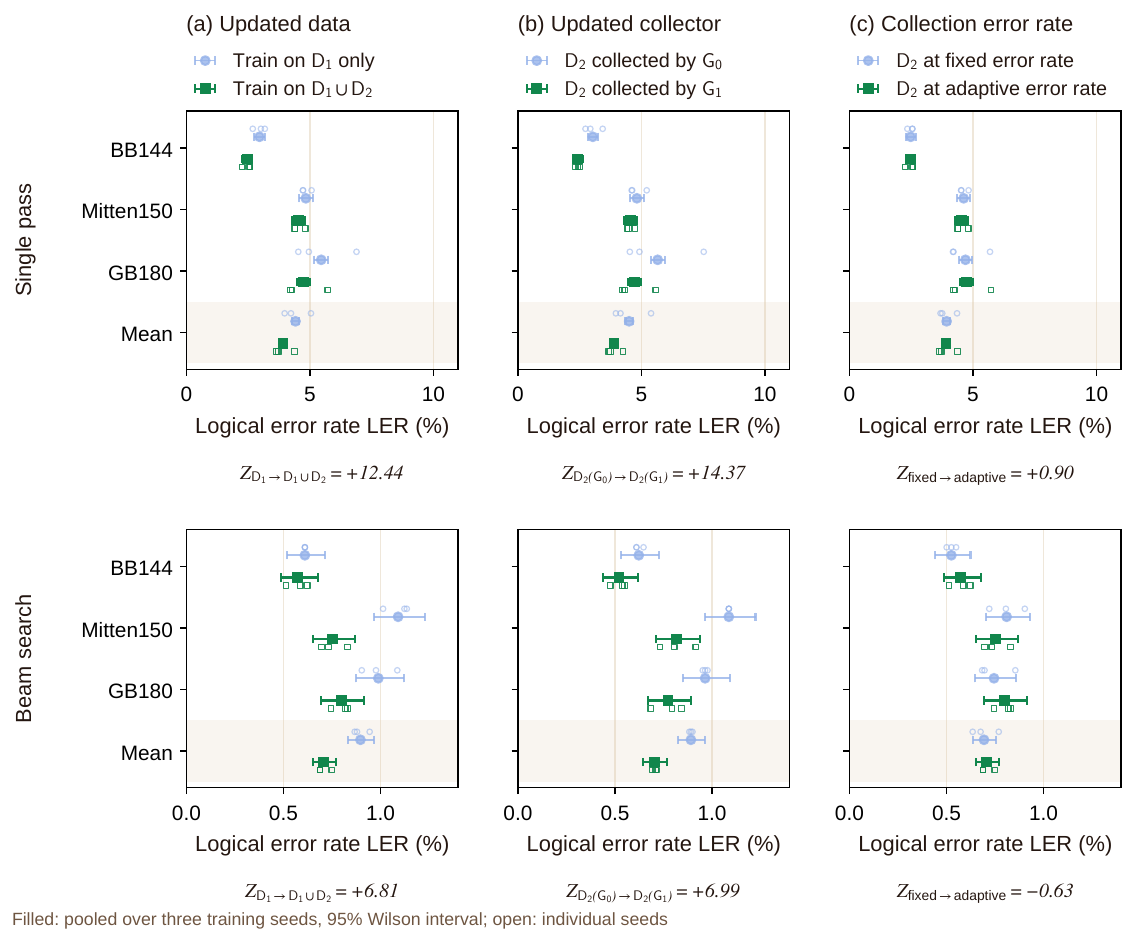}
  \caption{Logical error rate of the trained students, single pass
    (top) and beam search (bottom). Each panel pools three training
    seeds. Filled points pool the seeds; open points are the
    individual seeds. The $95\%$ Wilson intervals summarize pooled
    shot-level variation; open points show variation across training runs. Scores
    follow \eqnref{eq:matched-z} from the first legend entry to the
    second. In panel (c) both sets are \Ddata{2} collected by
    \Gmodel{1}, one at a fixed error rate and one at an adaptive
    error rate. An escape counts as a logical success.}
  \label{fig:factory-data-value}
\end{figure}

\begin{itemize}[leftmargin=*]
  \item \emph{Updated data.} If later data are better, they should
    train a better student. Both students start from \Gmodel{1} and take
    the same number of updates. One trains on \Ddata{1} only; the other,
    \Gmodel{2}, trains on $\Ddata{1}\cup\Ddata{2}$. The new data lower the logical error
    rate on every target task in a single pass
    (\(Z=+12.44\) on the equal-task mean) and on $\M_{150}$ and
    \GBeighteen\ under beam search. The \BBgross\ beam-search
    comparison is unresolved
    (Figure~\ref{fig:factory-data-value}(a)).
  \item \emph{Updated collection procedure.} How does updating collection
    affect training value? Both students are \Gmodel{2}, and both also
    train on the same \Ddata{1}. We compare \Ddata{2} collected by
    frozen \Gmodel{0} with \Ddata{2} collected by \Gmodel{1}. Collection uses
    the same noise distribution, error-rate settings and search budget.
    The two collections contain equal numbers of root and pinned records
    within each sampling stratum. The \Ddata{2} collected by \Gmodel{1}
    trains a decoder with a lower logical error rate on every target
    task in both modes, in all three runs
    (Figure~\ref{fig:factory-data-value}(b); single-pass
    \(Z=+14.37\)).
  \item \emph{Collection difficulty.} Does raising the collection
    error rate matter beyond who collects? Both students are \Gmodel{2}
    and both train on $\Ddata{1}\cup\Ddata{2}$ collected by \Gmodel{1}.
    One \Ddata{2} keeps the collection error-rate distribution used for
    \Ddata{1}; the other raises the collection error rates for the stronger
    collector. The comparison does not resolve a difference
    (single-pass \(Z=+0.90\); beam search \(Z=-0.63\);
    Figure~\ref{fig:factory-data-value}(c)).
\end{itemize}

In these experiments, later data and the updated collector provide
greater training value. Holding the collection difficulty fixed, rather
than adapting it, does not produce a resolvable difference at the
tested setting.

\subsection{Comparison with Classical Decoders}
\label{app:factory-classical}

\begin{figure}[!htbp]
  \centering
  \includegraphics[width=\textwidth]{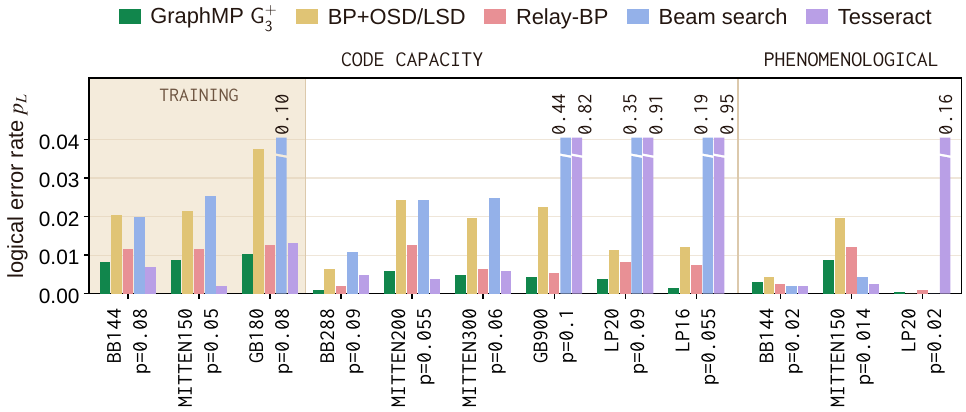}
  \caption{GraphMP \Gsearch{3} against classical decoders on
    $12$ decoding tasks at the nominal error-rate setting under each task; the same tasks and realizations as
    Figure~\ref{fig:sl-ler}, except for shorter Tesseract prefixes
    ($256$ realizations on code-capacity \GBnine, \LPtwenty\ and
    \LPsixteen, and $128$ on phenomenological \LPtwenty).
    The three shaded tasks form the model factory's
    target portfolio and supply its training data. The other tasks are
    unseen during training and evaluate transfer beyond that portfolio.
    Baseline configurations are defined in
    Section~\ref{app:bp-baselines}. Bars above $0.04$ are
    truncated.}
  \label{fig:eval-gen3}
\end{figure}

Figure~\ref{fig:eval-gen3} places GraphMP \Gsearch{3}, from the first
of the five model factory runs, next to BP+OSD
(BP+LSD on the phenomenological tasks), Relay-BP, the beam search
decoder of~\cite{ye2025}, and Tesseract~\cite{beni2025tesseract} on
$12$ decoding tasks, each at one nominal error-rate setting. The
comparisons use the same $2048$ error realizations per task, with
shorter Tesseract runs on the largest tasks.

On the three target tasks, the progression from \Gmodel{0} to \Gmodel{3} shows that the
model factory produces a decoder for the portfolio supplied to it. At \Gmodel{3} the
single-pass decoder has a lower logical error rate than the BP teacher
on most of the $12$ tasks, including unseen tasks that change the code
size, the code construction, or the noise model. These unseen tasks
measure an additional property of the product: the decoder transfers
beyond the target portfolio without further training. Beam search
further closes the gap between the single-pass decoder and the
classical decoders: the search-enhanced decoder \Gsearch{3} is broadly
competitive with Relay-BP and has a lower measured logical error rate
than BP+OSD on every code-capacity task shown. The ordering remains
task-dependent; Tesseract and BP-based beam search are stronger on
phenomenological $\M_{150}$. These comparisons concern the configured
budgets and the sampled realizations.

\paragraph{Statistical support for the benchmark comparison.}
Table~\ref{tab:portfolio-statistics} reports logical error rates and
pointwise $95\%$ Wilson intervals for the GraphMP and BP+OSD/LSD arms
of Figure~\ref{fig:eval-gen3}. Each decoder is evaluated on the same
$2048$ error realizations per task. The paired tests use the discordant
outcomes of the two decoders and a two-sided exact binomial test.
The comparison family consists of the $12$ taskwise GraphMP--BP+OSD/LSD
comparisons in the table; the reported $p$-values apply Holm's adjustment
within that family. These intervals describe evaluation-sampling
variation for the fixed models. Variation across independently trained
models is reported separately in Table~\ref{tab:generation-gains}.
All nine code-capacity comparisons favor GraphMP after adjustment.
The differences on phenomenological \BBgross\ and \LPtwenty\ remain
unresolved; this outcome establishes no equivalence between the decoders.

\begin{table}[!htbp]
\centering
\small

\begin{tabular}{lccc}
\toprule
Task & \shortstack{GraphMP \Gsearch{3}\\$p_L$ [$95\%$ CI]}
 & \shortstack{BP+OSD/LSD\\$p_L$ [$95\%$ CI]}
 & \shortstack{Holm-adjusted\\paired $p$}\\
\midrule
\multicolumn{4}{l}{Code-capacity tasks}\\
\BBgross & \shortstack{$0.0083$\\$[0.00519,0.0133]$} & \shortstack{$0.0205$\\$[0.0152,0.0276]$} & $7.65\times10^{-5}$\\
$\M_{150}$ & \shortstack{$0.00879$\\$[0.00557,0.0139]$} & \shortstack{$0.0215$\\$[0.016,0.0287]$} & $6.36\times10^{-4}$\\
\GBeighteen & \shortstack{$0.0103$\\$[0.00672,0.0156]$} & \shortstack{$0.0376$\\$[0.0302,0.0467]$} & $3.33\times10^{-16}$\\
\BBtwogross & \shortstack{$0.000977$\\$[0.000268,0.00355]$} & \shortstack{$0.00635$\\$[0.00371,0.0108]$} & $0.0047$\\
$\M_{200}$ & \shortstack{$0.00586$\\$[0.00335,0.0102]$} & \shortstack{$0.0244$\\$[0.0186,0.032]$} & $1.62\times10^{-8}$\\
$\M_{300}$ & \shortstack{$0.00488$\\$[0.00265,0.00897]$} & \shortstack{$0.0195$\\$[0.0144,0.0265]$} & $6.24\times10^{-7}$\\
\GBnine & \shortstack{$0.00439$\\$[0.00231,0.00833]$} & \shortstack{$0.0225$\\$[0.0169,0.0298]$} & $8.62\times10^{-9}$\\
\LPtwenty & \shortstack{$0.00391$\\$[0.00198,0.00769]$} & \shortstack{$0.0112$\\$[0.0075,0.0168]$} & $0.0047$\\
\LPsixteen & \shortstack{$0.00146$\\$[0.000498,0.0043]$} & \shortstack{$0.0122$\\$[0.00828,0.018]$} & $2.38\times10^{-5}$\\
\midrule
\multicolumn{4}{l}{Phenomenological tasks}\\
\BBgross & \shortstack{$0.00293$\\$[0.00134,0.00638]$} & \shortstack{$0.00439$\\$[0.00231,0.00833]$} & $1$\\
$\M_{150}$ & \shortstack{$0.00879$\\$[0.00557,0.0139]$} & \shortstack{$0.0195$\\$[0.0144,0.0265]$} & $0.0047$\\
\LPtwenty & \shortstack{$0.000488$\\$[0.0000862,0.00276]$} & \shortstack{$0$\\$[0,0.00187]$} & $1$\\
\bottomrule
\end{tabular}
\caption{Statistical support for the fixed-model benchmark in
  Figure~\ref{fig:eval-gen3}. Entries give logical error probabilities
  and pointwise $95\%$ Wilson intervals, including an upper bound
  when no logical error is observed. BP+OSD is used on the code-capacity
  tasks and BP+LSD on the phenomenological tasks. Paired $p$-values are
  adjusted across the $12$ comparisons shown; this family excludes
  the other decoder arms.}
\label{tab:portfolio-statistics}
\end{table}

For the search-enhanced GraphMP model in this same benchmark,
logical error and joint failure rates coincide on all nine code-capacity
tasks and on phenomenological \BBgross. On phenomenological $\M_{150}$,
the logical error rate is $0.00879$ and the joint failure rate is
$0.00928$; on phenomenological \LPtwenty, they are $0.000488$ and
$0.00244$, respectively. These portfolio checks complement the
five-run recovery diagnostics in Table~\ref{tab:factory-recovery}.

\subsection{Logical Error Rate}
\label{app:factory-curves}

\begin{figure}[!htbp]
  \centering
  \includegraphics[width=\textwidth]{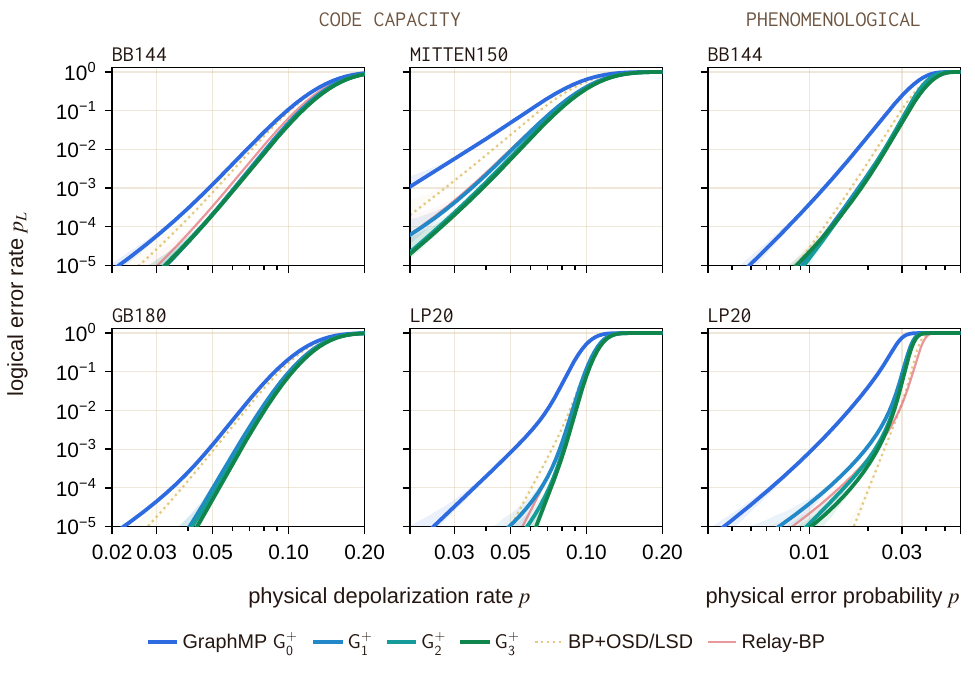}
  \caption{Logical error rate against physical error rate on the six
    decoding tasks of Figure~\ref{fig:gen-progress}, for GraphMP
    \Gsearch{0} to \Gsearch{3},
    BP+OSD (BP+LSD on phenomenological tasks), and Relay-BP. Curves are
    fitted to the measured points; bands are pointwise $90\%$
    parametric-bootstrap intervals conditional on the fit family.}
  \label{fig:ler-curves}
\end{figure}

Figure~\ref{fig:ler-curves} extends the comparison from one error rate
per task to the full curve of logical error rate against physical
error rate, for GraphMP \Gsearch{0} to \Gsearch{3} together with BP+OSD and
Relay-BP. The GraphMP curves evaluate one of the five model factory runs over
a physical-error grid on each task.
On \GBeighteen\ the grid extends down to $p=0.02$, and the lowest
points of \Gsearch{0} are sampled with up to $4\times10^{6}$ shots
each, so that its curve rests on measured points at the low end.
The curves show the main improvement from \Gsearch{0} to
the later generations, which approach Relay-BP over the measured
range. Differences among the later generations are smaller and
depend on the task and physical error rate. The independent runs in
Section~\ref{app:generations} show how generation gains vary with training.
The curves are fits of a smooth model to the measured points; below
the lowest measured rate they are extrapolations, and the bands
show the uncertainty of the fit, which widens where logical errors
become rare.

The bands are calculated from $200$ bootstrap parameter draws
for each curve. They quantify parameter uncertainty within the
chosen fit family and omit model discrepancy. In particular, the
code-capacity \LPtwenty\ BP+OSD fit is flagged for lack of fit by
a goodness-of-fit test ($p=0.0398$). Its smooth curve and
band should therefore be read alongside the measured rates; the
band alone gives an incomplete account of uncertainty for that
comparison. Figure~\ref{fig:main-cost-curves} restricts its fitted
curves to the measured range.

\subsection{Decoding Time}
\label{app:factory-time}

\begin{figure}[!htbp]
  \centering
  \includegraphics[width=\textwidth]{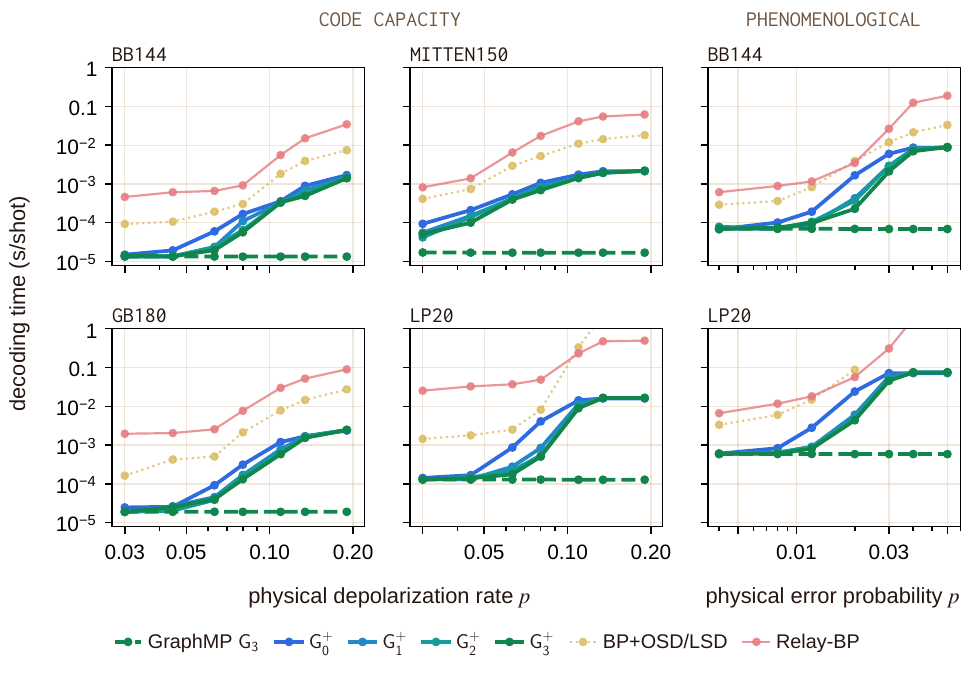}
  \caption{Decoding time per shot on the six tasks of
    Figure~\ref{fig:ler-curves}. GraphMP runs on one GPU with the
    inference implementation described in the text
    (single-pass \Gmodel{3} dashed, search-enhanced \Gsearch{0} to \Gsearch{3} solid,
    beam search in decoding mode); BP+OSD, BP+LSD, and Relay-BP run
    on one CPU core with one thread. Points above $1$~s per shot are
    clipped, and BP+LSD is omitted on \LPtwenty\ above $p=0.03$,
    where it is excluded from the accuracy evaluation as well.}
  \label{fig:latency}
\end{figure}

Figure~\ref{fig:latency} reports the decoding time
of Appendix~\ref{app:search} on the six tasks of
Figure~\ref{fig:ler-curves}, for the single-pass decoder \Gmodel{3}, the
search-enhanced decoders \Gsearch{0} to \Gsearch{3}, and the classical
decoders.

\paragraph{Measurement.} GraphMP decodes on one NVIDIA B300 GPU dedicated to that
measurement, running PyTorch~2.10.0 with CUDA~13.0. The classical decoders
run single-threaded on one CPU core. Decoding time is the incremental
wall time per shot in a stream of independently sampled syndromes,
including transfer of syndromes from host memory and return of the
predicted errors to host memory. After warm-up, we time streams of
different lengths and estimate the additional time per shot, separating
it from fixed call overhead. The shortest of three repetitions is
retained. Above the pseudo-threshold, selected points use one timed
repetition. All decoders receive the same sampled syndromes for each
task and physical error rate. These measurements characterize the
sustained cost of decoding a stream; model training and GPU compilation
precede the timed calls.

\paragraph{GPU inference.} GraphMP projects node states before edge
aggregation, sums incoming messages in a fixed order, and executes its
message-passing rounds through compiled kernels captured in a CUDA
execution graph. Matrix products use TF32 tensor-core arithmetic;
states, decisions, syndrome checks, and candidate scores retain single
or higher precision. Beam search stores and processes the shots whose
single-pass prediction leaves a residual syndrome, synchronizing with
the host once per search step.
The accuracy evaluations use single-precision matrix products.
Paired numerical checks on the six nominal evaluation settings give
mean logical error rates within $0.011$ percentage points of those
accuracy evaluations, over five runs and four generations in either
decoding mode.

\paragraph{Results.}

\begin{itemize}[leftmargin=*]
  \item \emph{Single pass.} Single-pass inference runs a fixed sequence
    of operations, so its decoding time is approximately independent
    of physical error rate. GraphMP has $10{,}753$
    model parameters, small compared with many learned
    decoders~\cite{bausch2024,astra2024,hu2025,park2025,gu2026,zenati2026},
    so a forward pass stays cheap.
  \item \emph{Search.} Beam search costs more, because it calls the
    model repeatedly. The extra cost grows with the physical error
    rate: when the physical error rate is high, the single-pass
    predicted error more often leaves a residual syndrome, so more
    shots enter beam search and the search goes deeper. From \Gmodel{0} to
    \Gmodel{3}, the decoder model factory strengthens the single-pass
    decoder. The stronger model is a better guide for beam search, so
    the search-enhanced decoder is more accurate
    (Figure~\ref{fig:gen-progress}); it also leaves a residual
    syndrome on fewer shots, so fewer shots enter the search and the
    search-enhanced decoder is faster.
  \item \emph{Classical decoders.} On all six tasks, search-enhanced
    \Gsearch{3} takes less time per shot than Relay-BP, BP+OSD, and
    BP+LSD at every measured point. The margin depends on the task
    and the physical error rate. The classical decoders run on
    one CPU core and GraphMP on one GPU; the curves describe the
    accuracy and decoding cost of these CPU and GPU configurations. At high
    physical error rates, search-enhanced and BP-based decoders slow
    down: almost every shot enters beam search, and belief propagation
    exhausts its iteration budget on almost every shot.
\end{itemize}

\subsection{Adaptation to Noise Model Drift}
\label{app:factory-adapt}

The preceding experiments assume that the noise model remains fixed.
On a quantum processor, however, the underlying noise model may drift
away from the distribution assumed when a decoder was configured or
trained. Classical decoders represent this assumption explicitly as a
prior distribution over error variables and use it to weight candidate
explanations of a syndrome. GraphMP learns a decoding strategy from
samples of the training noise distribution; its predictions depend on
the decoding graph, syndrome, and trained model parameters. This
difference raises a practical question: can continued model factory training
adapt GraphMP after the underlying noise model changes?

\paragraph{Experiment.}
At \Gmodel{3} we replace the depolarizing noise model, with independent $X$,
$Y$, and $Z$ sources of probability $p/3$ each, by an XZ noise model,
with independent $X$ and $Z$ sources of probability $p/2$ each.
The net nonidentity Pauli probability changes from $p-p^2/3$ to
$p-p^2/4$; simultaneous $X$ and $Z$ sources can still produce a $Y$
Pauli. This controlled shift preserves the nominal error-rate setting
and decoding graph.
This shift tests adaptation to a specified change in noise. GraphMP
continues to receive only the decoding
graph and
syndrome. The classical decoders, and the candidate selector used when
beam search is evaluated as a decoder, retain the depolarizing prior;
we also evaluate the classical decoders with an updated prior as an
informed reference. Starting from \Gmodel{3}, the model factory collects successful
trajectories from the XZ noise model and trains \Gmodel{4}, \Gmodel{5}, and \Gmodel{6} in
three successive rounds. In each round, successful trajectories are
collected on the three target tasks and used to train the next
generation. Collection accepts a leaf whose final prediction is
syndrome-consistent and logically correct, and retains the selected
path's root and ancestor states. In decoding
mode, candidate selection continues to use the original DEM prior.
\LPtwenty\ undergoes the same noise-model shift but
contributes no training records. We evaluate it only after training to
test whether the adaptation learned on the target portfolio
transfers to an unseen decoding task. All decoders are evaluated on
the same $8192$ realizations
per task at the nominal error-rate settings of Figure~\ref{fig:gen-progress}.

\begin{figure}[!htbp]
  \centering
  \includegraphics[width=\textwidth]{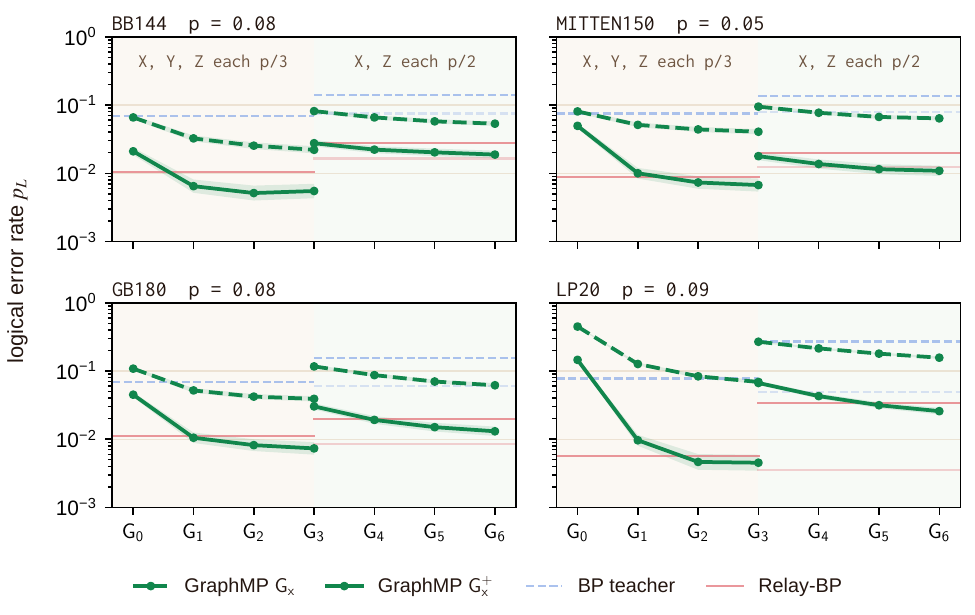}
  \caption{Generation progress after the noise model changes from
    the depolarizing noise model to the XZ noise model at \Gmodel{3}. The
    nominal error-rate setting and decoding graph remain fixed. Dashed: single-pass
    decoder GraphMP \Gmodel{x}; solid: search-enhanced decoder GraphMP
    \Gsearch{x}; bands are $90\%$ Wilson intervals on $8192$ realizations
    per task. The BP teacher and Relay-BP step at \Gmodel{3} to their logical
    error rate under the original prior; the faint lines give their
    rate under the updated prior. The model factory collects on the first
    three target tasks only; \LPtwenty\ is unseen during adaptation.}
  \label{fig:adapt}
\end{figure}

\paragraph{Results.}
The noise-model shift initially raises the logical error rate of
every fixed decoder. The classical decoders retain a prior that no
longer matches the XZ noise model. GraphMP \Gmodel{3} receives no prior, but
its training data were generated under the depolarizing noise model.
Continued model factory training reverses this loss. Every
post-shift generation lowers the logical error rate on every target
task, in both modes, on all three runs of this recipe. \Gmodel{4} is
the largest step, and the later steps are smaller. By \Gmodel{6},
single-pass GraphMP is below the BP teacher with the updated prior on
\BBgross\ and $\M_{150}$, and close to it on \GBeighteen.
\Gsearch{4} reaches stale-prior Relay-BP on the three target tasks.
\Gsearch{6} reaches updated-prior Relay-BP on $\M_{150}$ only.
The adaptation also transfers to the unseen task on \LPtwenty:
from \Gmodel{3} to \Gmodel{6} the single-pass logical error rate falls from
$26.9\%$ to $15.7\%$, and the search-enhanced rate falls from $6.7\%$ to
$2.6\%$. This is below stale-prior Relay-BP at $3.4\%$, although
updated-prior Relay-BP remains lower at $0.35\%$. The unseen task
therefore shows that the adaptation transfers beyond the target
portfolio, while leaving a gap to Relay-BP supplied with the updated
prior.

\paragraph{Which data enable adaptation?}
To isolate the contribution of the changed noise distribution, we
continue from the same \Gmodel{3} under two fresh collections:
depolarizing trajectories and XZ trajectories. Both collections use
that model as the collector, with the same search budget, acceptance
rule and collection error-rate distribution. Their root and pinned
records are matched in number within each sampling stratum. Both
students take $20\,000$ updates with the same optimizer and loss.
The comparison is repeated with three independent training seeds,
using the final checkpoints. Evaluation covers the changed noise
model and retention on the original one.

\begin{figure}[!htbp]
  \centering
  \includegraphics[width=\textwidth]{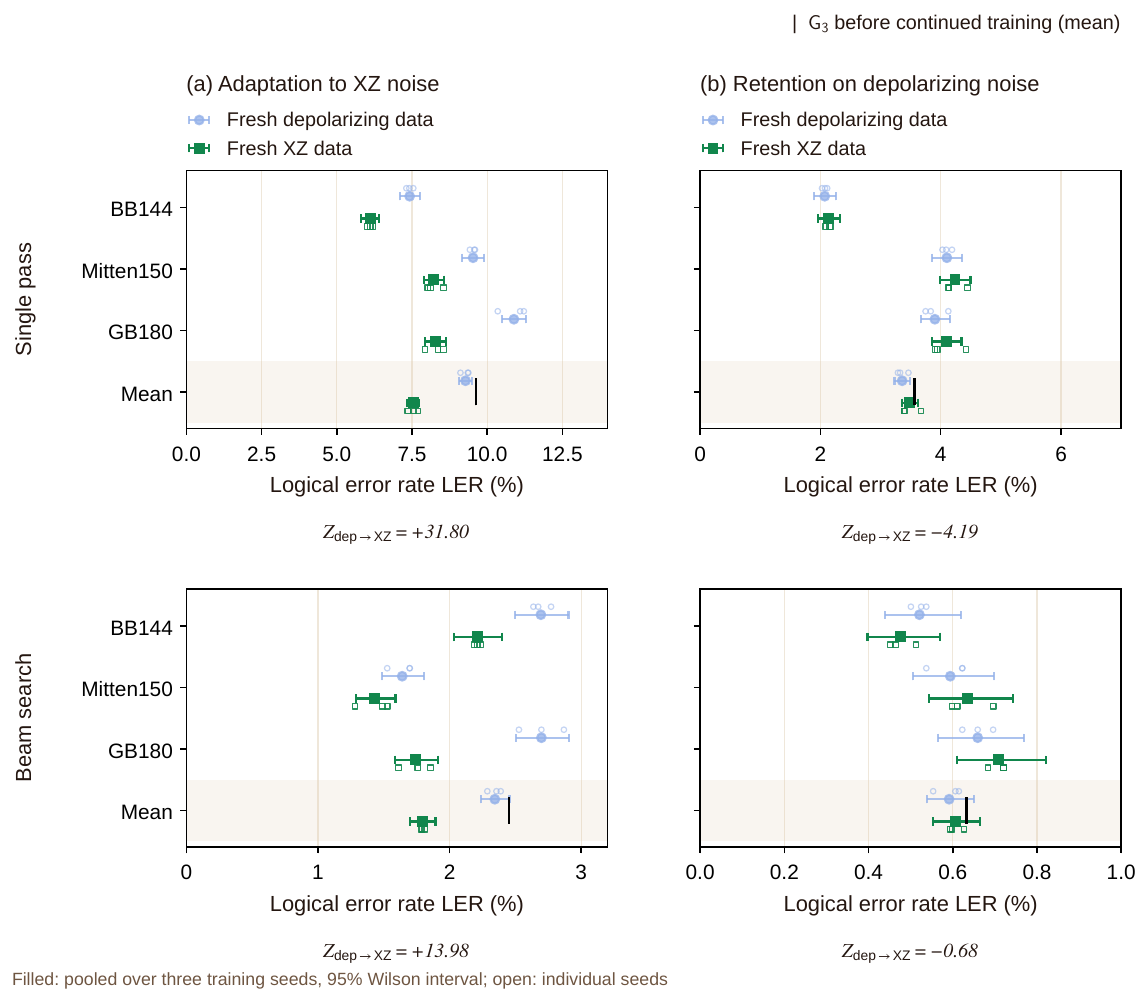}
  \caption{Training data for noise adaptation. Starting from the
    same \Gmodel{3} collector, equal-budget continuations train on fresh
    depolarizing data or fresh XZ data. Logical error rate is shown
    for a single pass (top) and for beam search (bottom), on the same
    $8192$ realizations per task under each noise model, averaged
    over three training seeds. Wilson intervals summarize the pooled
    shot-level variation; the mean weights the three target tasks equally. Each panel reports \(Z_{A\to B}\) of
    \eqnref{eq:matched-z} from the depolarizing-data student to the
    XZ-data student. The unmodified \Gmodel{3} mean is marked for reference.
    An escape counts as a logical success.}
  \label{fig:factory-noise-data}
\end{figure}

Figure~\ref{fig:factory-noise-data} shows that XZ-noise training data
improve decoding on every target task relative to equally many fresh
depolarizing trajectories
(\(Z_{\mathrm{dep}\to\mathrm{XZ}}=+31.80\) in a single pass, and
$+13.98$ under beam search).
The equal-task mean logical error rate falls by $18.8\%$
in the single-pass comparison and by $23.4\%$ under beam search.
This adaptation has a retention cost on the single pass: under the
original depolarizing noise, the student trained on XZ-noise data has a higher
logical error rate
(\(Z_{\mathrm{dep}\to\mathrm{XZ}}=-4.19\)); the corresponding
beam-search comparison is unresolved
(\(Z=-0.68\)). Thus the data's training value
depends on the noise environment. Both datasets are fresh collections
from the same \Gmodel{3} collector. Logical outcomes are supplied by
simulation here; obtaining equivalent feedback from dedicated device
experiments remains the hardware extension described in
Appendix~\ref{app:trajectory-collection}.

\clearpage
{\raggedright
\putbib[refs]
}
\end{bibunit}

\end{document}